\documentclass[aps,prl,twocolumn,superscriptaddress,a4paper]{revtex4-1}
\usepackage[dvipdfmx]{graphicx}

\usepackage{amsmath,amsthm,amssymb}
\usepackage{amsmath,bm}
\usepackage{color}
\usepackage{ifthen}
\usepackage{mathrsfs}

\usepackage[svgnames,psnames]{xcolor}

\usepackage{multirow,bigdelim}
\newcommand{\1}{\mbox{1}\hspace{-0.25em}\mbox{l}}

\def\rrangle{\rangle\!\rangle}
\def\llangle{\langle\!\langle}

\def\spopf#1{\mathscr{#1}}
\def\matf#1{\mathcal{#1}}

\newcommand{\mytitle}{
Non-Hermitian Mott Skin Effect
}
\newlength{\figwidth}
\newlength{\figlarge}
\begin{document}
%%%%%%%%%%%%%%%%%%%%%%%%%%%%%%%%%%%%%%%%%%%%%%%%%%%%%%%%%%%%%%%%%%%%%%%
\title{
\mytitle
}
%%%%%%%%%%%%%%%%%%%%%%%%%%%%%%%%%%%%%%%%%%%%%%%%%%%%%%%%%%%%%%%%%%%%%%%
\author{
Tsuneya Yoshida
}
\affiliation{%
Department of Physics, Kyoto University, Kyoto 606-8502, Japan
}%
\author{
Song-Bo Zhang
}
\affiliation{%
International Center for Quantum Design of Functional Materials (ICQD), Hefei National Research Center for Physical Sciences at the Microscale, University of Science and Technology of China, Hefei, Anhui 230026, China
}%
\author{
Titus Neupert
}
\affiliation{%
Department of Physics, University of Z\"urich, Winterthurerstrasse 190, 8057, Z\"urich, Switzerland
}%
\author{
Norio Kawakami
}
\affiliation{%
Department of Physics, Kyoto University, Kyoto 606-8502, Japan
}%
\affiliation{%
Department of Materials Engineering Science, Osaka University, Toyonaka 560-8531, Japan
}%
\affiliation{%
Department of Physics, Ritsumeikan University, Kusatsu, Shiga 525-8577, Japan
}%

%%%%%%%%%%%%%%%%%%%%
%%%%%%%%%%%%%%%%%%%%%%%%%%%%%%%%%%%%%%%%%%%%%%%%%%%%%%%%%%%%%%%%%%%%%%%
\date{\today}
%%%%%%%%%%%%%%%%%%%%%%%%%%%%%%%%%%%%%%%%%%%%%%%%%%%%%%%%%%%%%%%%%%%%%%%
\begin{abstract}
We propose a novel type of skin effects in non-Hermitian quantum many-body systems which we dub a non-Hermitian Mott skin effect.
This phenomenon is induced by the interplay between strong correlations and the non-Hermitian point-gap topology.
The Mott skin effect induces extreme sensitivity to the boundary conditions only in the spin degree of freedom (i.e., the charge distribution is not sensitive to boundary conditions), which is in sharp contrast to the ordinary non-Hermitian skin effect in non-interacting systems.
Concretely, we elucidate that a bosonic non-Hermitian chain exhibits the Mott skin effect in the strongly correlated regime by closely examining an effective Hamiltonian. 
The emergence of the Mott skin effect is also supported by numerical diagonalization of the bosonic chain. 
The difference between the ordinary non-Hermitian skin effect and the Mott skin effect is also reflected in the time-evolution of physical quantities; under the time-evolution spin accumulation is observed while the charge distribution remains spatially uniform.
\end{abstract}
%***
%} 
%%%%%%%%%%%%%%%%%%%%%%%%%%%%%%%%%%%%%%%%%%%%%%%%%%%%%%%%%%%%%%%%%%%%%%%
%%%%%%%%%%%%%%%%%%%%%%%%%%%%%%%%%%%%%%%%%%%%%%%%%%%%%%%%%%%%%%%%%%%%%%%
\maketitle
%%%%%%%%%%%%%%%%%%%%%%%%%%%%%%%%%%%%%%%%%%%%%%%%%%%%%%%%%%%%%%%%%%%%%%%

%%%%%%
\textit{Introduction--.}
%%%%%%
%
Since the discovery of topological insulators, extensive efforts have been devoted to understanding topological aspects of condensed matter systems~\cite{TI_review_Hasan10,TI_review_Qi10,Thouless_PRL1982,Halperin_PRB82,Hatsugai_PRL93,Kitaev_chain_01,Kane_2DZ2_PRL05,Kane_Z2TI_PRL05_2,Qi_TFT_PRB08}. 
While topological insulators are originally reported for free fermions, it has turned out that the interplay between strong correlations and non-trivial topology triggers further exotic phenomena~\cite{Tsui_FQHEExp_PRL82,Laughlin_FQHE_PRL83,Jain_FQHE_PRL89,Haldane_FQHEpesudo_PRL83,Haldane_TopoDeg_PRL85,Wen_TopoOrder_SciAdv95,Kitaev_ToricCode_Elsevier03,Wen_Textbook04,Levin_LevinWenModel_PRB05,Kitaev_ToricCode_AnnPhys06,Kitaev_KitaevHoney_AnnPhys06,Tang_FChern_PRL11,Sun_FChern_PRL11,Neupert_FChern_PRL11,Sheng_FChern_NComm12,Regnalt_FChen_PRX11,Bergholtz_FChern_IntJModPhysB13,Z_to_Zn_Fidkowski_PRB10,Turner_ZtoZ8_PRB11,Fidkowski_1Dclassificatin_PRB11,gu_supercohomology,YaoRyu_Z_to_Z8_2013,Ryu_Z_to_Z8_2013,Qi_Z_to_Z8_2013}. 
For instance, strong correlations can induce fractional quantum Hall states~\cite{Tsui_FQHEExp_PRL82,Laughlin_FQHE_PRL83,Wen_TopoOrder_SciAdv95,Tang_FChern_PRL11,Sun_FChern_PRL11,Neupert_FChern_PRL11,Sheng_FChern_NComm12,Regnalt_FChen_PRX11,Bergholtz_FChern_IntJModPhysB13}. 
Furthermore, topological Mott insulators exhibit the unique bulk-edge correspondence due to the interplay between correlations and the non-trivial topology~\cite{Pesin_TMI_NatPhys2010,Manmana_Chiral1D_PRB12,Yoshida_TMI1D_PRL14}. 
Namely, corresponding to the non-trivial topology in the bulk, gapless edge modes emerge only in the spin excitation spectrum (i.e., the charge excitation spectrum is gapped even around edges).

Along with the above progress, the topological band theory of non-Hermitian systems has been developed~\cite{Bergholtz_Review19,Ashida_nHReview_AdvPhys20,Gong_class_PRX18,KKawabata_TopoUni_NatComm19,Kawabata_gapped_PRX19,Zhou_gapped_class_PRB19,Lieu_Liouclass_PRL20,Hatano_PRL96,Hatano_PRB97,CMBender_PRL98,Fukui_nH_PRB98,Ashida_nHbHubb_PRA16,Hu_nH_PRB11,Esaki_nH_PRB11,Sato_nHPTEP12,Diehl_DissCher_NatPhys11,Bardyn_DissCher_NJP2013,Rivas_DissCher_PRB13,Zhu_nHPT_PRA14,Budich_DissCher_PRA15,Lieu_nHSSH_PRB2018,Rui_nH_PRB19,Denner_ETI_NatComm21,Nakamura_BBCptG_arXiv22,Yokomizo_BBC_PRL19,kawabata_NBlochBBC_PRB20,Tonielli_nHTQFT_PRL20,Kawabata_TQFTSkin_PRL20,Sayyad_nHTQFT_PRR22,Chang_nHEntSpec_PRR,Hsieh_nHCritical_SciPost23}
 and revealed unique phenomena due to the non-Hermitian point-gap topology which do not have Hermitian counterparts~\cite{HShen2017_non-Hermi,VKozii_nH_arXiv17,Yoshida_EP_DMFT_PRB18,Zhou_ObEP_Science18,Yang_Descri_PRL21,Budich_SPERs_PRB19,Okugawa_SPERs_PRB19,Zhou_SPERs_Optica19,Yoshida_SPERs_PRB19,Yoshida_nHReview_PTEP20,Kawabata_gapless_PRL19,Carlstrom_nHknot_PRB19,YXuPRL17_exceptional_ring,Delplace_Resul_PRL21,Mandal_EP3_PRL21}.
A representative example is the non-Hermitian skin effect~\cite{TELeePRL16_Half_quantized,Alvarez_nHSkin_PRB18,SYao_nHSkin-1D_PRL18,SYao_nHSkin-2D_PRL18,KFlore_nHSkin_PRL18,EEdvardsson_PRBnHSkinHOTI_PRB19,Borgnia_ptGapPRL2020,Lee_Skin19,Zhang_BECskin19,Okuma_BECskin19,Okuma_nHSkinReview_AnnRev23,Xiao_nHSkin_Exp_NatPhys19,Helbig_ExpSkin_19,Liang_nHSkinColdAtom_PRL22,Yoshida_MSkinPRR20,Okugawa_HOSkin_PRB20,Kawabata_HOSkin_PRB20,Fu_HOSkin_arXiv2020,Okuma_BECpg_PRL19,Song_LSkin_PRL19,Haga_LSkin_PRL21,Yang_LSkin_PRR22,Hwang_SkinJunction_arXiv23} 
which results in the novel bulk-edge correspondence unique to non-Hermitian systems; because of the non-trivial point-gap topology in the bulk, the eigenstates and eigenvalues exhibit extreme sensitivity to the presence or absence of boundaries~\cite{Zhang_BECskin19,Okuma_BECskin19}. 
Especially, in one-dimensional systems, most of eigenstates are localized only around one of the edges under open boundary conditions (OBC), while they extend to the bulk under periodic boundary conditions (PBC).
The above localized eigenstates are known as skin modes. 

The above progress on correlated systems and the non-interacting non-Hermitian topology 
naturally leads us to the following crucial question: \textit{how do strong correlations affect the non-Hermitian skin effects?}
The significance of this issue is further enhanced by resent advances in experiments with cold atoms~\cite{Tomita_Zeno_SciAdv17,Tomita_2BdyLoss_PRA19,Takasu_nHPTcoldAtom_PTEP2020} and quantum circuits~\cite{Ma_LossQuantumCircuits_Nature2019} where both dissipation and correlations can be introduced. 
Correlation effects on the non-Hermitian topological properties have been studied extensively~\cite{Yoshida_nHFQH19,Yoshida_nHFQHJ_PRR20,Guo_nHToric_PRB20,Matsumoto_nHToric_PRL20,Zhang_nHToric_Natcomm20,Guo_nHToric_EPL2020,Shackleton_nHFracton_PRR20,Yang_EPKitaev_PRL21,Wang_SteadyToricCode_arXiv23,Zhang_nHTMI_PRB20,Liu_nHTMI_RPB20,Xu_nHBM_PRB20,Pan_PTHubb_oQS_PRA20,Xi_nHcohomology_SciBull21,Yoshida_PtGpZtoZ2_PRB21,Yoshida_reduction1Dptgp_PRB22,Yoshida_reductionEP_PRB23,Orito_CorrSkin_PRB22,Shen_CorrSkin_CommPhys22,Mu_MbdySkin_PRB20,Lee_MbdySkin_PRB21,Zhang_CorrSkin_PRB22,Kawabata_CorrSkin_PRB22,Alsallom_CorrSkin_PRR22,Faugno_corrnHSkin_PRL22,Hamanaka_InteractionLSE_arXiv23,Tsubota_CorrInv_PRB22,Sayyad_corr1DKitaev_PRR2023,Chen_nHSpinChainPRL2023,Sayyad_nHChiral_arXiv2023,Sayyad_Transfe_arXiv2023}, but not on non-Hermitian skin effects, which is particularly the case for bosonic systems.

In this paper, we address this question for a one-dimensional bosonic system and discover a novel type of skin effects, a \textit{non-Hermitian Mott skin effect}, which induces striking skin modes. 
Namely, the non-trivial point-gap topology results in the skin modes in which only the spin degree of freedom is involved (i.e., charges are distributed uniformly even under OBC). 
This behavior is in sharp contrast to non-Hermitian skin effects in non-interacting systems where bosons are localized around the edge. 
We elucidate the emergence of the Mott skin effect by examining an effective spin model in the strong correlation regime, which hosts skin modes and possesses the non-trivial point-gap topology characterized by the spin winding number. We also support the emergence of the Mott skin effect by employing numerical diagonalization. 
Our numerical analysis also elucidates unique real-time dynamics induced by the Mott skin effect; dynamical spin accumulation is observed while the charge distribution remains spatially uniform.

%%%%%%
\textit{Model--.}
%%%%%%
Let us consider a one-dimensional chain of interacting bosons. The Hamiltonian reads
%%%%%%
\begin{eqnarray}
\label{eq: H}
\hat{H}   &=& \hat{H}_0 +\hat{H}_{\mathrm{int}}, \\
\label{eq: H_0}
\hat{H}_0 &=& \sum_{j\sigma} ( -t_{\mathrm{R} \sigma} \hat{b}^\dagger_{j+1\sigma} \hat{b}_{j\sigma} - t_{\mathrm{L}\sigma} \hat{b}^\dagger_{j\sigma} \hat{b}_{j+1\sigma} ), \\
\label{eq: H_int}
\hat{H}_{\mathrm{int}} &=& -iV \sum_{j\sigma} \hat{n}_{j\sigma}(\hat{n}_{j\sigma}-1) - iU \sum_{j} \hat{n}_{j\uparrow} \hat{n}_{j\downarrow},
\end{eqnarray}
%%%%%%
where $\hat{b}^\dagger_{j\sigma}$ ($\hat{b}_{j\sigma}$) creates (annihilates) a boson at site $j=0,1,2\ldots,L-1$ and in the spin state $\sigma=\uparrow,\downarrow$~\cite{pseudo-spin_ftnt}.
The non-reciprocal hopping integrals are denoted by $t_{\mathrm{R}\uparrow}=t_{\mathrm{L}\downarrow}=t_+$ and $t_{\mathrm{L}\uparrow}=t_{\mathrm{R}\downarrow}=t_-$ with real numbers $t_+$ and $t_-$.
The operator $\hat{n}_{j\sigma}$ denotes the number operator of bosons of spin $\sigma$ at site $j$; $\hat{n}_{j\sigma}:=\hat{b}^\dagger_{j\sigma}\hat{b}_{j\sigma}$.
The first (second) term of $\hat{H}_{\mathrm{int}}$ describes the on-site interaction between bosons with the same (opposite) spin.
The parameters $V$ and $U$ are non-negative numbers.
Without loss of generality, we assume the relation $t_+>t_-$ throughout this paper.

%%%%%%
\textit{Symmetry and a topological invariant--.}
%%%%%%
The above model preserves charge $\mathrm{U}(1)$ symmetry and spin $\mathrm{U}(1)$ symmetry, as indicated by
%%%%%%
\begin{eqnarray}
\label{eq: charge U1}
{} [\hat{H},\hat{N}]&=&0,\\
\label{eq: spin U1}
{} [\hat{H},\hat{S}^z]&=&0,
\end{eqnarray}
%%%%%%
with $\hat{N}=\hat{N}_\uparrow+\hat{N}_\downarrow$ and $2\hat{S}^z=\hat{N}_\uparrow-\hat{N}_\downarrow$.
Here, $\hat{N}_\sigma$ denotes the number operator of bosons in the spin state $\sigma$; $\hat{N}_\sigma=\sum_j \hat{n}_{j\sigma}$ ($\sigma=\uparrow,\downarrow$).
Therefore, the Hamiltonian can be block-diagonalized into $N_\uparrow$ and $N_\downarrow$ sectors. 
We denote eigenvalues of $\hat{N}_\uparrow$ and $\hat{N}_\downarrow$ as $N_\uparrow$ and $N_\downarrow$, respectively.

In order to characterize the point-gap topology~\cite{DefsOfptGap_ftnt}, we introduce the many-body spin winding number
%%%%%%
\begin{eqnarray}
\label{eq: Ws}
W_{\mathrm{s}} &=& \int^{2\pi}_0 \frac{d\theta_{\mathrm{s}}}{2\pi i}  \frac{\partial}{\partial \theta_{\mathrm{s}}}  \log \left[ \mathrm{det} \left( \hat{H}_{[N_\uparrow,N_\downarrow]}(\theta_{\mathrm{s}})-E_{\mathrm{ref}} \right) \right], 
\end{eqnarray}
%%%%%%
which is a variant of the previously introduced many-body winding number~\cite{Zhang_CorrSkin_PRB22,Kawabata_CorrSkin_PRB22,Yoshida_reduction1Dptgp_PRB22}.
Here, in order to compute $W_{\mathrm{s}}$, we have imposed twisted boundary conditions
where the twist angle of the down-spin state $\theta_\downarrow$ is opposite to that of the up-spin state $\theta_\uparrow$~\cite{stwist_ftnt}; $\hat{b}^\dagger_{0\sigma}\hat{b}_{L-1\sigma} \to e^{i \theta_\sigma }\hat{b}^\dagger_{0\sigma}\hat{b}_{L-1\sigma}$ with 
$(\theta_\uparrow,\theta_\downarrow)=(\theta_{\mathrm{s}},-\theta_{\mathrm{s}})$ and the twist angle $\theta_{\mathrm{s}}$.
For the Fock space specified by $[N_\uparrow, N_\downarrow]$, the block-diagonalized Hamiltonian is denoted by $\hat{H}_{[N_\uparrow,N_\downarrow]}(\theta_{\mathrm{s}})$.
We note that the spin U(1) symmetry is essential for the skin effect; breaking the spin $\mathrm{U}(1)$ symmetry destroys the skin effect even for the non-interacting case (see Sec.~\ref{sec: 1bdy app} of Supplemental Material~\cite{supple}).

%%%%%%
\textit{Analysis in the strong coupling regime--.}
%%%%%%
%
Before presenting numerical results, we derive an effective model which provides an intuition for the Mott skin effect.
In the strong coupling regime (i.e., $V, U\gg t_+, t_-$), longest lived states of the system are described by an effective spin model, i.e., eigenstates shift in the negative direction of the imaginary axis if bosons occupy the same site.
Applying the second order perturbation theory yields the following effective spin model for the states where each site is occupied by one boson
%%%%%%
\begin{eqnarray}
\label{eq: Hspin}
\hat{H}_{\mathrm{spin}}(\theta_{\mathrm{s}})&=& \sum^{L-2}_{j=0}\left[ J_z \hat{S}^z_{j+1}\hat{S}^z_{j} +J_+\hat{S}^+_{j+1}\hat{S}^-_{j}+J_-\hat{S}^-_{j+1}\hat{S}^+_{j} \right] \nonumber \\
                  &&+ J_+e^{2i\theta_{\mathrm{s}}}\hat{S}^+_{0}\hat{S}^-_{L-1}+J_-e^{-2i\theta_{\mathrm{s}}}\hat{S}^-_{0}\hat{S}^+_{L-1}+E_0, \nonumber \\
\end{eqnarray}
%%%%%%
with $J_z=-4i(t_+t_-)(\frac{1}{V}-\frac{1}{U})$, $J_{\pm}=-2i\frac{t^2_\pm}{U}$, and $E_0=-iL(t_+t_-)(\frac{1}{V}+\frac{1}{U})$.
Here, $\hat{S}^\mu_j$ ($\mu=x,y,z$) denotes spin operators at site $j$. The spin raising and lowering operators are defined as $S^{\pm}_j=S^x_j\pm iS^y_j$.
In terms of creation and annihilation operators of bosons, the spin operators are written as $\hat{S}^z_{j}=(\hat{n}_{j\uparrow}-\hat{n}_{j\downarrow})/2$, $\hat{S}^+_j= \hat{b}^\dagger_{j\uparrow}\hat{b}_{j\downarrow}$ and $\hat{S}^-_j= \hat{b}^\dagger_{j\downarrow}\hat{b}_{j\uparrow}$ for the subspace where each site is occupied by one boson.
Details of the derivation are provided in Sec.~\ref{sec: Hspin all app} of Supplemental Material~\cite{supple}.
The above Hamiltonian~(\ref{eq: Hspin}) preserves the spin $\mathrm{U(1)}$ symmetry and can be block-diagonalized with $S^z$, the eigenvalue of $\hat{S}^z=\sum_{j}\hat{S}^z_j$.

Applying the Jordan-Wigner transformation elucidates that the effective model~{(\ref{eq: Hspin})} gives rise to the Mott skin effect
(for the detailed derivation, see Sec.~\ref{sec: Hspin all app} of Supplemental Material~\cite{supple}); the spin model [Eq.~(\ref{eq: Hspin})] can be mapped to the following spinless fermion model
%%%%%%
\begin{eqnarray}
\label{eq: Hspin_f}
\hat{H}_{\mathrm{spin}}(\theta)&=& \sum^{L-2}_{j=0} \left( J_+ \hat{f}^\dagger_{j+1}\hat{f}_j + J_- \hat{f}^\dagger_{j}\hat{f}_{j+1} \right)  \nonumber \\
                          && - (-1)^{\hat{N}^{\mathrm{f}}}\left( J_+ e^{2i\theta_{\mathrm{s}}} \hat{f}^\dagger_{0}\hat{f}_{L-1} + J_-e^{-2i\theta_{\mathrm{s}}} \hat{f}^\dagger_{L-1}\hat{f}_0 \right) \nonumber \\
                          &&+J_z \sum^{L-1}_{j=0} \hat{n}^{\mathrm{f}}_{j+1}\hat{n}^{\mathrm{f}}_{j} -J_z \left( \hat{N}^{\mathrm{f}}-\frac{L}{4} \right)+E_0,
\end{eqnarray}
%%%%%%
with $\hat{N}^{\mathrm{f}}=\sum_{j}\hat{n}^{\mathrm{f}}_j$, $\hat{S}^+_{j}=e^{i\pi \hat{N}^{<}_j}\hat{f}^\dagger_j$, and $\hat{S}^z_j=(\hat{n}^{\mathrm{f}}_j-\frac{1}{2})$. 
Here, $\hat{N}^{<}_j$ and $\hat{n}^{\mathrm{f}}_j$ are defined as $\hat{N}^{<}_j=\sum^{L-1}_{j=0}\hat{n}^{\mathrm{f}}_j$ and $\hat{n}^{\mathrm{f}}_j=\hat{f}^\dagger_j\hat{f}_j$.
Operators $\hat{f}^\dagger_j$ ($\hat{f}_j$) create (annihilate) a spinless fermion at site $j$.

In particular, for the subspace with $S^z=1-\frac{L}{2}$ (i.e., $[N_\uparrow,N_\downarrow]=[1,L-1]$), there exists only one fermion created by $\hat{f}^\dagger_j$.
Therefore, the above Hamiltonian is simplified as
%%%%%%
\begin{eqnarray}
\label{eq: Hspin_f simple}
\hat{H}_{\mathrm{spin}}(\theta)&=& \sum^{L-2}_{j=0} \left( J_+ \hat{f}^\dagger_{j+1}\hat{f}_j + J_- \hat{f}^\dagger_{j}\hat{f}_{j+1} \right)  \nonumber \\
                          && +\left( J_+ e^{2i\theta_{\mathrm{s}}} \hat{f}^\dagger_{0}\hat{f}_{L-1} + J_-e^{-2i\theta_{\mathrm{s}}} \hat{f}^\dagger_{L-1}\hat{f}_0 \right) \nonumber \\
                          && +J_z(\frac{L}{4}-1)+E_0.
\end{eqnarray}
%%%%%%
This model is nothing but the Hatano-Nelson chain~\cite{Hatano_PRL96,Hatano_PRB97} which exhibits the skin effect. 
Specifically, substituting the Hamiltonian~(\ref{eq: Hspin_f simple}) to $\hat{H}_{[1,L-1]}(\theta_{\mathrm{s}})$, we obtain the winding number $W_{\mathrm{s}}=2$ for the subspace with $[N_\uparrow,N_\downarrow]=[1,L-1]$.
Here, we have set the reference energy $E_{\mathrm{ref}}$ located inside of the loop formed by the eigenvalues.
We recall that $iJ_+>iJ_-$ holds for $t_+>t_-$ [see below Eq.~(\ref{eq: Hspin})].
Corresponding to this non-trivial point-gap topology, skin modes emerge only around the right edge which are described by fermions created by $\hat{f}^\dagger_j$.
We note that the eigenvalues are aligned on the imaginary axis because $J_+$ and $J_-$ are purely imaginary~\cite{complexHN_ftnt}.

Recalling the relation between spin operators and operators $\hat{f}^\dagger_j$, we can conclude that the system exhibits the Mott skin effect. Namely, only the spin degree of freedom is involved in the skin modes induced by the non-trivial point-gap topology.

%%%%%%
\textit{Numerical results: non-interacting case--.}
%%%%%%
%
In order to numerically analyze this model, we employ exact diagonalization.
Unless otherwise noted, we set $(t_+,t_-)=(1,0.1)$.

In the non-interacting case, the system is decomposed into two bosonic Hatano-Nelson models where the hopping in the right (left) direction is dominant for $\sigma=\uparrow$ ($\sigma=\downarrow$).
Such non-reciprocal hoppings result in the ordinary non-Hermitian skin effect as discussed below.
In the following, we analyze the many-body Hamiltonian~(\ref{eq: H}) for the Fock space with $(N_\uparrow,N_\downarrow)=(1,L-1)$ with $L=6$ 
(for an analysis of the one-body Hamiltonian, see Sec.~\ref{sec: 1bdy app} of Supplemental Material~\cite{supple}).

Figure~\ref{fig: Boson Mbdy U0V0}(a) displays the spectral flow of $\hat{H}(\theta_\mathrm{s})$ for $0 \leq  \theta_{\mathrm{s}} \leq 2\pi$ where $E_{m}(\theta_{\mathrm{s}})$ $(m=0,1,2,\ldots)$~\cite{labelE_ftnt} are eigenvalues of $\hat{H}(\theta_{\mathrm{s}})$.
In this figure, we can see that the spectral flow forms the loop structure, which indicates the point-gap topology characterized by the many-body spin winding number taking a non-zero value for $E_{\mathrm{ref}}=-0.01i$ (for more details, see Sec.~\ref{sec: Mbdy W app} of Supplemental Material~\cite{supple}).
%%%%%%%%%%%%%%%%%%%%%%%%%
\begin{figure}[!h]
%
%%%%%%
\begin{minipage}{1\hsize}
\begin{center}
\includegraphics[width=1\hsize,clip]{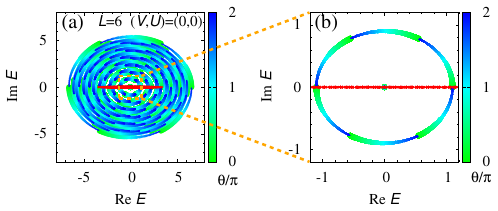}
\end{center}
\end{minipage}
%%%%%%
%
\begin{minipage}{0.49\hsize}
\begin{center}
\includegraphics[width=1\hsize,clip]{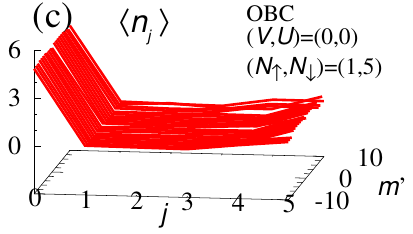}
\end{center}
\end{minipage}
\begin{minipage}{0.49\hsize}
\begin{center}
\includegraphics[width=1\hsize,clip]{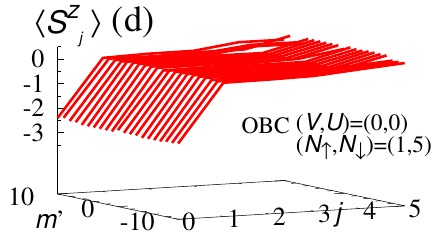}
\end{center}
\end{minipage}
\caption{
Energy eigenvalues and expectation values for $V=U=0$.
(a): Energy eigenvalues under twisted boundary conditions and OBC (red), respectively.
(b): A magnified version of the range $-1.2 \leq \mathrm{Im}E \leq 1.2$.
The data obtained under OBC are represented by red dots.
With increasing $\theta_{\mathrm{s}}$ from $0$ to $2\pi$, the eigenvalues wind the origin of the complex plane, which indicates $W_{\mathrm{s}}$ takes a non-zero value for $E_{\mathrm{ref}}=-0.01i$.
(c) [(d)]: Expectation values $\langle \hat{n}_j \rangle$ [$\langle \hat{S}^z_j \rangle$] at each site under OBC.
The expectation values are computed from right eigenvectors of the many-body Hamiltonian whose eigenvalues $E_{m'}$ are located around the origin of complex plane. 
Specifically, $m'$ is defined as $m'=m-756$ (for the definition of $m$, see footnote~\cite{labelE_ftnt}). 
We note that the other states show essentially the same behaviors. For $U=V=0$, the eigenvalues are aligned on the real axis, and the eigenvalues are numerically zero for $m=756$.
These data are obtained for subspace with $(N_{\uparrow},N_{\downarrow})=(1,5)$ and for $(t_+,t_-)=(1,0.1)$ and $L=6$.
}
\label{fig: Boson Mbdy U0V0}
\end{figure}
%%%%%%%%%%%%%%%%%%%%%%%%%

Corresponding to this non-trivial point-gap topology, the spectrum shows extreme sensitivity to the boundary conditions.
Imposing OBC significantly changes the spectrum of $\hat{H}$; as shown in Fig.~\ref{fig: Boson Mbdy U0V0}(a), the eigenvalues under OBC are aligned on the real axis in contrast to the eigenvalues under PBC. 
Eigenstates also show such sensitivity to the boundary conditions~\cite{Garbe_BosonicSkin_arXiv23}.
In order to show this, we compute the expectation values of $\hat{n}_j=\sum_{\sigma}\hat{n}_{j\sigma}$
%%%%%%
\begin{eqnarray}
\label{eq: <n_j>}
\langle \hat{n}_j \rangle
&=& 
\frac{ {}_\mathrm{R}\langle \Psi_m | \hat{n}_j |\Psi_m \rangle_{\mathrm{R}} } {  {}_\mathrm{R}\langle \Psi_m |\Psi_m \rangle_{\mathrm{R}}  },
\end{eqnarray}
%%%%%%
with $|\Psi_m \rangle_{\mathrm{R}}$ ($m=0,1,2,\ldots$) being right eigenvectors of the many-body Hamiltonian, $\hat{H}|\Psi_m \rangle_{\mathrm{R}}=E_m |\Psi_m \rangle_{\mathrm{R}}$. 
As displayed in Fig.~\ref{fig: Boson Mbdy U0V0}(c), the bosons are localized around edges under OBC in contrast to the case of PBC
(the data for PBC is provided in Sec.~\ref{sec: N Sz PBC app} of Supplemental Material~\cite{supple}).

Due to the localization of bosons, spin polarization is observed only in the presence of boundaries. 
Figure~\ref{fig: Boson Mbdy U0V0}(c) displays the expectation values of $\hat{S}^z_j=(\hat{n}_{j\uparrow}-\hat{n}_{j\downarrow})/2$. 
Spin polarization is observed under OBC in contrast to the case of PBC. 

As discussed above, in the non-interacting cases, the system shows the non-Hermitian skin effect which results in extreme sensitivity of the eigenvalues and eigenstates to the presence or absence of the boundaries.
Accordingly, the localization of bosons is observed only under OBC.
%%%%%%%%%%%%%%%%%%%%%%%%%
\begin{figure}[!t]
%
%%%%%%
\begin{minipage}{1\hsize}
\begin{center}
\includegraphics[width=1\hsize,clip]{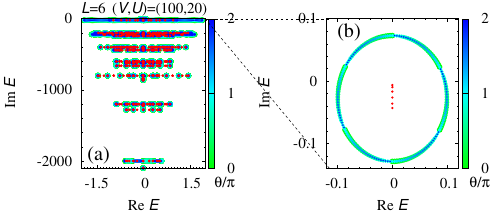}
\end{center}
\end{minipage}
%%%%%%
%
\begin{minipage}{0.49\hsize}
\begin{center}
\includegraphics[width=1\hsize,clip]{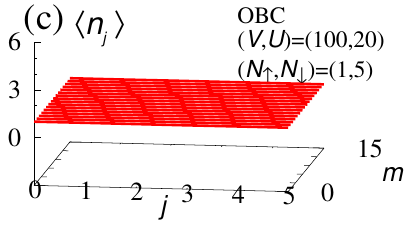}
\end{center}
\end{minipage}
\begin{minipage}{0.49\hsize}
\begin{center}
\includegraphics[width=1\hsize,clip]{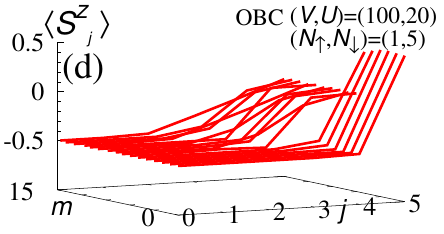}
\end{center}
\end{minipage}
\caption{
Energy eigenvalues and expectation values for $(V,U)=(100,20)$.
(a): Energy eigenvalues under twisted boundary conditions and OBC, respectively.
(b): A magnified version of the range $-0.14 \leq \mathrm{Im}E \leq 0.1$. In these figures the data obtained under OBC are represented by red dots.
With increasing $\theta_{\mathrm{s}}$ from $0$ to $2\pi$, the eigenvalues wind the origin of the complex plane, which indicates $W_{\mathrm{s}}=2$ for $E_{\mathrm{ref}}=-0.01i$.
(c) [(d)]: Expectation values $\langle \hat{n}_j \rangle$ [$\langle \hat{S}^z_j \rangle$] at each site under OBC.
Here, the expectation values are plotted for eigenstates whose eigenvalues~\cite{labelE_ftnt} are denoted by red dots in panel~(b).
As is the case of $V=U=0$, the expectation values are computed from the right eigenvectors (see Fig.~\ref{fig: Boson Mbdy U0V0}).
These data are obtained for $(t_+,t_-)=(1,0.1)$ and $L=6$.
}
\label{fig: Boson Mbdy U20V100}
\end{figure}
%%%%%%%%%%%%%%%%%%%%%%%%%

%%%%%%
\textit{Numerical results: interacting case--.}
%%%%%%
%
Now, we demonstrate that the interplay between strong correlations and non-reciprocal hoppings induces the Mott skin effect.
Namely, the non-trivial point-gap topology results in extreme sensitivity to the boundary conditions only in the spin degree of freedom
The spin-charge separation plays an essential role in the emergence of the Mott skin effect.
In the following, we focus on the Fock space with $(N_\uparrow,N_\downarrow)=(1,L-1)$~\cite{Nup2Ndow4_ftnt}.
The Mott skin effect is also observed for other subspaces of half-filling (see Sec.~\ref{sec: (Nup,Ndow)=(2,4) app} of Supplemental Material~\cite{supple}).

Figures~\ref{fig: Boson Mbdy U20V100}(a)~and~\ref{fig: Boson Mbdy U20V100}(b) display the spectral flow of $\hat{H}(\theta_{\mathrm{s}})$ for $(V,U)=(100,20)$ with increasing $\theta_{\mathrm{s}}$ from $0$ to $2\pi$.
The interactions $V$ and $U$ shift most of the loops in the negative direction of the imaginary axis [see Fig.~\ref{fig: Boson Mbdy U20V100}(a)].
However, one of the loops remains around the origin of the complex plane [see Fig.~\ref{fig: Boson Mbdy U20V100}(b)]. This is because bosons do not feel the on-site interactions unless multiple bosons occupy the same site.

The states remaining around the origin of the complex plane exhibit the Mott skin effect.
Because of the loop structure, the many-body spin winding number takes $W_{\mathrm{s}}=2$ for $E_{\mathrm{ref}}=-0.01i$ 
(for more details, see Sec.~\ref{sec: Mbdy W app} of Supplemental Material~\cite{supple}).
Corresponding to the non-trivial point-gap topology, eigenstates exhibit extreme sensitivity to the presence or absence of boundaries.
As shown in Fig.~\ref{fig: Boson Mbdy U20V100}(b), the eigenvalues are aligned on the imaginary axis under OBC in contrast to the case of PBC where eigenvalues are distributed on the complex plane~\cite{Ptb_param_ftnt}.

Such sensitivity to the boundary conditions is also observed for eigenstates. 
However, only the spin degree of freedom is involved in the sensitivity of the boundary conditions, which is in sharp contrast to the non-interacting case.
Figure~\ref{fig: Boson Mbdy U20V100}(c) displays the expectation values $\langle \hat{n}_j \rangle$. 
As shown in this figure, strong correlations prevent bosons from localizing around the edges, which suppresses the sensitivity to the boundary conditions for the charge degree of freedom.
Instead, the spin degree of freedom exhibits extreme sensitivity to the boundary conditions.
Figure~\ref{fig: Boson Mbdy U20V100}(d) displays the expectation values $\langle \hat{S}^z_j \rangle$.
As shown in this figure, spin polarization is observed under OBC in contrast to the case of PBC 
(the data for PBC are provided in Sec.~\ref{sec: N Sz PBC app} of Supplemental Material~\cite{supple}).

The above results reveal that the system exhibits the Mott skin effect resulting in extreme sensitivity to the boundary conditions only in the spin degree of freedom (i.e., such sensitivity is not observed in the charge degree of freedom). 
The emergence of the Mott skin effect is due to the interplay between strong correlations and non-Hermitian point-gap topology.
We note that the Mott skin effect is observed also for the subspace with $(N_{\uparrow},N_{\downarrow})=(2,4)$ and $L=6$
(for more details see Sec.~\ref{sec: (Nup,Ndow)=(2,4) app} of Supplemental Material~\cite{supple}).

%%%%%%
\textit{Real-time dynamics--.}
%%%%%%
%
As seen above, the Mott skin effect involves only the spin degree of freedom in contrast to the skin effect in the non-interacting case. 
This difference is also reflected in the time-evolution of physical quantities.

%%%%%%%%%%%%%%%%%%%%%%%%%
\begin{figure}[!t]
\begin{minipage}{0.49\hsize}
\begin{center}
\includegraphics[width=1\hsize,clip]{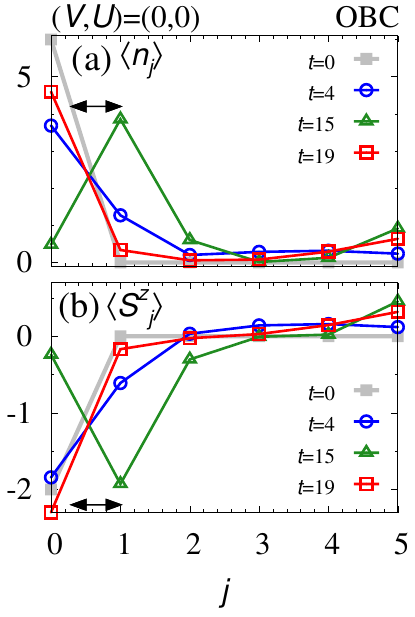}
\end{center}
\end{minipage}
\begin{minipage}{0.49\hsize}
\begin{center}
\includegraphics[width=1\hsize,clip]{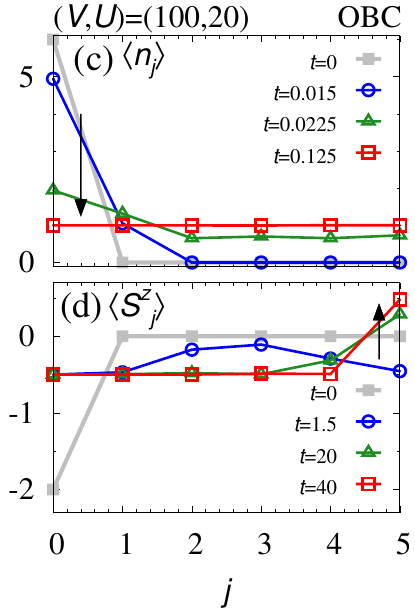}
\end{center}
\end{minipage}
\caption{
(a) and (c) [(b) and (d)]: The time-evolution of the expectation values $\langle \hat{n}_j(t)\rangle$ [$\langle \hat{S}^z_j(t)\rangle $] under OBC [see Eq.~(\ref{eq: n(t)})].
Panels (a) and (b) [(c) and (d)] display data for $(V,U)=(0,0)$ [$(V,U)=(100,20)$].
These data are obtained for $(t_+,t_-)=(1,0.1)$ and $L=6$.
}
\label{fig: TimeEvol OBC}
\end{figure}
%%%%%%%%%%%%%%%%%%%%%%%%%

Let us start with the non-interacting case.
Figure~\ref{fig: TimeEvol OBC}(a) displays the expectation values
%%%%%%
\begin{eqnarray}
\label{eq: n(t)}
\langle \hat{n}_j(t) \rangle &=& \frac{\langle \Phi(t) | \hat{n}_j | \Phi(t) \rangle} {\langle \Phi(t)| \Phi(t) \rangle}, 
\end{eqnarray}
%%%%%%
with $|\Phi(t)\rangle =e^{-i\hat{H}t}|\Phi(0)\rangle$ and $t$ being time~\cite{nHSeq_ftnt}.
The initial state is chosen so that all of the bosons occupy site $j=0$; $\sqrt{(L-1)!}|\Phi(0)\rangle=\hat{b}^\dagger_{0\uparrow}(\hat{b}^\dagger_{0\downarrow})^{L-1}|0\rangle$.
Figure~\ref{fig: TimeEvol OBC}(a) indicates that the bosons remain localized around the left edge due to the skin effect~\cite{osci_ftnt}.
Figure~\ref{fig: TimeEvol OBC}(b) indicates that the above localization is also observed in the expectation values $\langle S^z_{j}(t)\rangle= \langle \Phi(t) | \hat{S}^z_j | \Phi(t) \rangle/ \langle \Phi(t)| \Phi(t) \rangle$. This is because bosons with spin up (spin down) are localized around the right (left) edge (see Sec.~\ref{sec: 1bdy H0 app} of Supplemental Material~\cite{supple}).
The above results demonstrate that in the non-interacting case, the dynamical properties of the spin degree of freedom are tied to those of charge degree of freedom.

Now, we turn to the real-time dynamics in the correlated case. 
In contrast to the non-interacting case, dynamics of the spin degree of freedom exhibit an essential difference from those of the charge degree of freedom.
Figure~\ref{fig: TimeEvol OBC}(c) displays the time-evolution of $\langle \hat{n}_j(t)\rangle$ for $(V,U)=(100,20)$.
As seen in Fig.~\ref{fig: TimeEvol OBC}(c), bosons immediately extend to the bulk. In particular, for $0.125 \lesssim t$, each site is occupied by one boson. 
This is because the states where each site is occupied by one boson have a longer lifetime $\tau \sim 1/\mathrm{Im}E_m$ than others. 
Contrary to the sudden change in the charge degree of freedom, the dynamics of the spin degree of freedom show gradual changes.
The time-evolution of $\langle \hat{S}^z_j \rangle$ is plotted in Fig.~\ref{fig: TimeEvol OBC}(d).
In this figure, dynamical spin accumulation is observed for $20 \lesssim t$.
This behavior is observed only under OBC (for data under PBC, see Sec.~\ref{sec: TimeEvol PBC app} of Supplemental Material~\cite{supple}).

As seen above, the Mott skin effect induces the dynamical spin accumulation while the charge degree of freedom remains spatially uniform.
These dynamical properties in the strongly correlated case are in sharp contrast to those in the non-interacting case.
The above dynamical properties are considered to be observed for open quantum systems even in the presence of quantum jumps.

%%%%%%
\textit{Conclusion--.}
%%%%%%
We have proposed a non-Hermitian Mott skin effect induced by the interplay between strong correlations and the non-Hermitian point-gap topology. 
In contrast to the ordinary non-Hermitian skin effect, the Mott skin effect results in extreme sensitivity to the boundary conditions only in the spin degree of freedom. 
We have demonstrated the emergence of the Mott skin effect by analyzing the bosonic non-Hermitian Hamiltonian with non-reciprocal hoppings and on-site interactions. 
The difference from the ordinary skin effect is also reflected in the dynamical properties; spin accumulation is observed while charge distribution remains spatially uniform.

We finish this paper with a remark on the relevance to open quantum systems described by the Lindblad equation~\cite{Lindblad_CommMathPhys1976,Gorini_JMathPhys1976,Breuer_textbook2007}.
In this paper, we have analyzed the non-Hermitian Hamiltonian as a toy model.
We note, however, that our Hamiltonian is relevant to an open quantum system having particle losses, and that the above dynamical spin accumulation can be observed even under the presence of the jump term (for more details, see Sec.~\ref{sec: OQS app} of Supplemental Material~\cite{supple}).
This fact implies that open quantum systems such as cold atoms with strong correlations may provide a feasible platform to observe the Mott skin effect.

\textit{Note added--.}
While finising this paper, we noticed Ref.~\cite{Kim_CollectiveSkin_arXiv23} posted on arXiv which has some overlap with our results.

%%%%%%%%%%%%%%
\textit{Acknowledgments--.}
%%%%%%%%%%%%%%
T.Y. thanks Manfred Sigrist for fruitful discussion.
T.Y. particulary thanks Shunsuke Furukawa and Yoshihito Kuno for fruitful discussions on the technical details of exact diagonalization.
T.Y. is grateful to the long term workshop YITP-T-23-01 held at YITP, Kyoto University, where a part of this work was done.
This work is supported by JSPS KAKENHI Grants No.~JP21K13850 and No.~JP22H05247.

%%%%%%%%%%%%%%%
%\bibliography{MottSkin}

%%%%%%%%%%%%%%%

\clearpage

\renewcommand{\thesection}{S\arabic{section}}
\renewcommand{\theequation}{S\arabic{equation}}
\setcounter{equation}{0}
\renewcommand{\thefigure}{S\arabic{figure}}
\setcounter{figure}{0}
\renewcommand{\thetable}{S\arabic{table}}
\setcounter{table}{0}
\makeatletter
\c@secnumdepth = 2
\makeatother

\onecolumngrid
\begin{center}
 {\large \textmd{Supplemental Materials:} \\[0.3em]
 {\bfseries 
 \mytitle
 }
 }
\end{center}

\setcounter{page}{1}

%%%%%%%%%%%%%%
\section{
Details of the effective spin model
}
\label{sec: Hspin all app}
%%%%%%%%%%%%%%

%%%%%%%%%%%%%%
\subsection{
Derivation of Equation~(\ref{eq: Hspin})
}
\label{sec: deri Heff app}
%%%%%%%%%%%%%%

Here, based on the second order perturbation theory, we derive the effective spin model [see Eq.~(\ref{eq: Hspin})] for $V, U\gg t_+,t_-$.
In the following, we impose PBC for the sake of simplicity.

Let us consider a Fock space where each site is occupied by a boson. 
For such a Fock space, the effective Hamiltonian can be written as
%%%%%%
\begin{eqnarray}
\label{eq: ptb gen app}
\hat{H}_{\mathrm{ptb}} &=& \hat{H}_{\mathrm{ptb},\mathrm{RL}} + \hat{H}_{\mathrm{ptb},\mathrm{LR}},\\
\hat{H}_{\mathrm{ptb},\mathrm{RL}}
&=&
\sum_{j \sigma s} 
\hat{P}
\left[
t_{\mathrm{R}\sigma}t_{\mathrm{L}s}
\hat{b}^\dagger_{j+1\sigma}\hat{b}_{j\sigma}
\frac{1}{E_{\mathrm{g}}-\hat{H}_{\mathrm{int}}}
\hat{b}^\dagger_{js}\hat{b}_{j+1s}
\right]
\hat{P},
\nonumber \\
\hat{H}_{\mathrm{ptb},\mathrm{LR}}
&=&
\sum_{j \sigma s} 
\hat{P}
\left[
t_{\mathrm{L}\sigma}t_{\mathrm{R}s}
\hat{b}^\dagger_{j\sigma}\hat{b}_{j+1\sigma}
\frac{1}{E_{\mathrm{g}}-\hat{H}_{\mathrm{int}}}
\hat{b}^\dagger_{j+1s}\hat{b}_{js}
\right]
\hat{P},
\nonumber
\end{eqnarray}
%%%%%%
where $\hat{P}$ is the projection operator of the Fock space where a boson occupies each site.
The ground state energy $E_{\mathrm{g}}$ takes zero $E_{\mathrm{g}}=\hat{P}\hat{H}_{\mathrm{int}}\hat{P}=0$.

By a straightforward calculation, we obtain
%%%%%%
\begin{eqnarray}
\hat{H}_{\mathrm{ptb},\mathrm{RL}}
&=& 
-i\sum_{j\sigma}
\frac{t_{\mathrm{R}\sigma}t_{\mathrm{L}\sigma}}{V} 
\hat{P}
\left(
\hat{n}_{j+1\sigma}\hat{n}_{j\sigma}
\right)
\hat{P}
\nonumber \\
&&
-i\sum_{j\sigma}
\frac{t_{\mathrm{R}\sigma}t_{\mathrm{L}\sigma}}{U} 
\hat{P}
\left(
\hat{n}_{j+1\sigma}\hat{n}_{j\bar{\sigma}}
\right)
\hat{P}
\nonumber \\
&&
-i\sum_{j\sigma}
\frac{t_{\mathrm{R}\sigma}t_{\mathrm{L}\bar{\sigma}}}{U} 
\hat{P}
\left(
\hat{b}^\dagger_{j+1\sigma}\hat{b}_{j+1\bar{\sigma}}\hat{b}^\dagger_{j\bar{\sigma}}\hat{b}_{j\sigma}
\right)
\hat{P},
\nonumber \\
\end{eqnarray}
%%%%%%
where $\bar{\sigma}$ takes $\downarrow$ ($\uparrow$) for $\sigma=\uparrow$ ($\sigma=\downarrow$).

By using the following relations
%%%%%%
\begin{eqnarray}
4\hat{S}^z_{j+1}\hat{S}^z_{j} +\hat{n}_{j+1}\hat{n}_{j}&=& 2(\hat{n}_{j+1\uparrow}\hat{n}_{j\uparrow}+\hat{n}_{j+1\downarrow}\hat{n}_{j\downarrow}), \\
-4\hat{S}^z_{j+1}\hat{S}^z_{j} +\hat{n}_{j+1}\hat{n}_{j}&=& 2(\hat{n}_{j+1\uparrow}\hat{n}_{j\downarrow}+\hat{n}_{j+1\downarrow}\hat{n}_{j\uparrow}),
\end{eqnarray}
%%%%%%
we obtain 
%%%%%%
\begin{eqnarray}
\hat{H}_{\mathrm{ptb},\mathrm{RL}}
&=& \sum_j\left[
-i\frac{t_+t_-}{2V}\hat{P}\left(4\hat{S}^z_{j+1}\hat{S}^z_{j}+\hat{n}_{j+1}\hat{n}_{j} \right)\hat{P}  \right. \nonumber \\ 
&&-i\frac{t_+t_-}{2U}\hat{P}\left(-4\hat{S}^z_{j+1}\hat{S}^z_{j}+\hat{n}_{j+1}\hat{n}_{j} \right)\hat{P} \nonumber \\
&& \left. -i\frac{t^2_+}{U}\hat{P}\left(\hat{S}^+_{j+1}\hat{S}^-_{j}\right)\hat{P} -i\frac{t^2_-}{U}\hat{P}\left(\hat{S}^-_{j+1}\hat{S}^+_{j}\right)\hat{P} \right], \nonumber \\
\end{eqnarray}
%%%%%%
where we have also chosen the parameters as $(t_{\mathrm{R}\uparrow},t_{\mathrm{R}\downarrow})=(t_+,t_-)$ and $(t_{\mathrm{L}\uparrow},t_{\mathrm{L}\downarrow})=(t_-,t_+)$.

In a similar way, we obtain
%%%%%%
\begin{eqnarray}
\hat{H}_{\mathrm{ptb},\mathrm{LR}}
&=& \sum_j\left[ 
   -i\frac{t_+t_-}{2V} \hat{P} \left( 4\hat{S}^z_{j+1}\hat{S}^z_{j} +\hat{n}_{j+1}\hat{n}_{j} \right) \hat{P} \right. \nonumber\\
&& -i\frac{t_+t_-}{2U} \hat{P} \left( -4\hat{S}^z_{j+1}\hat{S}^z_{j} +\hat{n}_{j+1}\hat{n}_{j} \right) \hat{P}  \nonumber\\
&& \left. -i\hat{P} \left( \frac{t^2_+}{U} \hat{S}^+_{j+1}\hat{S}^-_{j} +\frac{t^2_-}{U} \hat{S}^-_{j+1}\hat{S}^+_{j} \right) \hat{P} \right].
\end{eqnarray}
%%%%%%

Therefore, the effective Hamiltonian is obtained as
%%%%%%
\begin{eqnarray}
\label{eq: Hspin PBC app}
\hat{H}_{\mathrm{ptb}}
&=& 
\sum_{j}\left[
J_+ \hat{S}^+_{j+1}\hat{S}^-_{j} +J_- \hat{S}^-_{j+1}\hat{S}^+_{j} +J_z \hat{S}^z_{j+1} \hat{S}^z_{j}
\right] +E_0, \nonumber \\
\end{eqnarray}
%%%%%%
with 
%%%%%%
\begin{eqnarray}
J_+ &=&-2i\frac{t^2_+}{U}, \\
J_- &=&-2i\frac{t^2_-}{U}, \\
J_z &=&-4it_+t_- (\frac{1}{V}-\frac{1}{U}), \\
E_0 &=&-4it_+t_- L (\frac{1}{V}+\frac{1}{U}). 
\end{eqnarray}
%%%%%%

Equation~(\ref{eq: Hspin PBC app}) corresponds to the spin model [Eq.~(\ref{eq: Hspin})].

Under twisted boundary conditions, the following replacement results in the effective Hamiltonian $\hat{H}_{\mathrm{spin}}(\theta_{\mathrm{s}})$:
%%%%%%
\begin{eqnarray}
\hat{S}^+_{0}\hat{S}^-_{L-1}&\to& e^{2i\theta} \hat{S}^+_{0}\hat{S}^-_{L-1}, \\
\hat{S}^-_{0}\hat{S}^+_{L-1}&\to& e^{-2i\theta} \hat{S}^-_{0}\hat{S}^+_{L-1}.
\end{eqnarray}
%%%%%%

Therefore, we obtain the spin model shown in Eq.~(\ref{eq: Hspin}).

%%%%%%%%%%%%%%
\subsection{
Derivation of Eq.~(\ref{eq: Hspin_f simple})
}
\label{sec: JW app}
%%%%%%%%%%%%%%

Applying the Jordan-Wigner transformation to the effective spin model [see Eq.~(\ref{eq: Hspin})] yields the spinless fermion model [see Eq.~(\ref{eq: Hspin_f})]. 

With the Jordan-Wigner transformation, $\hat{S}^+_{j}$ and $\hat{S}^z_j$ are rewritten as
%%%%%%
\begin{eqnarray}
\hat{S}^+_{j} &=& e^{i\pi \hat{N}^{<}_j}\hat{f}^\dagger_j, \\
\hat{S}^z_j &=& (\hat{n}^{\mathrm{f}}_j-\frac{1}{2}),
\end{eqnarray}
with creation (annihilation) operators of spinless fermions $\hat{f}^\dagger_{j}$ ($\hat{f}_{j}$).
Here, the operators are defined as $\hat{N}^{<}_j = \sum^{j-1}_{i=0}\hat{n}^{\mathrm{f}}_j$, $\hat{N}^{\mathrm{f}} = \sum_{j}\hat{n}^{\mathrm{f}}_j$, and $\hat{n}^{\mathrm{f}}_j = \hat{f}^\dagger_j\hat{f}_j$.

Thus, each term of the spin model is transformed as follows:
for $0 \leq j <L-1$,
%%%%%%
\begin{eqnarray}
\hat{S}^+_{j+1}\hat{S}^-_{j}&=& \hat{f}^\dagger_{j+1} \hat{f}_{j}, \\
\hat{S}^-_{j+1}\hat{S}^+_{j}&=& \hat{f}^\dagger_{j} \hat{f}_{j+1}, 
\end{eqnarray}
%%%%%%
and for $j=L-1$,
%%%%%%
\begin{eqnarray}
\hat{S}^+_{j+1}\hat{S}^-_{j}&=& \hat{f}^\dagger_0e^{i\pi \hat{N}^{<}_{L-1}} \hat{f}_{L-1}  \nonumber \\
                            &=& \hat{f}^\dagger_0e^{i\pi ( \hat{N}^{\mathrm{f}} +\hat{n}^{\hat{f}}_{L-1})} \hat{f}_{L-1} \nonumber \\
                            &=& -e^{i\pi \hat{N}^{\mathrm{f}}}\hat{f}^\dagger_{0} e^{i\pi \hat{n}^{\hat{f}}_{L-1}} \hat{f}_{L-1} \nonumber \\
                            &=& -e^{i\pi \hat{N}^{\mathrm{f}}}\hat{f}^\dagger_{0} \hat{f}_{L-1}, \\
\hat{S}^-_{j+1}\hat{S}^+_{j}&=& \hat{f}_{0} e^{i\pi \hat{N}^{<}_{L-1} } \hat{f}^\dagger_{L-1} \nonumber \\
                            &=& \hat{f}_{0} e^{i\pi (\hat{N}^{\mathrm{f}}+\hat{n}^{\hat{f}}_{L-1} ) } \hat{f}^\dagger_{L-1} \nonumber \\
                            &=& -e^{i\pi \hat{N}^{\mathrm{f}} } \hat{f}^\dagger_{L-1}\hat{f}_{0}.
\end{eqnarray}
%%%%%%

In a similar way, we obtain
%%%%%%
\begin{eqnarray}
\hat{S}^z_{j+1}\hat{S}^z_{j}&=& (\hat{n}^{\mathrm{f}}_{j+1}-\frac{1}{2}) (\hat{n}^{\mathrm{f}}_{j}-\frac{1}{2}),
\end{eqnarray}
%%%%%%
for $0\leq j \leq L-1$.

Therefore, the effective spin model is rewritten as
%%%%%%
\begin{eqnarray}
\label{eq: Hspin_f app}
\hat{H}_{\mathrm{spin}}(\theta)&=& \sum^{L-2}_{j=0} \left( J_+ \hat{f}^\dagger_{j+1}\hat{f}_j + J_- \hat{f}^\dagger_{j}\hat{f}_{j+1} \right)  \nonumber \\
                          && - (-1)^{\hat{N}^{\mathrm{f}}}\left( J_+ e^{2i\theta_{\mathrm{s}}} \hat{f}^\dagger_{0}\hat{f}_{L-1} + J_-e^{-2i\theta_{\mathrm{s}}} \hat{f}^\dagger_{L-1}\hat{f}_0 \right) \nonumber \\
                          &&+J_z \sum^{L-1}_{j=0} \hat{n}^{\mathrm{f}}_{j+1}\hat{n}^{\mathrm{f}}_{j} -J_z(\hat{N}^{\mathrm{f}}-\frac{L}{4})+E_0.
\end{eqnarray}
%%%%%%
In particular, for the Fock space with $S^z=1-\frac{L}{2}$ (i.e., $[N_\uparrow,N_\downarrow]=[1,L-1]$), there exists only one fermion created by $\hat{f}^\dagger_j$.
In this case, the following terms are simplified as
%%%%%%
\begin{eqnarray}
\sum^{L-1}_{j=0} \hat{n}^{\mathrm{f}}_{j+1}\hat{n}^{\mathrm{f}}_{j} &\to& 0 \\
\hat{N}^{\mathrm{f}}-\frac{L}{4} &\to& 1- -\frac{L}{4}.
\end{eqnarray}
%%%%%%
Therefore, we obtain Eq.~(\ref{eq: Hspin_f simple}) for the subspace with $S^z=1-\frac{L}{2}$.

%%%%%%%%%%%%%%
\section{
Properties of the one-body Hamiltonian
}
\label{sec: 1bdy app}
%%%%%%%%%%%%%%

%%%%%%%%%%%%%%
\subsection{
Analysis of the one-body Hamiltonian of $\hat{H}_0$
}
\label{sec: 1bdy H0 app}
%%%%%%%%%%%%%%

By analyzing the first quantized Hamiltonian, we see that the system shows the non-Hermitian skin effect at the non-interacting case.
The Hamiltonian~(\ref{eq: H_0}) is written as
%%%%%%
\begin{eqnarray}
\hat{H}_{0}&=& \sum_{k} h_{\sigma}(k) \hat{b}^\dagger_{k\sigma} \hat{b}_{k\sigma}
\end{eqnarray}
%%%%%%
with $\hat{b}^\dagger_{k\sigma}=\frac{1}{\sqrt{L}}\sum_{j}e^{-ikj}\hat{b}^\dagger_{j\sigma}$ and $0 \leq k <2\pi$.
Here, we have imposed PBC.
The first quantized Hamiltonian $h_{\uparrow}(k)$ and $h_{\downarrow}(k)$ are written as $h_{\uparrow}(k)=t_+e^{ik}+t_-e^{-ik}$ and $h_{\downarrow}(k)=t_-e^{ik}+t_+e^{-ik}$, respectively.

Now, we introduce the spin winding number $w_{\mathrm{s}}$ characterizing the point-gap topology of the first quantized Hamiltonian
%%%
$h(k)=
\left(
\begin{array}{cc}
h_{\uparrow}(k) & 0 \\
0  & h_{\downarrow}(k)
\end{array}
\right).
$
%%%
The spin winding number $w_{\mathrm{s}}$ is defined as
%%%%%%
\begin{eqnarray}
w_{\mathrm{s}}(\epsilon_{\mathrm{ref}})&=& \int^{2\pi}_{0} \frac{dk}{2\pi i} \partial_k \mathrm{tr} [ s_z \mathrm{log} ( h(k) -\epsilon_{\mathrm{ref}}\1_{2\times2}) ],
\end{eqnarray}
%%%%%%
with $\epsilon_{\mathrm{ref}}\in \mathbb{C}$, $\1_{2\times2}$ being the $2\times2$ identity matrix, and 
$s^z=\frac{1}{2}
\left(
\begin{array}{cc}
1 & 0 \\
0 & -1
\end{array}
\right)
$.
The derivative with respect to $k$ is denoted by $\partial_{k}$.

%%%%%%%%%%%%%%%%%%%%%%%%%
\begin{figure}[!h]
\begin{minipage}{0.5\hsize}
\begin{center}
\includegraphics[width=1\hsize,clip]{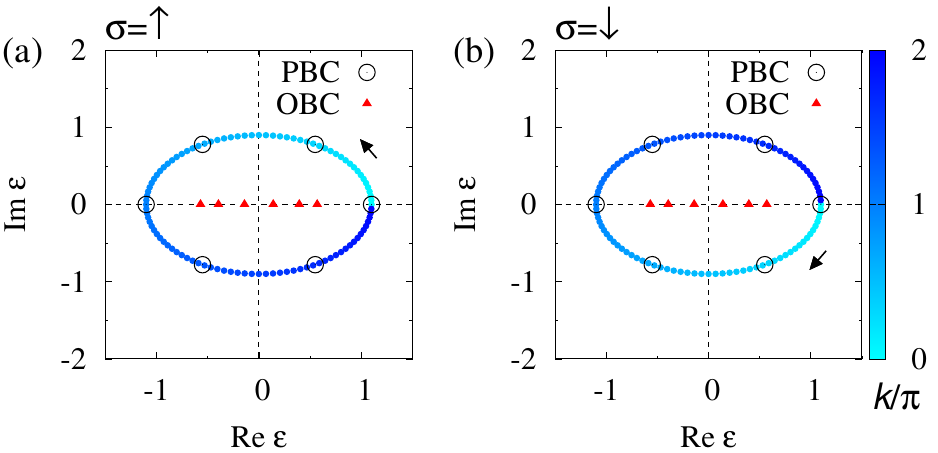}
\end{center}
\end{minipage}
\begin{minipage}{0.5\hsize}
\begin{center}
\includegraphics[width=1\hsize,clip]{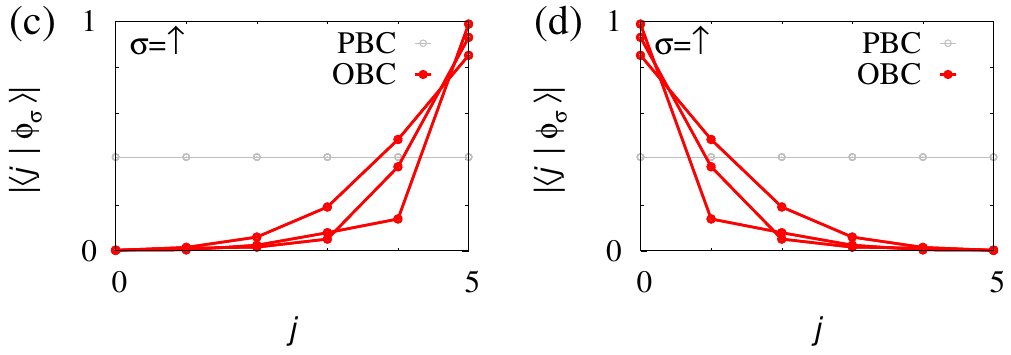}
\end{center}
\end{minipage}
\caption{
(a) [(b)]: Eigenvalues of the one-body Hamiltonian for $\sigma=\uparrow$ [$\sigma=\downarrow$].
Open circles (closed triangle) denote the data under PBC (OBC) for $L=6$.
Closed circles denote the data under PBC for $L=100$. The color of the closed circes represent the momentum $k$.
(c) [(d)]: Amplitude of the right eigenvectors for $\sigma=\uparrow$ [$\sigma=\downarrow$].
The gray (red) lines denote the data under PBC (OBC) for $L=6$.
}
\label{fig: Boson 1bdy}
\end{figure}
%%%%%%%%%%%%%%%%%%%%%%%%%

Figures~\ref{fig: Boson 1bdy}(a)~and~\ref{fig: Boson 1bdy}(b) display the eigenvalues of the one-body Hamiltonian $\epsilon_{k\sigma}=h_\sigma(k)$ under PBC.
This figure indicates that the spin winding number takes $1$ for $\epsilon_{\mathrm{ref}}=0$.

This non-trivial point-gap topology induces extreme sensitivity of eigenvalues and eigenstates to the presence or absence of boundaries.
Figures~\ref{fig: Boson 1bdy}(a)~and~\ref{fig: Boson 1bdy}(b) indicate that the eigenvalues are aligned on the real axis under OBC in contrast to the case of PBC.
Correspondingly, the system under OBC exhibits the skin modes localized around the boundaries [see Figs.~\ref{fig: Boson 1bdy}(c)~and~\ref{fig: Boson 1bdy}(d)].

The above results indicate that the first quantized Hamiltonian shows the non-Hermitian skin effect.
We finish this part with a remark on the realness of eigenvalues under OBC which can be understood by applying an imaginary gauge transformation.
Introducing operators
%%%%%%
\begin{eqnarray}
\bar{\hat{a}}_{j}&=& e^{-\phi j} \hat{c}^\dagger_j, \\
\hat{a}_{j}&=& e^{\phi j} \hat{c}_j,
\end{eqnarray}
%%%%%%
we can rewrite the non-interacting Hamiltonian $\hat{H}_{0}$ under OBC
%%%%%%
\begin{eqnarray}
\label{eq: SimTrans HN app}
\hat{H}_{0}
&=& -t_{\mathrm{a}} \sum_{j=0,\cdots,L-2} [ \bar{\hat{a}}_{j+1\sigma}\hat{a}_{j\sigma} + \bar{\hat{a}}_{j\sigma}\hat{a}_{j+1\sigma} ].
\end{eqnarray}
%%%%%%
with $t_{\mathrm{a}}=\sqrt{ t_{\mathrm{R}}t_{\mathrm{L}} }$ and $\phi=\frac{1}{2}\ln \left( \frac{t_L}{t_R} \right)$.
Here $\bar{\hat{a}}_j$ and $\hat{a}_j$ satisfy the anti-commutation relation $\{ \hat{a}_i, \bar{\hat{a}}_j \}=\delta_{ij}$.
Equation~(\ref{eq: SimTrans HN app}) can be regarded as the Hermitian Hamiltonian. Thus, all of the eigenvalues are real.

%%%%%%%%%%%%%%
\subsection{
Effects of spin U(1) breaking
}
\label{sec: spin U(1) breaking 1bdy app}
%%%%%%%%%%%%%%

We demonstrate that breaking spin U(1) symmetry may destroy the skin effect.
In order to see this, let us add a spin-mixing term to $\hat{H}_0$ [Eq.~(\ref{eq: H_0})]
%%%%%%
\begin{eqnarray}
\label{eq: H_0 U1 break app}
\hat{H}'_0 &=& \hat{H}_0 + i\Delta \sum_{j}(\hat{b}^\dagger_{j\uparrow} \hat{b}_{j\downarrow}+\hat{b}^\dagger_{j\downarrow} \hat{b}_{j\uparrow}),
\end{eqnarray}
%%%%%%
with a real-value $\Delta$ making the second term anti-Hermitian.

Applying the Fourier transformation, we can rewrite $\hat{H}'_{0}$ as
%%%%%%
\begin{eqnarray}
\hat{H}'_{0}&=& \sum_{k} \hat{b}^\dagger_{k\sigma} h'_{\sigma\sigma'}(k) \hat{b}_{k\sigma'},
\end{eqnarray}
%%%%%%
where 
we have imposed PBC.
The first quantized Hamiltonian $h'(k)$ is written as
%%%
\begin{eqnarray}
h'(k)=
\left(
\begin{array}{cc}
h_{\uparrow} & i\Delta \\
i\Delta  & h_{\downarrow}
\end{array}
\right)
\end{eqnarray}
%%%
with $h_{\uparrow}(k)=t_+e^{ik}+t_-e^{-ik}$ and $h_{\downarrow}(k)=h^*_{\uparrow}(k)$.

Eigenvalues of $h'(k)$ are 
%%%%%%
\begin{eqnarray}
\epsilon'_{\pm}(k)&=& (t_+ +t_-)\cos k \pm i\sqrt{ (t_+ -t_-)^2\sin^2 k +\Delta^2}.
\end{eqnarray}
%%%%%%
For $\Delta >0$, the imaginary part of $\epsilon'_{\pm}$ remains finite, which makes the system trivial in terms of the point-gap topology.

\begin{figure}[!h]
\begin{minipage}{0.45\hsize}
\begin{center}
\includegraphics[width=1\hsize,clip]{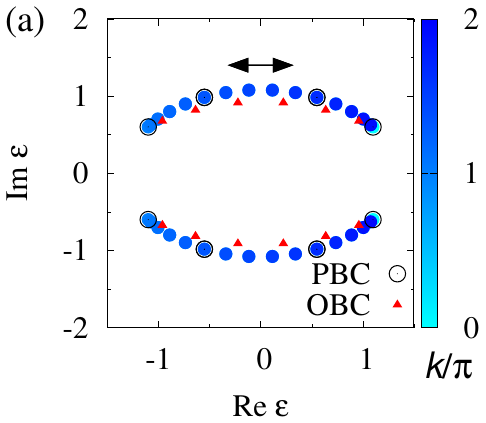}
\end{center}
\end{minipage}
\begin{minipage}{0.45\hsize}
\begin{center}
\includegraphics[width=1\hsize,clip]{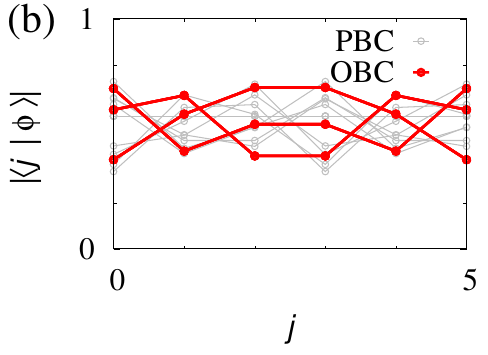}
\end{center}
\end{minipage}
\caption{
(a): Eigenvalues of the one-body Hamiltonian $h'(k)$.
Open circles (closed triangle) denote the data under PBC (OBC) for $L=6$.
Closed circles denote the data under PBC for $L=30$. The color of the closed circles represent the momentum $k$.
(b): Amplitude of the right eigenvectors.
The gray (red) lines denote the data under PBC (OBC) for $L=6$.
These data are obtained for $t_+=1$, $t_-=0.1$, and $\Delta=0.6$.
}
\label{fig: spin U(1) breaking 1bdy app}
\end{figure}

Figure~\ref{fig: spin U(1) breaking 1bdy app}(a) displays the destruction of the loop structure observed for $\Delta=0$ [see Figs.~\ref{fig: Boson 1bdy}(a)~and~\ref{fig: Boson 1bdy}(b)].
Correspondingly, skin modes observed for $\Delta=0$ [see Fig.~\ref{fig: Boson 1bdy}(c)~and~\ref{fig: Boson 1bdy}(d)] disappear, which can be seen in Fig~\ref{fig: spin U(1) breaking 1bdy app}(b).
%%%%%%%%%%%%%%%%%%%%%%%%%
%%%%%%%%%%%%%%%%%%%%%%%%%

\begin{figure}[!h]
\begin{minipage}{0.45\hsize}
\begin{center}
\includegraphics[width=1\hsize,clip]{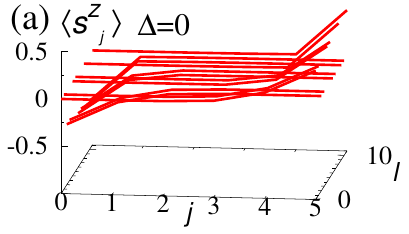}
\end{center}
\end{minipage}
\begin{minipage}{0.45\hsize}
\begin{center}
\includegraphics[width=1\hsize,clip]{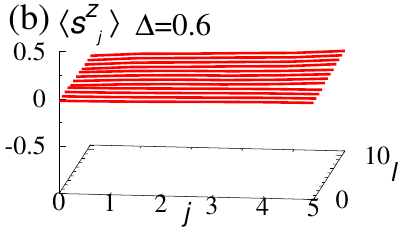}
\end{center}
\end{minipage}
\caption{
Expectation values $\langle u_{\mathrm{R}l}  | s^z_j | u_{\mathrm{R}l} \rangle/ \langle u_{\mathrm{R}l} | u_{\mathrm{R}l} \rangle $ at each site $j$ under OBC. 
Panel~(a) [(b)] displays the data for $\Delta=0$ [$\Delta=0.6$].
}
\label{fig: 1bdy szj  app}
\end{figure}

Due to the destruction of the skin effect, expectation values 
$
\langle u_{\mathrm{R}l}  | s^z_j | u_{\mathrm{R}l} \rangle/ \langle u_{\mathrm{R}l} | u_{\mathrm{R}l} \rangle
$
at each site are also suppressed by the spin-mixing term [see Fig.~\ref{fig: 1bdy szj  app}]. Here $s^z_j$ denotes $z$-component of the first quantized spin operator at site $j$.
Right eigenstates of the first quantized Hamiltonian $h'$ are denoted by $| u_{\mathrm{R}l} \rangle $ with $l=1,2,\ldots,2L$ labeling the eigenstates.

We note that the expectation values of spin $\langle u_{\mathrm{R}l}|s^z_j| u_{\mathrm{R}l} \rangle/\langle u_{\mathrm{R}l}| u_{\mathrm{R}l} \rangle $ is small but finite. However, this fact does not indicates the Mott skin effect because the expectation values of spin become exactly zero under continuous deformation keeping the point-gap: decreasing $t_+$ to $t_-$ for $\Delta >0$.
We recall that the definition of the Mott skin effect is satisfying the following two conditions. (1) The many-body spin winding number is finite. (2) The eigenvalues and eigenstates exhibit extreme sensitivity to boundary conditions but only the spin degree of freedom is involved to the sensitivity of the eigenstates.

%%%%%%%%%%%%%%%%%%%%%%%%%
%%%%%%%%%%%%%%%%%%%%%%%%%

%%%%%%%%%%%%%%
\section{
Details of computation of $W_{\mathrm{s}}$
}
\label{sec: Mbdy W app}
%%%%%%%%%%%%%%

Let us discuss the computation of $W_{\mathrm{s}}$ in details.
Under twisted boundary conditions, the non-interacting Hamiltonian is written as
%%%%%%
\begin{eqnarray}
\label{eq: H0theta app}
\hat{H}_0(\theta_{\mathrm{s}})&=& \sum_{\sigma=\uparrow,\downarrow}\left[ \sum^{L-2}_{j=0}
\left(
-t_{\mathrm{R} \sigma} \hat{b}^\dagger_{j+1\sigma} \hat{b}_{j\sigma} - t_{\mathrm{L}\sigma} \hat{b}^\dagger_{j\sigma} \hat{b}_{j+1\sigma}
\right) \right. \nonumber \\
&&
\left.
-t_{\mathrm{R} \sigma}e^{i\theta_{\sigma}} \hat{b}^\dagger_{0\sigma} \hat{b}_{L-1\sigma} - t_{\mathrm{L}\sigma}e^{i\theta_{\sigma}} \hat{b}^\dagger_{L-1\sigma} \hat{b}_{0\sigma}
\right]
\end{eqnarray}
%%%%%%
with $\theta_{\uparrow}=\theta_{\mathrm{s}}$ and $\theta_{\downarrow}=-\theta_{\mathrm{s}}$.
As mentioned below Eq.~(\ref{eq: H_0}), $t_{\mathrm{R}\sigma}$ and $t_{\mathrm{L}\sigma}$ are chosen as $t_{\mathrm{R}\uparrow}=t_{\mathrm{L}\downarrow}=t_+$ and $t_{\mathrm{L}\uparrow}=t_{\mathrm{R}\downarrow}=t_-$.

%%%%%%%%%%%%%%%%%%%%%%%%%
\begin{figure}[!h]
\begin{minipage}{0.5\hsize}
\begin{center}
\includegraphics[width=1\hsize,clip]{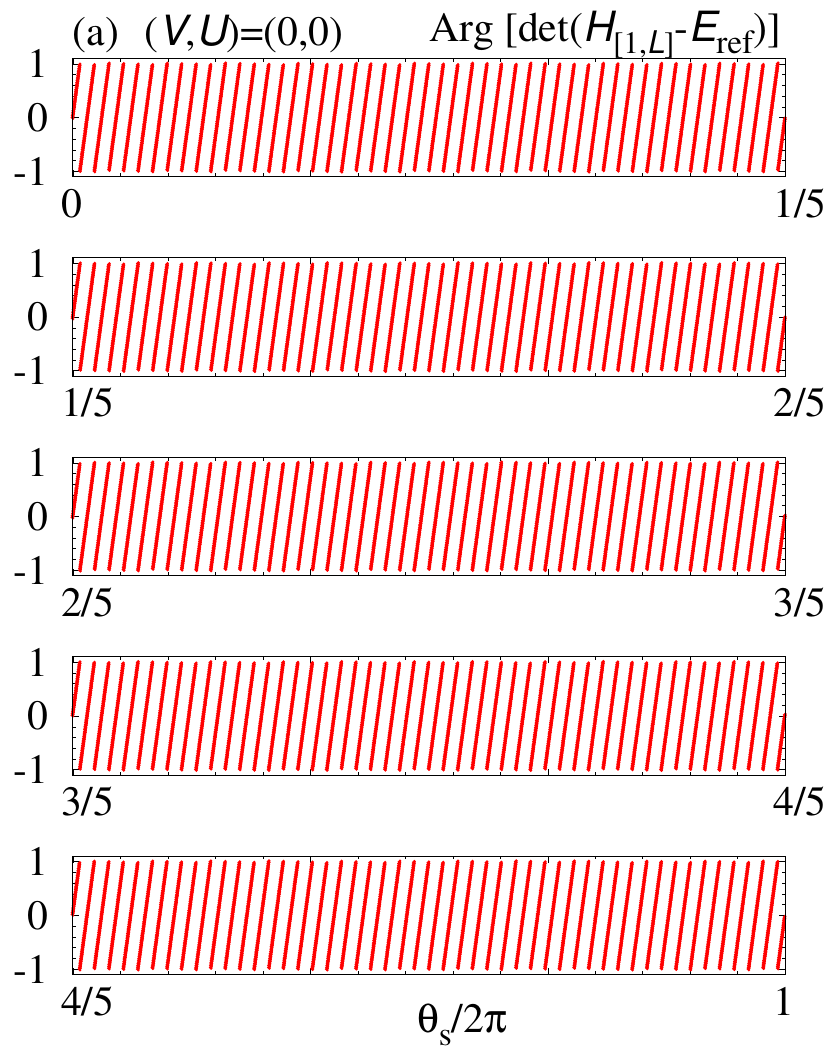}
\end{center}
\end{minipage}
\begin{minipage}{0.5\hsize}
\begin{center}
\includegraphics[width=1\hsize,clip]{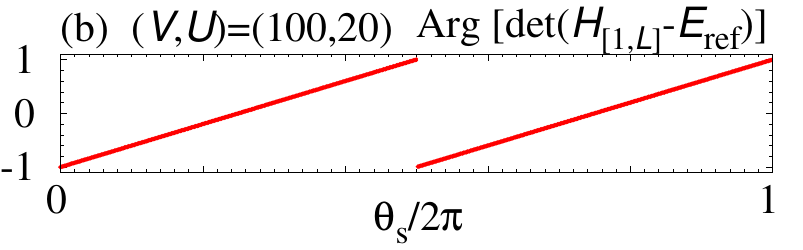}
\end{center}
\end{minipage}
\caption{
Argument $\mathrm{Arg}[\mathrm{det}(\hat{H}_{{[1,L-1]}}-E_{\mathrm{ref}})]$ as functions of $\theta_{\mathrm{s}}$ with $E_{\mathrm{ref}}=-0.01i$.
Panel~(a)~[(b)] displays the data for $(V,U)=(0,0)$ [(100,20)].
}
\label{fig: Arg_Mbdy_H app}
\end{figure}
%%%%%%%%%%%%%%%%%%%%%%%%%

Equation~(\ref{eq: Ws}) indicates that the many-body winding number $W_{\mathrm{s}}$ can be extracted from $\theta_{\mathrm{s}}$ dependence of the argument, $\mathrm{Arg}[\mathrm{det}(\hat{H}_{{[1,L-1]}}-E_{\mathrm{ref}})]$.
Figure~\ref{fig: Arg_Mbdy_H app}(a) displays the argument as a function of $\theta$ for $(V,U)=(0,0)$ and $E_{\mathrm{ref}}=-0.01i$.
This figure indicates that the winding number takes $W_{\mathrm{s}}=245$ for $E_{\mathrm{ref}}=-0.01i$ in the non-interacting case.
Figure~\ref{fig: Arg_Mbdy_H app}(b) displays the argument as a function of $\theta$ for $(V,U)=(100,20)$ and $E_{\mathrm{ref}}=-0.01i$.
This figure indicates that the winding number takes $W_{\mathrm{s}}=2$ for $E_{\mathrm{ref}}=-0.01i$ for the correlated case.

%%%%%%%%%%%%%%
\section{
Expectation values for PBC
}
\label{sec: N Sz PBC app}
%%%%%%%%%%%%%%

\begin{figure}[!h]
\begin{minipage}{0.49\hsize}
\begin{center}
\includegraphics[width=0.75\hsize,clip]{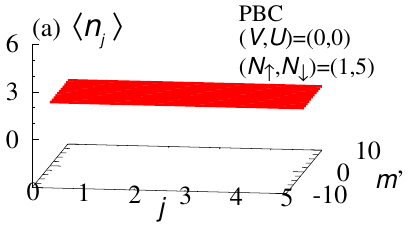}
\end{center}
\end{minipage}
\begin{minipage}{0.49\hsize}
\begin{center}
\includegraphics[width=0.75\hsize,clip]{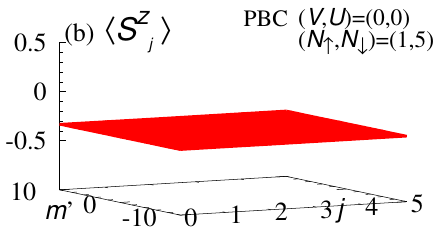}
\end{center}
\end{minipage}
\caption{
The expectation values for $(V,U)=(0,0)$ under PBC.
Panel~(a)~[(b)] displays $\langle \hat{n}_j\rangle$ [$\langle \hat{S}^z_j\rangle$].
The other eigenstates also show essentially the same behaviors.
These data are obtained for $(t_+,t_-)=(1,0.1)$ and $L=6$.
These figures are plotted in a similar way as Figs.~\ref{fig: Boson Mbdy U0V0}(c)~and~\ref{fig: Boson Mbdy U0V0}(d).
}
\label{fig: N Sz PBC V0U0 app}
\end{figure}

We provide the expectation values under PBC.
Under PBC, the expectation values $\langle \hat{n}_j\rangle$ and $\langle \hat{S}^z_j\rangle$ are homogeneous for the non-interacting case (see Fig.~\ref{fig: N Sz PBC V0U0 app}),

%%%%%%%%%%%%%%%%%%%%%%%%%
%%%%%%%%%%%%%%%%%%%%%%%%%

\begin{figure}[!h]
\begin{minipage}{0.49\hsize}
\begin{center}
\includegraphics[width=0.75\hsize,clip]{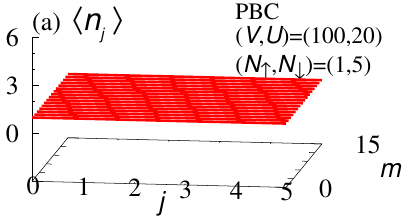}
\end{center}
\end{minipage}
\begin{minipage}{0.49\hsize}
\begin{center}
\includegraphics[width=0.75\hsize,clip]{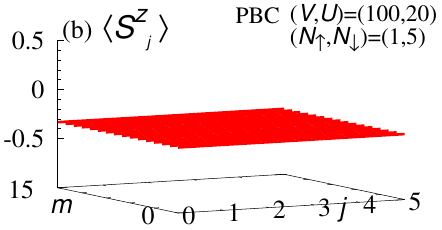}
\end{center}
\end{minipage}
\caption{
The expectation values for $(V,U)=(100,20)$ under PBC.
Panel~(a)~[(b)] displays $\langle \hat{n}_j\rangle$ [$\langle \hat{S}^z_j\rangle$].
The other eigenstates also show essentially the same behaviors.
These data are obtained for $(t_+,t_-)=(1,0.1)$ and $L=6$.
These figures are plotted in a similar way as Figs.~\ref{fig: Boson Mbdy U20V100}(c)~and~\ref{fig: Boson Mbdy U20V100}(d).
The definition of $m$ labeling eigenstates is given in footnote ~\cite{labelE_ftnt}.
}
\label{fig: N Sz PBC V100U20}
\end{figure}

These properties do not change even in the presence of the interactions.
Figure~\ref{fig: N Sz PBC V100U20} displays the expectation values $\langle \hat{n}_j\rangle$ and $\langle \hat{S}^z_j\rangle$ under PBC for $(V,U)=(100,20)$.
%%%%%%%%%%%%%%%%%%%%%%%%%
%%%%%%%%%%%%%%%%%%%%%%%%%

%%%%%%%%%%%%%%
\section{
Results for $(N_{\uparrow},N_{\downarrow})=(2,4)$
}
\label{sec: (Nup,Ndow)=(2,4) app}
%%%%%%%%%%%%%%

Here, we provide the results for subspace with $(N_{\uparrow},N_{\downarrow})=(2,4)$ and $L=6$ which indicates that the Mott skin effect is still observed. 

\begin{figure}[!h]
\begin{minipage}{0.49\hsize}
\begin{center}
\includegraphics[width=0.75\hsize,clip]{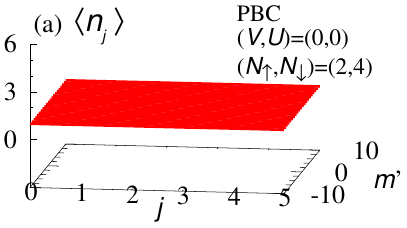}
\end{center}
\end{minipage}
\begin{minipage}{0.49\hsize}
\begin{center}
\includegraphics[width=0.75\hsize,clip]{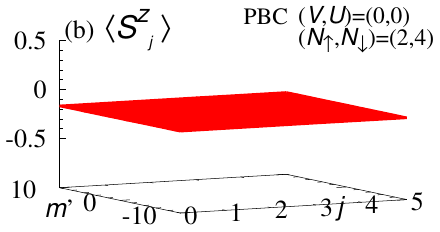}
\end{center}
\end{minipage}
\begin{minipage}{0.49\hsize}
\begin{center}
\includegraphics[width=0.75\hsize,clip]{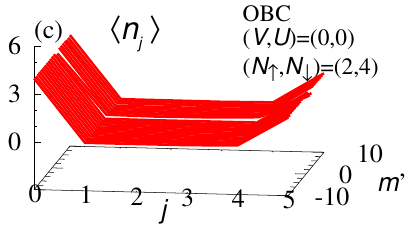}
\end{center}
\end{minipage}
\begin{minipage}{0.49\hsize}
\begin{center}
\includegraphics[width=0.75\hsize,clip]{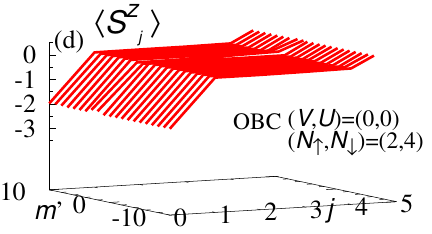}
\end{center}
\end{minipage}
\caption{
The expectation values for the subspace with $(N_{\uparrow},N_{\downarrow})=(2,4)$ and for $(V,U)=(0,0)$.
Panels~(a)~and~(b) [(c)~and~(d)] are data under PBC [OBC].
Panels~(a)~and~(c) [(b)~and~(d)] display $\langle \hat{n}_j \rangle$ [$\langle \hat{S}^z_j \rangle$].
The expectation values are plotted for eigenstates whose eigenvalues $E_{m'}$ are located around the origin of the complex plane. Specifically, $m'$ is defined as $m'=m-1321$ (for the definition of $m$, see footnote~\cite{labelE_ftnt}). 
For OBC, the eigenvalues are aligned on the real axis, and the eigenvalue is numerically zero for $m=1321$ ($E_{1321}\sim 0$).
We note that the other states show essentially the same behaviors. 
The data are obtained for $(t_+,t_-)=(1,0.1)$ and $L=6$. 
}
\label{fig: U0V0_Nup2Ndow4 PBC-OBC app}
\end{figure}

Let us start with the non-interacting case where the ordinary non-Hermitian skin effect is observed. Specifically, under PBC, the expectation values $\langle \hat{n}_j \rangle $ and $\langle \hat{S}^z_j \rangle $ are homogeneous [see Figs.~\ref{fig: U0V0_Nup2Ndow4 PBC-OBC app}(a)~and~\ref{fig: U0V0_Nup2Ndow4 PBC-OBC app}(b)]. 
Imposing OBC results in the localization of eigenstates which is reflected in both expectation values $\langle \hat{n}_j \rangle $ and $\langle \hat{S}^z_j \rangle $ [see Figs.~\ref{fig: U0V0_Nup2Ndow4 PBC-OBC app}(c)~and~\ref{fig: U0V0_Nup2Ndow4 PBC-OBC app}(d)].

%%%%%%%%%%%%%%%%%%%%%%%%%
%%%%%%%%%%%%%%%%%%%%%%%%%

\begin{figure}[!h]
\begin{minipage}{0.49\hsize}
\begin{center}
\includegraphics[width=0.75\hsize,clip]{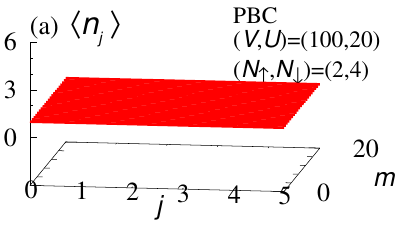}
\end{center}
\end{minipage}
\begin{minipage}{0.49\hsize}
\begin{center}
\includegraphics[width=0.75\hsize,clip]{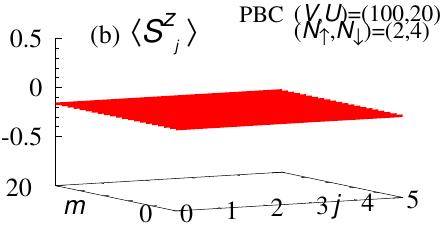}
\end{center}
\end{minipage}
\begin{minipage}{0.49\hsize}
\begin{center}
\includegraphics[width=0.75\hsize,clip]{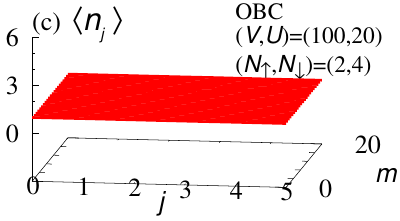}
\end{center}
\end{minipage}
\begin{minipage}{0.49\hsize}
\begin{center}
\includegraphics[width=0.75\hsize,clip]{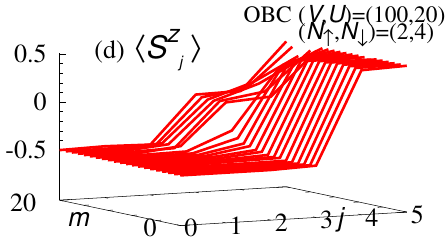}
\end{center}
\end{minipage}
\caption{
The expectation values for the subspace with $(N_{\uparrow},N_{\downarrow})=(2,4)$ and for $(V,U)=(100,20)$.
Panels~(a)~and~(b) [(c)~and~(d)] are data under PBC.
Panels~(a)~and~(c) [(b)~and~(d)] display $\langle \hat{n}_j \rangle$ [$\langle \hat{S}^z_j \rangle$].
These figures are plotted in a similar way as Figs.~\ref{fig: Boson Mbdy U0V0}(c)~and~\ref{fig: Boson Mbdy U0V0}(d).
The data are obtained for $(t_+,t_-)=(1,0.1)$ and $L=6$.
}
\label{fig: U20V100_Nup2Ndow4 PBC-OBC app}
\end{figure}

In contrast to the non-interacting case, for $(V,U)=(100,20)$, the emergence of the skin modes is observed only in the spin degree of freedom. 
Firstly, we note that under PBC, the expectation values $\langle \hat{n}_j \rangle $ and $\langle \hat{S}^z_j \rangle $ are homogeneous [see Figs.~\ref{fig: U20V100_Nup2Ndow4 PBC-OBC app}(a)~and~\ref{fig: U20V100_Nup2Ndow4 PBC-OBC app}(b)]. 
Imposing OBC results in localization in the spin degree of freedom although such localization is not observed in the charge degree of freedom [see Figs.~\ref{fig: U20V100_Nup2Ndow4 PBC-OBC app}(c)~and~\ref{fig: U20V100_Nup2Ndow4 PBC-OBC app}(d)].

%%%%%%%%%%%%%%%%%%%%%%%%%
%%%%%%%%%%%%%%%%%%%%%%%%%

The above results indicate that the Mott skin effect is observed even for the subspace with $(N_{\uparrow},N_{\downarrow})=(2,4)$ and $L=6$.

%%%%%%%%%%%%%%
\section{
Time-evolution under PBC
}
\label{sec: TimeEvol PBC app}
%%%%%%%%%%%%%%
We discuss the time-evolution of the expectation values $\langle \hat{n}_j(t)\rangle$ and $\langle \hat{S}^z_j(t) \rangle $ under PBC.
In a similar way to the case of OBC, we compute the expectation values $\langle \hat{n}_j(t)\rangle$ and $\langle \hat{S}^z_j(t) \rangle $ [see Eq.~(\ref{eq: n(t)})].
The initial state is chosen so that all the bosons occupy site $j=0$.

\begin{figure}[!h]
\begin{minipage}{0.49\hsize}
\begin{center}
\includegraphics[width=0.75\hsize,clip]{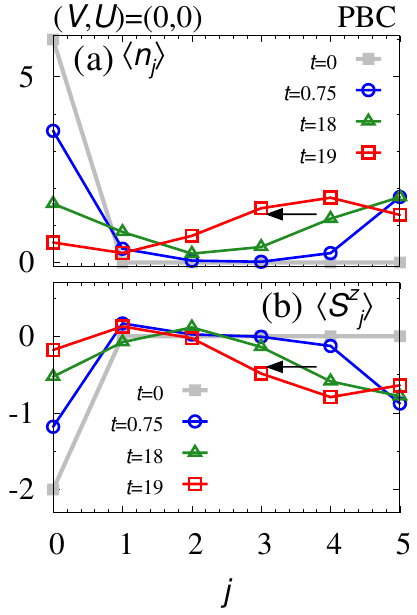}
\end{center}
\end{minipage}
\begin{minipage}{0.49\hsize}
\begin{center}
\includegraphics[width=0.75\hsize,clip]{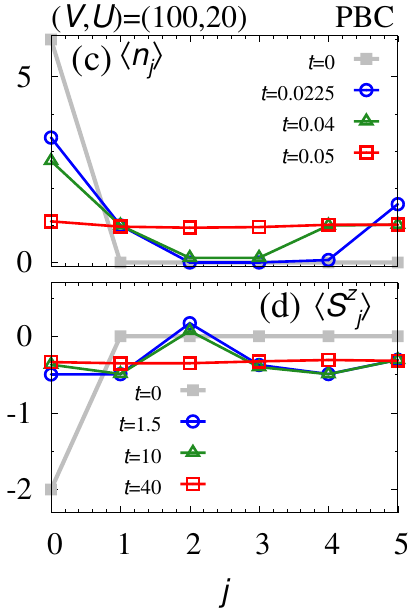}
\end{center}
\end{minipage}
\caption{
The time-evolution of the expectation values $\langle \hat{n}_j(t)\rangle$ and $\langle \hat{S}^z_j(t) \rangle $ under PBC [see Eq.~(\ref{eq: n(t)})]
These figures are plotted in a similar way to Fig.~\ref{fig: TimeEvol OBC}
These data are obtained for $(t_+,t_-)=(1,0.1)$ and $L=6$.
}
\label{fig: TimeEvol PBC}
\end{figure}

Figures~\ref{fig: TimeEvol PBC}(a)~and~\ref{fig: TimeEvol PBC}(b) display the data for the non-interacting case.
In contrast to the case of OBC, bosons do not localize around the left edge $j=0$ due to the absence of the skin modes.
In addition, the bosons propagate to left, which is attributed to the fact that the eigenstates with the longest lifetime $\tau$ ($\tau\sim1/\mathrm{Im}E$) are composed of bosons propagating to left.
%%%%%%%%%%%%%%%%%%%%%%%%%
%%%%%%%%%%%%%%%%%%%%%%%%%

Figures~\ref{fig: TimeEvol PBC}(c)~and~\ref{fig: TimeEvol PBC}(d) display the data for $(V,U)=(100,20)$.
In this case, increasing time $t$, the expectation values $\langle \hat{n}_j(t)\rangle$ and $\langle \hat{S}^z_j(t) \rangle $ become homogeneous.
This is because the eigenstate with the longest lifetime $\tau$ ($\tau\sim1/\mathrm{Im}E$) is the state described by a fermion with zero momentum in terms of the effective fermion model~(\ref{eq: Hspin_f simple}).

%%%%%%%%%%%%%%
\section{
An open quantum system and relation to Eq.~(\ref{eq: H}).
}
\label{sec: OQS app}
%%%%%%%%%%%%%%

Our toy model [Eq.~(\ref{eq: H})] is relevant to an open quantum system with dissipation. Specifically, the toy model corresponds to an effective Hamiltonian describing dynamics without quantum jumps. 
In addition, the dynamical spin accumulation discussed in the main text [see Fig.~\ref{fig: TimeEvol OBC}] is also observed even in the presence of quantum jumps.

%%%%%%%%%%%%%%
\subsection{
Model
}
\label{sec: L model app}
%%%%%%%%%%%%%%

Consider an open quantum system whose dynamics follows the Lindblad equation~\cite{Lindblad_CommMathPhys1976,Gorini_JMathPhys1976,Breuer_textbook2007}
%%%%%%
\begin{eqnarray}
i\partial_t\hat{\rho}(t) &=& \spopf{L}[\hat{\rho}(t)], \\
\spopf{L}[\hat{\rho}(t)] &=& [\hat{H}_\mathrm{sys}, \hat{\rho}(t)] +i\sum_{\alpha} \left( \hat{L}_\alpha \hat{\rho}(t) \hat{L}^\dagger_\alpha -\frac{1}{2} \{ \hat{L}^\dagger_\alpha \hat{L}_\alpha,  \hat{\rho}(t) \}  \right),
\end{eqnarray}
%%%%%%
where $\hat{L}_{\alpha}$ are Lindblad operators describing dissipation, and $\alpha$ specifies types of dissipation as well as site and spin.
The density matrix of the system is denoted by $\rho$.
Square brackets and curly brackets denote $[A,B]=AB-BA$ and $\{A,B\}=AB+BA$ with operators $A$ and $B$.
The Hermitian Hamiltonian of the system is denoted by
%%%%%%
\begin{eqnarray}
\hat{H}_\mathrm{sys} &=& t_0 \sum_{j\sigma} ( \hat{b}^\dagger_{j+1\sigma} \hat{b}_{j\sigma} +\hat{b}^\dagger_{j\sigma} \hat{b}_{j+1\sigma} ) +V_{\mathrm{R}} \sum_{j\sigma} \hat{n}_{j\sigma}(\hat{n}_{j\sigma}-1) +U_{\mathrm{R}} \sum_{j}\hat{n}_{j\uparrow} \hat{n}_{j\downarrow}.
\end{eqnarray}
%%%%%%
Parameters $t_0$, $V_\mathrm{R}$, and $U_\mathrm{R}$ take real values, respectively.
Here, we suppose that the system is under the one- and two-body losses defined as
%%%%%%
\begin{eqnarray}
\hat{L}_{1j\sigma} &=& \sqrt{\gamma_1}(\hat{b}_{j\sigma} -i \hat{b}_{j'\sigma}), \\
\hat{L}_{2j\sigma} &=& \sqrt{\gamma_2}\hat{b}_{j\sigma}\hat{b}_{j\sigma}, \\
\hat{L}_{3j\sigma} &=& \sqrt{\gamma_3}\hat{b}_{j\uparrow}\hat{b}_{j\downarrow}\delta_{\sigma,\uparrow},
\end{eqnarray}
%%%%%%
with $j'=j+\mathrm{sgn}(\sigma)$ and $\gamma_l$ ($l=1,2,3$) taking non-negative values.
We have introduced $\mathrm{sgn}(\sigma)$ and $\delta_{\sigma,\uparrow}$ defined as follows:
$\mathrm{sgn}(\sigma)$ taking $1$ ($-1$) for $\sigma=\uparrow$ ($\downarrow$);
$\delta_{\sigma,\uparrow}$ takes $1$ ($0$) for $\sigma=\uparrow$ ($\downarrow$).

Focusing on the dynamics without quantum jumps, the time-evolution is described by the following effective Hamiltonian
%%%%%%
\begin{eqnarray}
\label{eq: Heff app}
\hat{H}_{\mathrm{eff}} &=& \hat{H}_{\mathrm{sys}} - i\sum_{lj\sigma}\frac{ \gamma_l }{2} \hat{L}^\dagger_{lj\sigma} \hat{L}_{lj\sigma}, \\
&=&
\sum_{j\sigma} \left[ \left(t_0+\frac{\mathrm{sgn}(\sigma)}{2}\gamma_1 \right) \hat{b}^\dagger_{j+1\sigma} \hat{b}_{j\sigma} + \left(t_0-\frac{\mathrm{sgn}(\sigma)}{2}\gamma_1 \right)  \hat{b}^\dagger_{j\sigma} \hat{b}_{j+1\sigma} \right] -i\gamma_1 \sum_{j\sigma}\hat{n}_{j\sigma} \nonumber \\ 
&&\quad+\left(V_{\mathrm{R}}-i\frac{\gamma_2}{2}\right) \sum_{j\sigma} \hat{n}_{j\sigma}(\hat{n}_{j\sigma}-1) +\left(U_{\mathrm{R}}-i\frac{\gamma_3}{2}\right) \sum_{j}\hat{n}_{j\uparrow} \hat{n}_{j\downarrow}
\end{eqnarray}
%%%%%%
which corresponds to our toy model [Eq.~(\ref{eq: H})]. Here, we have imposed PBC.

%%%%%%%%%%%%%%
\subsection{
Analysis of $\hat{H}_{\mathrm{eff}}$
}
\label{sec: Heff app}
%%%%%%%%%%%%%%

In this section, we confirm that $\hat{H}_{\mathrm{eff}}$ [Eq.~(\ref{eq: Heff app})] exhibits the Mott skin effect.

\begin{figure}[!h]
%
%%%%%%
\begin{minipage}{1\hsize}
\begin{center}
\includegraphics[width=1\hsize,clip]{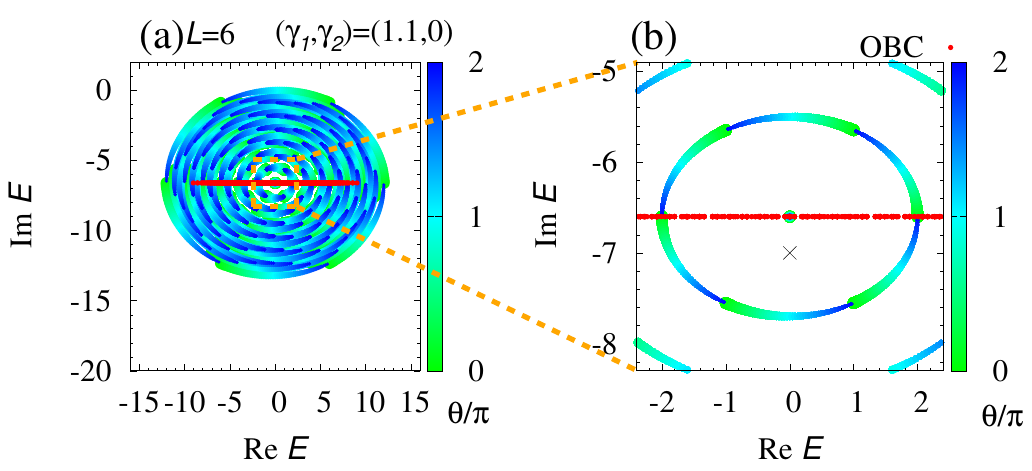}
\end{center}
\end{minipage}
%%%%%%
%
\begin{minipage}{0.49\hsize}
\begin{center}
\includegraphics[width=1\hsize,clip]{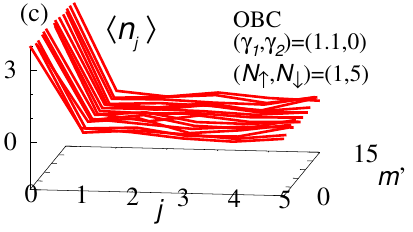}
\end{center}
\end{minipage}
\begin{minipage}{0.49\hsize}
\begin{center}
\includegraphics[width=1\hsize,clip]{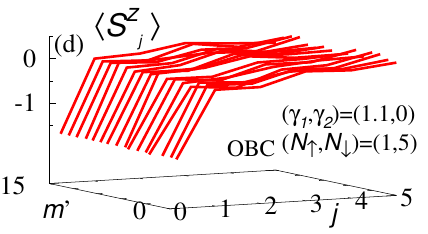}
\end{center}
\end{minipage}
\caption{
Energy eigenvalues of $\hat{H}_{\mathrm{eff}}$ [Eq.~(\ref{eq: Heff app})] and expectation values for $(\gamma_1,\gamma_2,\gamma_3)=(1.1,0,0)$.
(a): Energy eigenvalues under twisted boundary conditions and OBC (red), respectively.
(b): A magnified version of the range $-8.3 \leq \mathrm{Im}E \leq -4.9$.
The data obtained under OBC are represented by red dots.
With increasing $\theta_{\mathrm{s}}$ from $0$ to $2\pi$, the eigenvalues wind the point denoted by the cross, which indicates $W_{\mathrm{s}}$ takes a non-zero value for $E_{\mathrm{ref}}=-7i$.
(c) [(d)]: Expectation values $\langle \hat{n}_j \rangle$ [$\langle \hat{S}^z_j \rangle$] at each site under OBC.
As is the case of Fig.~\ref{fig: Boson Mbdy U0V0}, the expectation values are computed from right eigenvectors of $\hat{H}_{\mathrm{eff}}$ whose eigenvalues are located around $E=-6.5i$.
Here, $m'$ labels these eigenvalues as $\mathrm{Re} E_{m'}<\mathrm{Re} E_{m'+1}$, $m'=m-756$. For $m=756$, the eigenvalue is numerically zero (for the definition of $m$ see footnote~\cite{labelE_ftnt}).
We note that the other states show essentially the same behaviors. 
These data are obtained for subspace with $(N_{\uparrow},N_{\downarrow})=(1,5)$ and for $t_0=1$, $V_{\mathrm{R}}=U_{\mathrm{R}}=0$, and $L=6$.
}
\label{fig: Heff ga11gb0gc0}
\end{figure}

Let us start with the non-interacting case where two-body losses and interactions are absent ($\gamma_2=\gamma_3=0$ and $U_{\mathrm{R}}=V_{\mathrm{R}}=0$).
Figures~\ref{fig: Heff ga11gb0gc0}(a)~and~\ref{fig: Heff ga11gb0gc0}(b) indicate that the many-body spin winding number takes a non-trivial value for $E_{\mathrm{ref}}=-7i$.
Correspondingly, the skin effect is observed for both charge and spin degrees of freedom [see Figs.~\ref{fig: Heff ga11gb0gc0}(c)~and~\ref{fig: Heff ga11gb0gc0}(d)].

%%%%%%%%%%%%%%%%%%%%%%%%%
%%%%%%%%%%%%%%%%%%%%%%%%%

\begin{figure}[!h]
%
%%%%%%
\begin{minipage}{1\hsize}
\begin{center}
\includegraphics[width=1\hsize,clip]{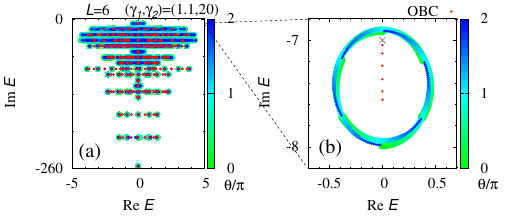}
\end{center}
\end{minipage}
%%%%%%
%
\begin{minipage}{0.49\hsize}
\begin{center}
\includegraphics[width=1\hsize,clip]{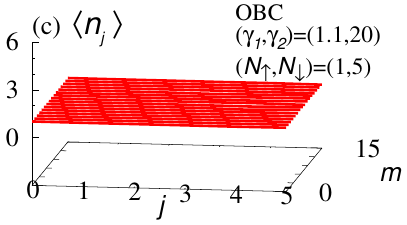}
\end{center}
\end{minipage}
\begin{minipage}{0.49\hsize}
\begin{center}
\includegraphics[width=1\hsize,clip]{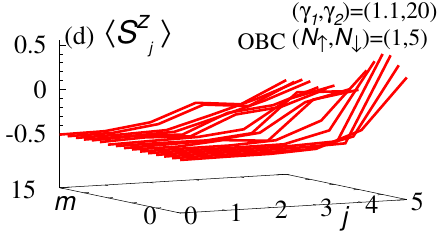}
\end{center}
\end{minipage}
\caption{
Energy eigenvalues of $\hat{H}_{\mathrm{eff}}$ [Eq.~(\ref{eq: Heff app})] and expectation values for $(\gamma_1,\gamma_2,\gamma_3)=(1.1,20,20)$.
(a): Energy eigenvalues under twisted boundary conditions and OBC, respectively.
(b): A magnified version of the range $-8.2 \leq \mathrm{Im}E \leq -6.8$. In these figures the data obtained under OBC are represented by red dots.
With increasing $\theta_{\mathrm{s}}$ from $0$ to $2\pi$, the eigenvalues wind the point denoted by the cross, which indicates $W_{\mathrm{s}}=2$ for $E_{\mathrm{ref}}=-7i$.
(c) [(d)]: Expectation values $\langle \hat{n}_j \rangle$ [$\langle \hat{S}^z_j \rangle$] at each site under OBC.
Here, the expectation values are plotted for eigenstates~\cite{labelE_ftnt} whose eigenvalues are denoted by red dots in panel~(b).
As is the case of Fig.~\ref{fig: Boson Mbdy U20V100}, the expectation values are computed from the right eigenvectors (see Fig.~\ref{fig: Boson Mbdy U0V0}).
These data are obtained for subspace with $(N_{\uparrow},N_{\downarrow})=(1,5)$ and for $t_0=1$, $V_{\mathrm{R}}=U_{\mathrm{R}}=0$, and $L=6$.
}
\label{fig: Heff ga11gb20gc20}
\end{figure}

In the presence of two-body losses ($\gamma_2$, and $\gamma_3$), $\hat{H}_{\mathrm{eff}}$ becomes non-Hermitian Hamiltonian with two-body interactions.
Figure~\ref{fig: Heff ga11gb20gc20} displays the results for $\gamma_2=\gamma_3=20$. 
In this case, two-body interactions due to dissipation induce the Mott skin effect.
Figures~\ref{fig: Heff ga11gb20gc20}(a)~and~\ref{fig: Heff ga11gb20gc20}(b) indicate that the many-body spin winding number $W_{\mathrm{s}}$ takes two for $E_{\mathrm{ref}}=-7i$.
Correspondingly, the skin effect is observed only for spin degree of freedom [see Figs.~\ref{fig: Heff ga11gb20gc20}(c)~and~\ref{fig: Heff ga11gb20gc20}(d)], indicating the emergence of the Mott skin effect.

%%%%%%%%%%%%%%%%%%%%%%%%%
%%%%%%%%%%%%%%%%%%%%%%%%%
%
%
The Mott skin effect discussed above is observed for other sets of parameters.
For systematic analysis, we introduce charge and spin polarizations, $P_{\mathrm{c}}$ and $P_{\mathrm{s}}$, which are computed as
%%%%%%
\begin{eqnarray}
P_{\mathrm{c}}&=& P_{\uparrow}+P_{\downarrow}, \\
P_{\mathrm{s}}&=& \frac{1}{2}(P_{\uparrow}-P_{\downarrow}),
\end{eqnarray}
%%%%%%
with
%%%%%%
\begin{eqnarray}
P_{\sigma} &=& {\textstyle \sum'_{m}} \left[ \sum_{j} \frac{ {}_\mathrm{R}\langle \Psi_m | j \hat{n}_{j\sigma} |\Psi_m \rangle_{\mathrm{R}} } {  {}_\mathrm{R}\langle \Psi_m |\Psi_m \rangle_{\mathrm{R}}  } \right].
\end{eqnarray}
%%%%%%
Here, $\sum'_{m}$ denotes summation taking over eigenstates whose imaginary part of eigenstates are 6-th largest; $0 \geq \mathrm{Im}E_0\geq \mathrm{Im}E_1 \geq \ldots \geq \mathrm{Im}E_5$.

\begin{figure}[!h]
%
%%%%%%
\begin{minipage}{0.45\hsize}
\begin{center}
\includegraphics[width=1\hsize,clip]{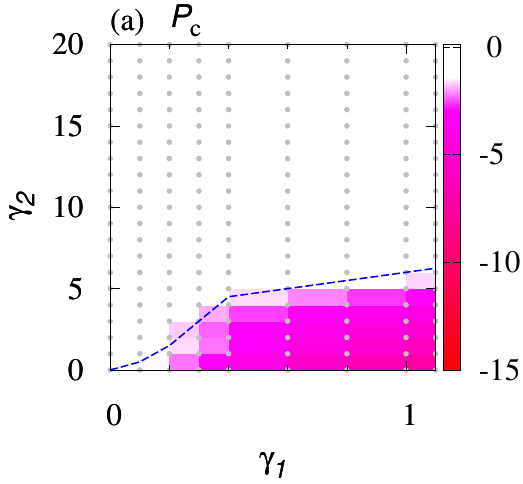}
\end{center}
\end{minipage}
%%%%%%
%
\begin{minipage}{0.45\hsize}
\begin{center}
\includegraphics[width=1\hsize,clip]{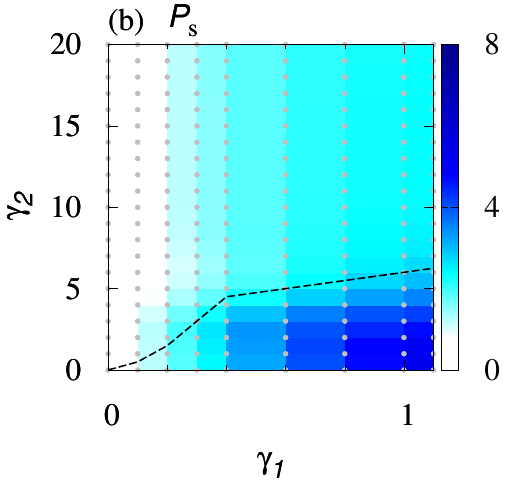}
\end{center}
\end{minipage}
\caption{
Charge and spin polarizations under OBC for $\gamma_{2}=\gamma_{3}$ and $V_{\mathrm{R}}=U_{\mathrm{R}}=0$.
(a): Charge polarization $P_{\mathrm{c}}$.
(b): Spin polarization $P_{\mathrm{s}}$.
In the region where only the spin degree of freedom exhibits the polarization, the Mott skin effect is observed.
Dashed line denotes the parameters where $P_{\mathrm{c}}$ approaches zero.
These data are obtained for subspace with $(N_{\uparrow},N_{\downarrow})=(1,5)$ and for $t_0=1$, and $L=6$.
}
\label{fig: Heff P}
\end{figure}

The charge polarization $P_{\mathrm{c}}$ (spin polarization $P_{\mathrm{s}}$) captures the localization of charge (spin) degrees of freedom around the edge.
Figrue~\ref{fig: Heff P} displays a color plot of polarizations under OBC against $\gamma_1$ and $\gamma_2$ for $\gamma_{2}=\gamma_{3}$ and $V_{\mathrm{R}}=U_{\mathrm{R}}=0$.
In this figure, we can see the region where $P_{\mathrm{c}}$ is almost zero but $P_{\mathrm{s}}$ is finite. In this region, the Mott skin effect is observed.

%%%%%%%%%%%%%%%%%%%%%%%%%
%%%%%%%%%%%%%%%%%%%%%%%%%

%%%%%%%%%%%%%%
\subsection{
Real-time dynamics
}
\label{sec: TimeEvol app}
%%%%%%%%%%%%%%

The Mott skin effect results in the dynamical spin accumulations for the open quantum system with the jump term $\hat{L}_{lj\sigma} \hat{\rho}(t) \hat{L}^\dagger_{lj\sigma}$.
We compute the time-evolution for the initial state
%%%%%%
\begin{eqnarray}
\label{eq: Phi ini uniform app}
|\Phi(t=0) \rangle &=& \hat{b}^\dagger_{0\downarrow}\hat{b}^\dagger_{1\downarrow}\hat{b}^\dagger_{2\uparrow}\hat{b}^\dagger_{3\downarrow}\hat{b}^\dagger_{4\downarrow}\hat{b}^\dagger_{5\downarrow} |0\rangle
\end{eqnarray}
%%%%%%
with $L=6$.
Here, as another initial state, one may consider a state where all of the particles occupy a specific site. 
In this case, the open quantum system immediately loses bosons due to on-site two-body losses when quantum jumps are taken into account.

\begin{figure}[!h]
\begin{minipage}{0.45\hsize}
\begin{center}
\includegraphics[width=1\hsize,clip]{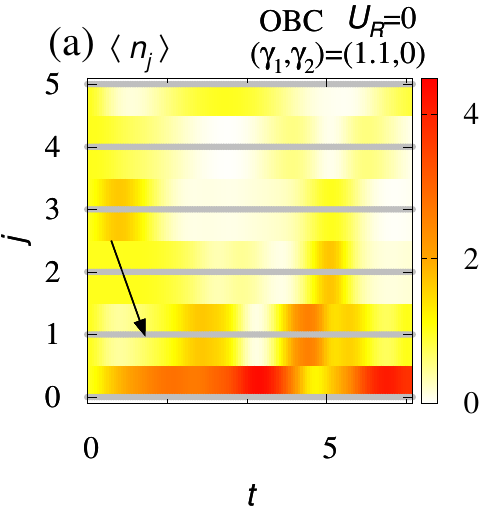}
\end{center}
\end{minipage}
\begin{minipage}{0.45\hsize}
\begin{center}
\includegraphics[width=1\hsize,clip]{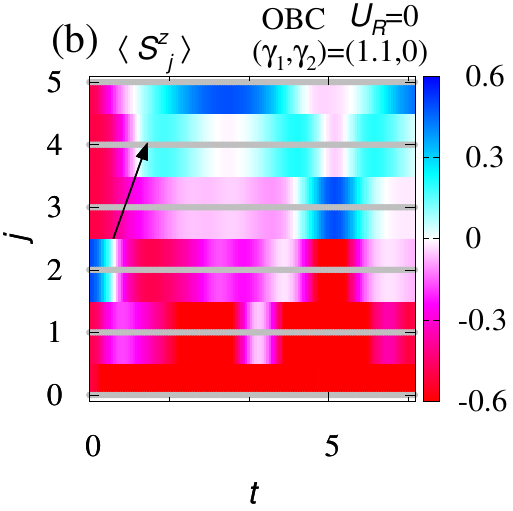}
\end{center}
\end{minipage}
\begin{minipage}{0.45\hsize}
\begin{center}
\includegraphics[width=1\hsize,clip]{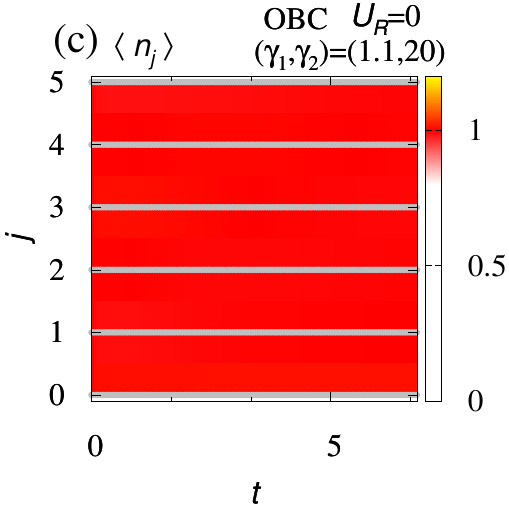}
\end{center}
\end{minipage}
\begin{minipage}{0.45\hsize}
\begin{center}
\includegraphics[width=1\hsize,clip]{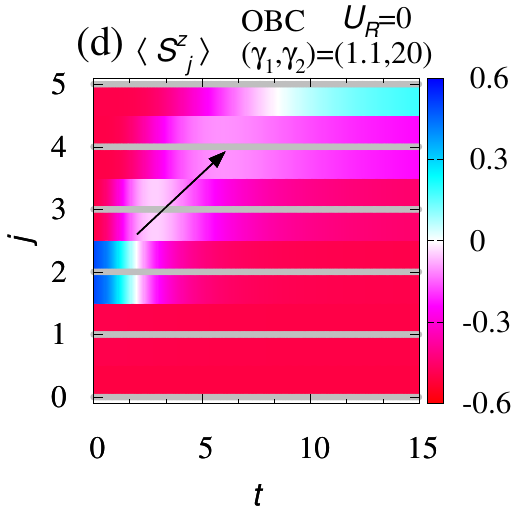}
\end{center}
\end{minipage}
\caption{
Time-evolution of expectation values with $\hat{H}_{\mathrm{eff}}$ under OBC for $\gamma_1=1.1$.
(a) [(b)]: $\langle \hat{n}_{j} \rangle$ [$\langle \hat{S}^z_{j} \rangle$] for $\gamma_2=\gamma_3=0$.
(c) [(d)]: $\langle \hat{n}_{j} \rangle$ [$\langle \hat{S}^z_{j} \rangle$] for $\gamma_2=\gamma_3=20$.
Gray dots denote the data points.
The data are obtained for $V_{\mathrm{R}}=U_{\mathrm{R}}=0$, $t_0=1$ and $L=6$.
}
\label{fig: Heff TimeEvol ga11UR0}
\end{figure}

We start with the real-time dynamics described by the effective Hamiltonian $\hat{H}_{\mathrm{eff}}$ [see Fig.~\ref{fig: Heff TimeEvol ga11UR0}].
In contrast to the case without two-body losses [see Figs.~\ref{fig: Heff TimeEvol ga11UR0}(a)~and~\ref{fig: Heff TimeEvol ga11UR0}(b)], the finite values of $\gamma_2$ and $\gamma_3$ results in unique dynamics due to the Mott skin effect; dynamical spin accumulation is observed while charge distribution remains spatially uniform[see Figs.~\ref{fig: Heff TimeEvol ga11UR0}(c)~and~\ref{fig: Heff TimeEvol ga11UR0}(d)].

Now, we discuss the real-time dynamics described by the full Liouvillian $\spopf{L}[\cdot]$.
In the presence of the jump term $\hat{L}_{lj\sigma} \hat{\rho}(t) \hat{L}^\dagger_{lj\sigma}$, the number of bosons in the spin state $\sigma$ ($N_{\sigma}$) is no longer a conserved quantity. 
Specifically, our open quantum system only with dissipation continues to lose particles. 
Even in this case, one can observe a signal of the Mott skin effect.

Before presenting the numerical results, we describe the details of the computation.
The expectation values are computed as
%%%%%%
\begin{eqnarray}
\langle \hat{O}_{j} \rangle&=& \mathrm{Tr} \left[ \hat{O}_{j} \hat{\rho}(t) \right] 
\end{eqnarray}
%%%%%%
with
$
\hat{O}_{j} = \hat{n}_j, \, \hat{S}^z_j.
$
Here, the time-evolution is computed by vectoring the density matrix $\hat{\rho}(t) \to |\rho(t)\rrangle$.
Corresponding to the vectorization, the superoperator called Liouvillian $\spopf{L}[\cdot]$ is rewritten as the matrix $\hat{\matf{L}}$; $\spopf{L}[\hat{\rho}(t)] \to \hat{\matf{L}} |\rho(t)\rrangle$.
With this form, we obtain $|\rho(t)\rrangle =\sum_{\mu} e^{\Lambda_{\mu} t}|R_\mu \rrangle \llangle L_\mu | \rho(0)\rrangle $ where $\Lambda_{\mu}$ $(\mu=0,1,2,\ldots)$ are the eigenvalues of the matrix $\hat{\matf{L}}$, and $\llangle L_\mu |$ ($|R_\mu \rrangle $) are the corresponding left (right) vectors.
The initial state is chosen to be $\hat{\rho}(0) =|\Phi(0)\rangle \langle \Phi(0)|$ with $|\Phi(0)\rangle$ defined in Eq.~(\ref{eq: Phi ini uniform app}).

\begin{figure}[!h]
\begin{minipage}{0.45\hsize}
\begin{center}
\includegraphics[width=1\hsize,clip]{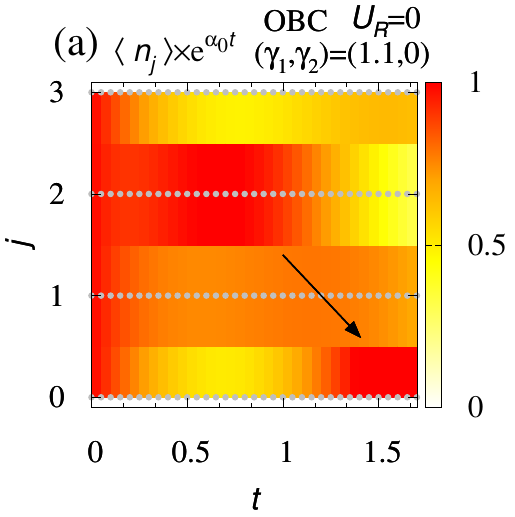}
\end{center}
\end{minipage}
\begin{minipage}{0.45\hsize}
\begin{center}
\includegraphics[width=1\hsize,clip]{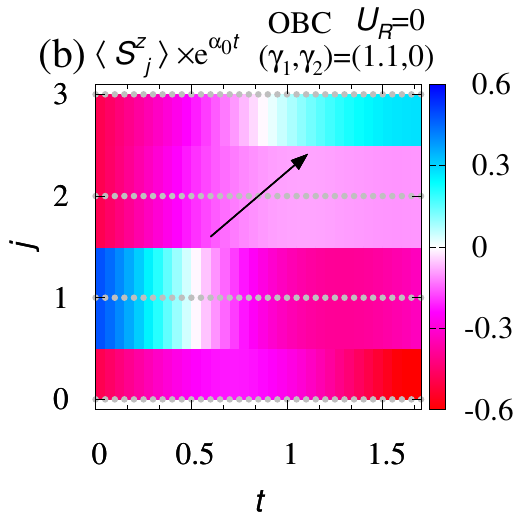}
\end{center}
\end{minipage}
\begin{minipage}{0.45\hsize}
\begin{center}
\includegraphics[width=1\hsize,clip]{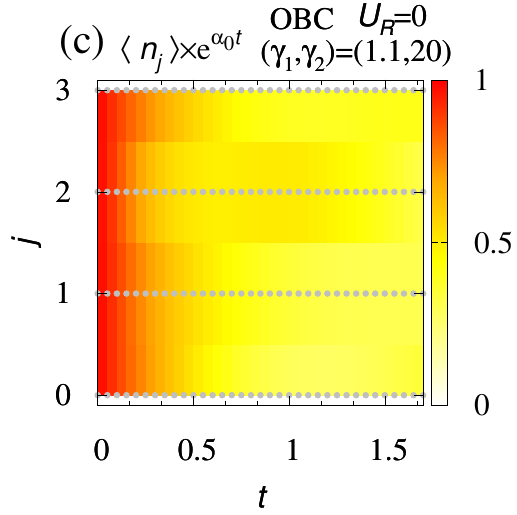}
\end{center}
\end{minipage}
\begin{minipage}{0.45\hsize}
\begin{center}
\includegraphics[width=1\hsize,clip]{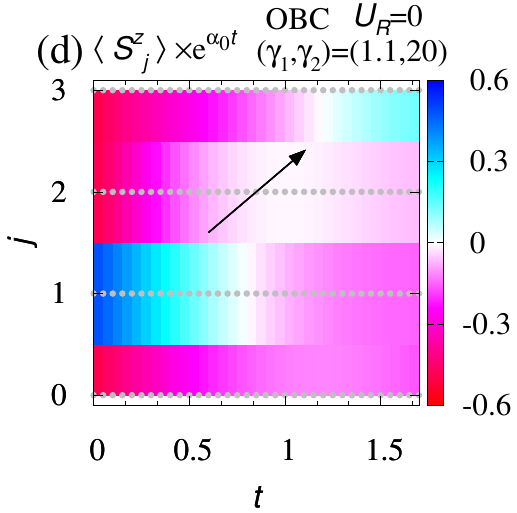}
\end{center}
\end{minipage}
\caption{
Time-evolution of expectation values with $\hat{\matf{L}}$ under OBC for $\gamma_1=1.1$.
(a) [(b)]: $\langle \hat{n}_{j} \rangle$ [$\langle \hat{S}^z_{j} \rangle$] for $\gamma_2=\gamma_3=0$.
(c) [(d)]: $\langle \hat{n}_{j} \rangle$ [$\langle \hat{S}^z_{j} \rangle$] for $\gamma_2=\gamma_3=20$.
Here, we have magnified the data with multiplying $e^{\alpha_0 t}$ with $\alpha_0=1.07295860$ because the particle number monotonically decreases for our purely dissipative system.
Gray points denote the data points.
The data are obtained for $V_{\mathrm{R}}=U_{\mathrm{R}}=0$, $t_0=1$ and $L=6$.
}
\label{fig: L TimeEvol ga11UR0}
\end{figure}

Figure~\ref{fig: L TimeEvol ga11UR0} displays the real-time dynamics of the full Liouvillian $\hat{\matf{L}}$ for $\gamma_1=1.1$.
For $\gamma_2=\gamma_3=0$, both charge and spin accumulations are observed in the real-time dynamics [Figs.~\ref{fig: L TimeEvol ga11UR0}(a)~and~\ref{fig: L TimeEvol ga11UR0}(b)].
In contrast to this behavior, turing on $\gamma_2$ and $\gamma_3$ induces the dynamical spin accumulation while the charge distribution is spatially uniform [Figs.~\ref{fig: L TimeEvol ga11UR0}(c)~and~\ref{fig: L TimeEvol ga11UR0}(d)], which is a signal of the Mott skin effect.

%%%%%%%%%%%%%%%%%%%%%%%%%
%%%%%%%%%%%%%%%%%%%%%%%%%

%%%%%%%%%%%%%%%%%%%%%%%%%
%%%%%%%%%%%%%%%%%%%%%%%%%

\begin{figure}[!h]
\begin{minipage}{0.45\hsize}
\begin{center}
\includegraphics[width=1\hsize,clip]{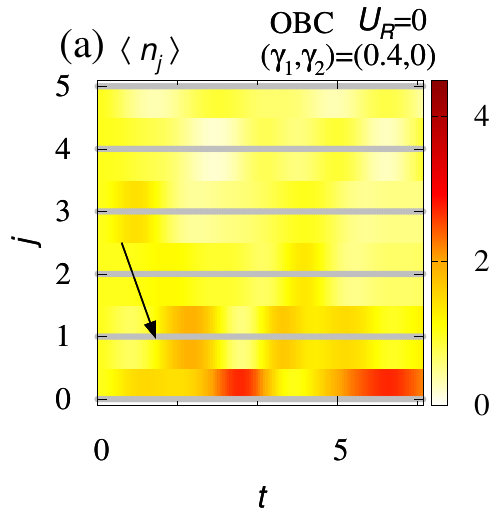}
\end{center}
\end{minipage}
\begin{minipage}{0.45\hsize}
\begin{center}
\includegraphics[width=1\hsize,clip]{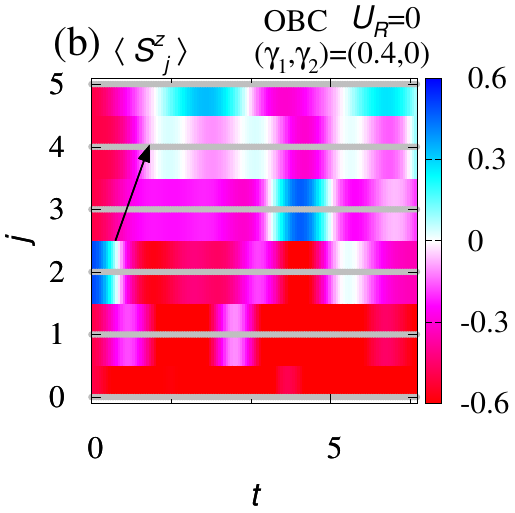}
\end{center}
\end{minipage}
\begin{minipage}{0.45\hsize}
\begin{center}
\includegraphics[width=1\hsize,clip]{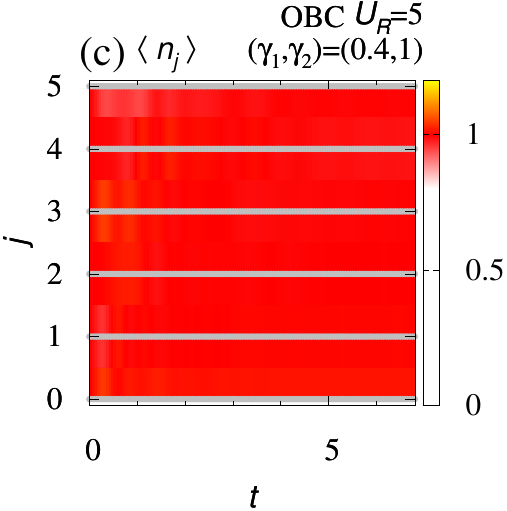}
\end{center}
\end{minipage}
\begin{minipage}{0.45\hsize}
\begin{center}
\includegraphics[width=1\hsize,clip]{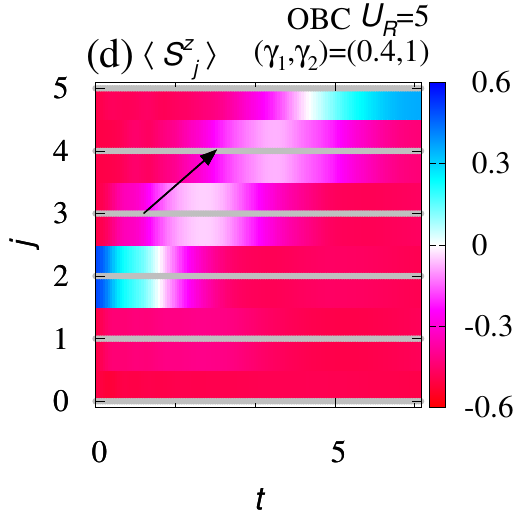}
\end{center}
\end{minipage}
\caption{
Time-evolution of expectation values with $\hat{H}_{\mathrm{eff}}$ under OBC for $\gamma_1=0.4$.
(a) [(b)]: $\langle \hat{n}_{j} \rangle$ [$\langle \hat{S}^z_{j} \rangle$] for $\gamma_2=\gamma_3=0$ and $V_{\mathrm{R}}=U_{\mathrm{R}}=0$.
(c) [(d)]: $\langle \hat{n}_{j} \rangle$ [$\langle \hat{S}^z_{j} \rangle$] for $\gamma_2=\gamma_3=1$ and $V_{\mathrm{R}}=U_{\mathrm{R}}=5$.
Gray dots denote the data points.
The data are obtained for $t_0=1$ and $L=6$.
}
\label{fig: Heff TimeEvol ga04varUR}
\end{figure}

\begin{figure}[!h]
\begin{minipage}{0.45\hsize}
\begin{center}

\includegraphics[width=1\hsize,clip]{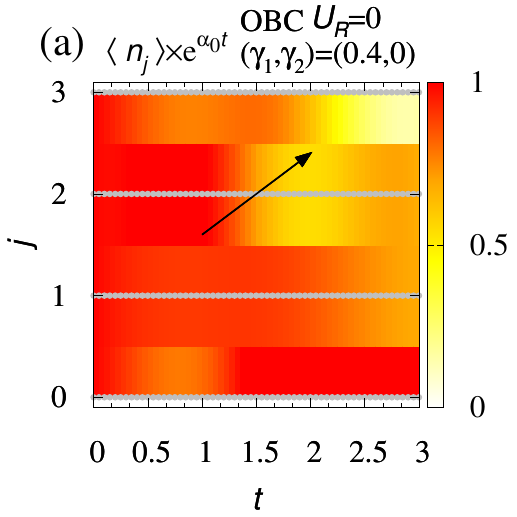}
\end{center}
\end{minipage}
\begin{minipage}{0.45\hsize}
\begin{center}
\includegraphics[width=1\hsize,clip]{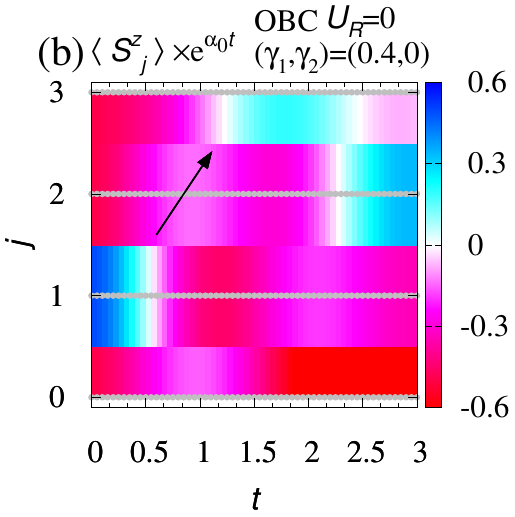}
\end{center}
\end{minipage}
\begin{minipage}{0.45\hsize}
\begin{center}
\includegraphics[width=1\hsize,clip]{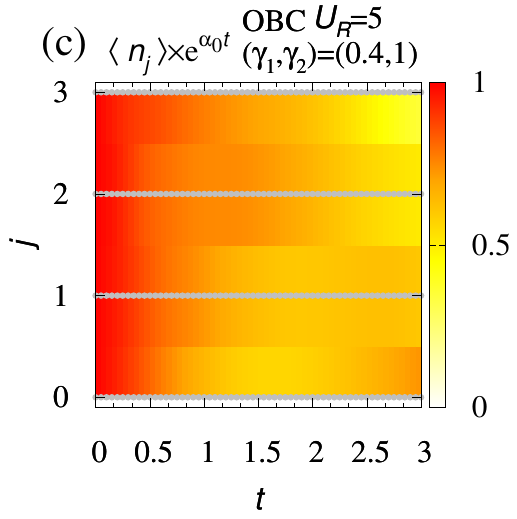}
\end{center}
\end{minipage}
\begin{minipage}{0.45\hsize}
\begin{center}
\includegraphics[width=1\hsize,clip]{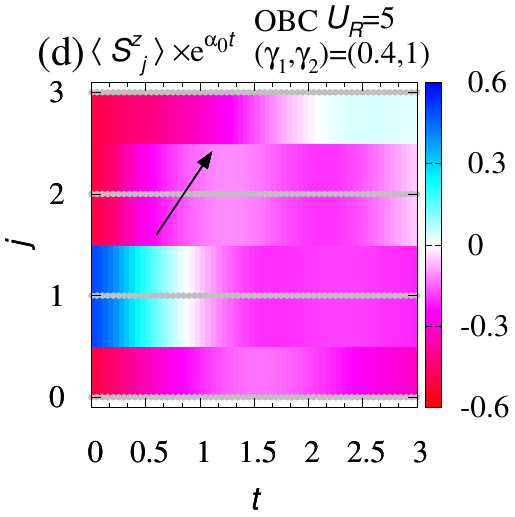}
\end{center}
\end{minipage}
\caption{
Time-evolution of expectation values with $\hat{\matf{L}}$ under OBC for $\gamma_1=0.4$.
(a) [(b)]: $\langle \hat{n}_{j} \rangle$ [$\langle \hat{S}^z_{j} \rangle$] for $\gamma_2=\gamma_3=0$ and $V_{\mathrm{R}}=U_{\mathrm{R}}=0$.
(c) [(d)]: $\langle \hat{n}_{j} \rangle$ [$\langle \hat{S}^z_{j} \rangle$] for $\gamma_2=\gamma_3=1$ and $V_{\mathrm{R}}=U_{\mathrm{R}}=5$.
Here, we have magnified the data with multiplying $e^{\alpha_0 t}$ with $\alpha_0=0.53647930414$ because the particle number monotonically decreases for our purely dissipative system.
Gray dots denote the data points.
The data are obtained for $t_0=1$ and $L=6$.
}
\label{fig: L TimeEvol ga04varUR}
\end{figure}

We finish this section with a remark on the case of small $\gamma_l$ $(l=1,2,3)$.
So far, we have discussed for $\gamma_2,\gamma_3 \gg  t_0=1$ and $V_{\mathrm{R}}=U_{\mathrm{R}}=0$.
Although decreasing $\gamma_l$ smears the dynamical spin accumulation, such a signal of Mott skin effect is observed for $\gamma_2=\gamma_3=t_0$ for finite $V_{\mathrm{R}}$ and $U_{\mathrm{R}}$ [see Figs.~\ref{fig: Heff TimeEvol ga04varUR}~and~\ref{fig: L TimeEvol ga04varUR}].

%%%%%%%%%%%%%%%%%%%%%%%%%
%%%%%%%%%%%%%%%%%%%%%%%%%

%%%%%%%%%%%%%%%%%%%%%%%%%
%%%%%%%%%%%%%%%%%%%%%%%%%


\begin{thebibliography}{155}%
\makeatletter
\providecommand \@ifxundefined [1]{%
 \@ifx{#1\undefined}
}%
\providecommand \@ifnum [1]{%
 \ifnum #1\expandafter \@firstoftwo
 \else \expandafter \@secondoftwo
 \fi
}%
\providecommand \@ifx [1]{%
 \ifx #1\expandafter \@firstoftwo
 \else \expandafter \@secondoftwo
 \fi
}%
\providecommand \natexlab [1]{#1}%
\providecommand \enquote  [1]{``#1''}%
\providecommand \bibnamefont  [1]{#1}%
\providecommand \bibfnamefont [1]{#1}%
\providecommand \citenamefont [1]{#1}%
\providecommand \href@noop [0]{\@secondoftwo}%
\providecommand \href [0]{\begingroup \@sanitize@url \@href}%
\providecommand \@href[1]{\@@startlink{#1}\@@href}%
\providecommand \@@href[1]{\endgroup#1\@@endlink}%
\providecommand \@sanitize@url [0]{\catcode `\\12\catcode `\$12\catcode
  `\&12\catcode `\#12\catcode `\^12\catcode `\_12\catcode `\%12\relax}%
\providecommand \@@startlink[1]{}%
\providecommand \@@endlink[0]{}%
\providecommand \url  [0]{\begingroup\@sanitize@url \@url }%
\providecommand \@url [1]{\endgroup\@href {#1}{\urlprefix }}%
\providecommand \urlprefix  [0]{URL }%
\providecommand \Eprint [0]{\href }%
\providecommand \doibase [0]{http://dx.doi.org/}%
\providecommand \selectlanguage [0]{\@gobble}%
\providecommand \bibinfo  [0]{\@secondoftwo}%
\providecommand \bibfield  [0]{\@secondoftwo}%
\providecommand \translation [1]{[#1]}%
\providecommand \BibitemOpen [0]{}%
\providecommand \bibitemStop [0]{}%
\providecommand \bibitemNoStop [0]{.\EOS\space}%
\providecommand \EOS [0]{\spacefactor3000\relax}%
\providecommand \BibitemShut  [1]{\csname bibitem#1\endcsname}%
\let\auto@bib@innerbib\@empty
%</preamble>
\bibitem [{\citenamefont {Hasan}\ and\ \citenamefont
  {Kane}(2010)}]{TI_review_Hasan10}%
  \BibitemOpen
  \bibfield  {author} {\bibinfo {author} {\bibfnamefont {M.~Z.}\ \bibnamefont
  {Hasan}}\ and\ \bibinfo {author} {\bibfnamefont {C.~L.}\ \bibnamefont
  {Kane}},\ }\href {\doibase 10.1103/RevModPhys.82.3045} {\bibfield  {journal}
  {\bibinfo  {journal} {Rev. Mod. Phys.}\ }\textbf {\bibinfo {volume} {82}},\
  \bibinfo {pages} {3045} (\bibinfo {year} {2010})}\BibitemShut {NoStop}%
\bibitem [{\citenamefont {Qi}\ and\ \citenamefont
  {Zhang}(2011)}]{TI_review_Qi10}%
  \BibitemOpen
  \bibfield  {author} {\bibinfo {author} {\bibfnamefont {X.-L.}\ \bibnamefont
  {Qi}}\ and\ \bibinfo {author} {\bibfnamefont {S.-C.}\ \bibnamefont {Zhang}},\
  }\href {\doibase 10.1103/RevModPhys.83.1057} {\bibfield  {journal} {\bibinfo
  {journal} {Rev. Mod. Phys.}\ }\textbf {\bibinfo {volume} {83}},\ \bibinfo
  {pages} {1057} (\bibinfo {year} {2011})}\BibitemShut {NoStop}%
\bibitem [{\citenamefont {Thouless}\ \emph {et~al.}(1982)\citenamefont
  {Thouless}, \citenamefont {Kohmoto}, \citenamefont {Nightingale},\ and\
  \citenamefont {den Nijs}}]{Thouless_PRL1982}%
  \BibitemOpen
  \bibfield  {author} {\bibinfo {author} {\bibfnamefont {D.~J.}\ \bibnamefont
  {Thouless}}, \bibinfo {author} {\bibfnamefont {M.}~\bibnamefont {Kohmoto}},
  \bibinfo {author} {\bibfnamefont {M.~P.}\ \bibnamefont {Nightingale}}, \ and\
  \bibinfo {author} {\bibfnamefont {M.}~\bibnamefont {den Nijs}},\ }\href
  {\doibase 10.1103/PhysRevLett.49.405} {\bibfield  {journal} {\bibinfo
  {journal} {Phys. Rev. Lett.}\ }\textbf {\bibinfo {volume} {49}},\ \bibinfo
  {pages} {405} (\bibinfo {year} {1982})}\BibitemShut {NoStop}%
\bibitem [{\citenamefont {Halperin}(1982)}]{Halperin_PRB82}%
  \BibitemOpen
  \bibfield  {author} {\bibinfo {author} {\bibfnamefont {B.~I.}\ \bibnamefont
  {Halperin}},\ }\href {\doibase 10.1103/PhysRevB.25.2185} {\bibfield
  {journal} {\bibinfo  {journal} {Phys. Rev. B}\ }\textbf {\bibinfo {volume}
  {25}},\ \bibinfo {pages} {2185} (\bibinfo {year} {1982})}\BibitemShut
  {NoStop}%
\bibitem [{\citenamefont {Hatsugai}(1993)}]{Hatsugai_PRL93}%
  \BibitemOpen
  \bibfield  {author} {\bibinfo {author} {\bibfnamefont {Y.}~\bibnamefont
  {Hatsugai}},\ }\href {\doibase 10.1103/PhysRevLett.71.3697} {\bibfield
  {journal} {\bibinfo  {journal} {Phys. Rev. Lett.}\ }\textbf {\bibinfo
  {volume} {71}},\ \bibinfo {pages} {3697} (\bibinfo {year}
  {1993})}\BibitemShut {NoStop}%
\bibitem [{\citenamefont {Kitaev}(2001)}]{Kitaev_chain_01}%
  \BibitemOpen
  \bibfield  {author} {\bibinfo {author} {\bibfnamefont {A.~Y.}\ \bibnamefont
  {Kitaev}},\ }\href {http://stacks.iop.org/1063-7869/44/i=10S/a=S29}
  {\bibfield  {journal} {\bibinfo  {journal} {Physics-Uspekhi}\ }\textbf
  {\bibinfo {volume} {44}},\ \bibinfo {pages} {131} (\bibinfo {year}
  {2001})}\BibitemShut {NoStop}%
\bibitem [{\citenamefont {Kane}\ and\ \citenamefont
  {Mele}(2005{\natexlab{a}})}]{Kane_2DZ2_PRL05}%
  \BibitemOpen
  \bibfield  {author} {\bibinfo {author} {\bibfnamefont {C.~L.}\ \bibnamefont
  {Kane}}\ and\ \bibinfo {author} {\bibfnamefont {E.~J.}\ \bibnamefont
  {Mele}},\ }\href {\doibase 10.1103/PhysRevLett.95.146802} {\bibfield
  {journal} {\bibinfo  {journal} {Phys. Rev. Lett.}\ }\textbf {\bibinfo
  {volume} {95}},\ \bibinfo {pages} {146802} (\bibinfo {year}
  {2005}{\natexlab{a}})}\BibitemShut {NoStop}%
\bibitem [{\citenamefont {Kane}\ and\ \citenamefont
  {Mele}(2005{\natexlab{b}})}]{Kane_Z2TI_PRL05_2}%
  \BibitemOpen
  \bibfield  {author} {\bibinfo {author} {\bibfnamefont {C.~L.}\ \bibnamefont
  {Kane}}\ and\ \bibinfo {author} {\bibfnamefont {E.~J.}\ \bibnamefont
  {Mele}},\ }\href {\doibase 10.1103/PhysRevLett.95.226801} {\bibfield
  {journal} {\bibinfo  {journal} {Phys. Rev. Lett.}\ }\textbf {\bibinfo
  {volume} {95}},\ \bibinfo {pages} {226801} (\bibinfo {year}
  {2005}{\natexlab{b}})}\BibitemShut {NoStop}%
\bibitem [{\citenamefont {Qi}\ \emph {et~al.}(2008)\citenamefont {Qi},
  \citenamefont {Hughes},\ and\ \citenamefont {Zhang}}]{Qi_TFT_PRB08}%
  \BibitemOpen
  \bibfield  {author} {\bibinfo {author} {\bibfnamefont {X.-L.}\ \bibnamefont
  {Qi}}, \bibinfo {author} {\bibfnamefont {T.~L.}\ \bibnamefont {Hughes}}, \
  and\ \bibinfo {author} {\bibfnamefont {S.-C.}\ \bibnamefont {Zhang}},\ }\href
  {\doibase 10.1103/PhysRevB.78.195424} {\bibfield  {journal} {\bibinfo
  {journal} {Phys. Rev. B}\ }\textbf {\bibinfo {volume} {78}},\ \bibinfo
  {pages} {195424} (\bibinfo {year} {2008})}\BibitemShut {NoStop}%
\bibitem [{\citenamefont {Tsui}\ \emph {et~al.}(1982)\citenamefont {Tsui},
  \citenamefont {Stormer},\ and\ \citenamefont {Gossard}}]{Tsui_FQHEExp_PRL82}%
  \BibitemOpen
  \bibfield  {author} {\bibinfo {author} {\bibfnamefont {D.~C.}\ \bibnamefont
  {Tsui}}, \bibinfo {author} {\bibfnamefont {H.~L.}\ \bibnamefont {Stormer}}, \
  and\ \bibinfo {author} {\bibfnamefont {A.~C.}\ \bibnamefont {Gossard}},\
  }\href {\doibase 10.1103/PhysRevLett.48.1559} {\bibfield  {journal} {\bibinfo
   {journal} {Phys. Rev. Lett.}\ }\textbf {\bibinfo {volume} {48}},\ \bibinfo
  {pages} {1559} (\bibinfo {year} {1982})}\BibitemShut {NoStop}%
\bibitem [{\citenamefont {Laughlin}(1983)}]{Laughlin_FQHE_PRL83}%
  \BibitemOpen
  \bibfield  {author} {\bibinfo {author} {\bibfnamefont {R.~B.}\ \bibnamefont
  {Laughlin}},\ }\href {\doibase 10.1103/PhysRevLett.50.1395} {\bibfield
  {journal} {\bibinfo  {journal} {Phys. Rev. Lett.}\ }\textbf {\bibinfo
  {volume} {50}},\ \bibinfo {pages} {1395} (\bibinfo {year}
  {1983})}\BibitemShut {NoStop}%
\bibitem [{\citenamefont {Jain}(1989)}]{Jain_FQHE_PRL89}%
  \BibitemOpen
  \bibfield  {author} {\bibinfo {author} {\bibfnamefont {J.~K.}\ \bibnamefont
  {Jain}},\ }\href {\doibase 10.1103/PhysRevLett.63.199} {\bibfield  {journal}
  {\bibinfo  {journal} {Phys. Rev. Lett.}\ }\textbf {\bibinfo {volume} {63}},\
  \bibinfo {pages} {199} (\bibinfo {year} {1989})}\BibitemShut {NoStop}%
\bibitem [{\citenamefont {Haldane}(1983)}]{Haldane_FQHEpesudo_PRL83}%
  \BibitemOpen
  \bibfield  {author} {\bibinfo {author} {\bibfnamefont {F.~D.~M.}\
  \bibnamefont {Haldane}},\ }\href {\doibase 10.1103/PhysRevLett.51.605}
  {\bibfield  {journal} {\bibinfo  {journal} {Phys. Rev. Lett.}\ }\textbf
  {\bibinfo {volume} {51}},\ \bibinfo {pages} {605} (\bibinfo {year}
  {1983})}\BibitemShut {NoStop}%
\bibitem [{\citenamefont {Haldane}(1985)}]{Haldane_TopoDeg_PRL85}%
  \BibitemOpen
  \bibfield  {author} {\bibinfo {author} {\bibfnamefont {F.~D.~M.}\
  \bibnamefont {Haldane}},\ }\href {\doibase 10.1103/PhysRevLett.55.2095}
  {\bibfield  {journal} {\bibinfo  {journal} {Phys. Rev. Lett.}\ }\textbf
  {\bibinfo {volume} {55}},\ \bibinfo {pages} {2095} (\bibinfo {year}
  {1985})}\BibitemShut {NoStop}%
\bibitem [{\citenamefont {Wen}(1995)}]{Wen_TopoOrder_SciAdv95}%
  \BibitemOpen
  \bibfield  {author} {\bibinfo {author} {\bibfnamefont {X.-G.}\ \bibnamefont
  {Wen}},\ }\href {\doibase 10.1080/00018739500101566} {\bibfield  {journal}
  {\bibinfo  {journal} {Advances in Physics}\ }\textbf {\bibinfo {volume}
  {44}},\ \bibinfo {pages} {405} (\bibinfo {year} {1995})}\BibitemShut
  {NoStop}%
\bibitem [{Kit(2003)}]{Kitaev_ToricCode_Elsevier03}%
  \BibitemOpen
  \href {\doibase https://doi.org/10.1016/S0003-4916(02)00018-0} {\bibfield
  {journal} {\bibinfo  {journal} {Annals of Physics}\ }\textbf {\bibinfo
  {volume} {303}},\ \bibinfo {pages} {2 } (\bibinfo {year} {2003})}\BibitemShut
  {NoStop}%
\bibitem [{\citenamefont {Wen}(2004)}]{Wen_Textbook04}%
  \BibitemOpen
  \bibfield  {author} {\bibinfo {author} {\bibfnamefont {X.-G.}\ \bibnamefont
  {Wen}},\ }\href@noop {} {\emph {\bibinfo {title} {Quantum field theory of
  many-body systems: from the origin of sound to an origin of light and
  electrons}}}\ (\bibinfo  {publisher} {OUP Oxford},\ \bibinfo {year}
  {2004})\BibitemShut {NoStop}%
\bibitem [{\citenamefont {Levin}\ and\ \citenamefont
  {Wen}(2005)}]{Levin_LevinWenModel_PRB05}%
  \BibitemOpen
  \bibfield  {author} {\bibinfo {author} {\bibfnamefont {M.~A.}\ \bibnamefont
  {Levin}}\ and\ \bibinfo {author} {\bibfnamefont {X.-G.}\ \bibnamefont
  {Wen}},\ }\href {\doibase 10.1103/PhysRevB.71.045110} {\bibfield  {journal}
  {\bibinfo  {journal} {Phys. Rev. B}\ }\textbf {\bibinfo {volume} {71}},\
  \bibinfo {pages} {045110} (\bibinfo {year} {2005})}\BibitemShut {NoStop}%
\bibitem [{\citenamefont {Kitaev}(2003)}]{Kitaev_ToricCode_AnnPhys06}%
  \BibitemOpen
  \bibfield  {author} {\bibinfo {author} {\bibfnamefont {A.}~\bibnamefont
  {Kitaev}},\ }\href {\doibase https://doi.org/10.1016/S0003-4916(02)00018-0}
  {\bibfield  {journal} {\bibinfo  {journal} {Annals of Physics}\ }\textbf
  {\bibinfo {volume} {303}},\ \bibinfo {pages} {2} (\bibinfo {year}
  {2003})}\BibitemShut {NoStop}%
\bibitem [{\citenamefont {Kitaev}(2006)}]{Kitaev_KitaevHoney_AnnPhys06}%
  \BibitemOpen
  \bibfield  {author} {\bibinfo {author} {\bibfnamefont {A.}~\bibnamefont
  {Kitaev}},\ }\href {\doibase https://doi.org/10.1016/j.aop.2005.10.005}
  {\bibfield  {journal} {\bibinfo  {journal} {Annals of Physics}\ }\textbf
  {\bibinfo {volume} {321}},\ \bibinfo {pages} {2} (\bibinfo {year} {2006})},\
  \bibinfo {note} {january Special Issue}\BibitemShut {NoStop}%
\bibitem [{\citenamefont {Tang}\ \emph {et~al.}(2011)\citenamefont {Tang},
  \citenamefont {Mei},\ and\ \citenamefont {Wen}}]{Tang_FChern_PRL11}%
  \BibitemOpen
  \bibfield  {author} {\bibinfo {author} {\bibfnamefont {E.}~\bibnamefont
  {Tang}}, \bibinfo {author} {\bibfnamefont {J.-W.}\ \bibnamefont {Mei}}, \
  and\ \bibinfo {author} {\bibfnamefont {X.-G.}\ \bibnamefont {Wen}},\ }\href
  {\doibase 10.1103/PhysRevLett.106.236802} {\bibfield  {journal} {\bibinfo
  {journal} {Phys. Rev. Lett.}\ }\textbf {\bibinfo {volume} {106}},\ \bibinfo
  {pages} {236802} (\bibinfo {year} {2011})}\BibitemShut {NoStop}%
\bibitem [{\citenamefont {Sun}\ \emph {et~al.}(2011)\citenamefont {Sun},
  \citenamefont {Gu}, \citenamefont {Katsura},\ and\ \citenamefont
  {Das~Sarma}}]{Sun_FChern_PRL11}%
  \BibitemOpen
  \bibfield  {author} {\bibinfo {author} {\bibfnamefont {K.}~\bibnamefont
  {Sun}}, \bibinfo {author} {\bibfnamefont {Z.}~\bibnamefont {Gu}}, \bibinfo
  {author} {\bibfnamefont {H.}~\bibnamefont {Katsura}}, \ and\ \bibinfo
  {author} {\bibfnamefont {S.}~\bibnamefont {Das~Sarma}},\ }\href {\doibase
  10.1103/PhysRevLett.106.236803} {\bibfield  {journal} {\bibinfo  {journal}
  {Phys. Rev. Lett.}\ }\textbf {\bibinfo {volume} {106}},\ \bibinfo {pages}
  {236803} (\bibinfo {year} {2011})}\BibitemShut {NoStop}%
\bibitem [{\citenamefont {Neupert}\ \emph {et~al.}(2011)\citenamefont
  {Neupert}, \citenamefont {Santos}, \citenamefont {Chamon},\ and\
  \citenamefont {Mudry}}]{Neupert_FChern_PRL11}%
  \BibitemOpen
  \bibfield  {author} {\bibinfo {author} {\bibfnamefont {T.}~\bibnamefont
  {Neupert}}, \bibinfo {author} {\bibfnamefont {L.}~\bibnamefont {Santos}},
  \bibinfo {author} {\bibfnamefont {C.}~\bibnamefont {Chamon}}, \ and\ \bibinfo
  {author} {\bibfnamefont {C.}~\bibnamefont {Mudry}},\ }\href {\doibase
  10.1103/PhysRevLett.106.236804} {\bibfield  {journal} {\bibinfo  {journal}
  {Phys. Rev. Lett.}\ }\textbf {\bibinfo {volume} {106}},\ \bibinfo {pages}
  {236804} (\bibinfo {year} {2011})}\BibitemShut {NoStop}%
\bibitem [{\citenamefont {Sheng}\ \emph {et~al.}(2011)\citenamefont {Sheng},
  \citenamefont {Gu}, \citenamefont {Sun},\ and\ \citenamefont
  {Sheng}}]{Sheng_FChern_NComm12}%
  \BibitemOpen
  \bibfield  {author} {\bibinfo {author} {\bibfnamefont {D.~N.}\ \bibnamefont
  {Sheng}}, \bibinfo {author} {\bibfnamefont {Z.-C.}\ \bibnamefont {Gu}},
  \bibinfo {author} {\bibfnamefont {K.}~\bibnamefont {Sun}}, \ and\ \bibinfo
  {author} {\bibfnamefont {L.}~\bibnamefont {Sheng}},\ }\href@noop {}
  {\bibfield  {journal} {\bibinfo  {journal} {Nature Communications}\ }\textbf
  {\bibinfo {volume} {2}},\ \bibinfo {pages} {389 EP } (\bibinfo {year}
  {2011})},\ \bibinfo {note} {article}\BibitemShut {NoStop}%
\bibitem [{\citenamefont {Regnault}\ and\ \citenamefont
  {Bernevig}(2011)}]{Regnalt_FChen_PRX11}%
  \BibitemOpen
  \bibfield  {author} {\bibinfo {author} {\bibfnamefont {N.}~\bibnamefont
  {Regnault}}\ and\ \bibinfo {author} {\bibfnamefont {B.~A.}\ \bibnamefont
  {Bernevig}},\ }\href {\doibase 10.1103/PhysRevX.1.021014} {\bibfield
  {journal} {\bibinfo  {journal} {Phys. Rev. X}\ }\textbf {\bibinfo {volume}
  {1}},\ \bibinfo {pages} {021014} (\bibinfo {year} {2011})}\BibitemShut
  {NoStop}%
\bibitem [{\citenamefont {Bergholtz}\ and\ \citenamefont
  {Liu}(2013)}]{Bergholtz_FChern_IntJModPhysB13}%
  \BibitemOpen
  \bibfield  {author} {\bibinfo {author} {\bibfnamefont {E.~J.}\ \bibnamefont
  {Bergholtz}}\ and\ \bibinfo {author} {\bibfnamefont {Z.}~\bibnamefont
  {Liu}},\ }\href {\doibase 10.1142/S021797921330017X} {\bibfield  {journal}
  {\bibinfo  {journal} {International Journal of Modern Physics B}\ }\textbf
  {\bibinfo {volume} {27}},\ \bibinfo {pages} {1330017} (\bibinfo {year}
  {2013})}\BibitemShut {NoStop}%
\bibitem [{\citenamefont {Fidkowski}\ and\ \citenamefont
  {Kitaev}(2010)}]{Z_to_Zn_Fidkowski_PRB10}%
  \BibitemOpen
  \bibfield  {author} {\bibinfo {author} {\bibfnamefont {L.}~\bibnamefont
  {Fidkowski}}\ and\ \bibinfo {author} {\bibfnamefont {A.}~\bibnamefont
  {Kitaev}},\ }\href {\doibase 10.1103/PhysRevB.81.134509} {\bibfield
  {journal} {\bibinfo  {journal} {Phys. Rev. B}\ }\textbf {\bibinfo {volume}
  {81}},\ \bibinfo {pages} {134509} (\bibinfo {year} {2010})}\BibitemShut
  {NoStop}%
\bibitem [{\citenamefont {Turner}\ \emph {et~al.}(2011)\citenamefont {Turner},
  \citenamefont {Pollmann},\ and\ \citenamefont {Berg}}]{Turner_ZtoZ8_PRB11}%
  \BibitemOpen
  \bibfield  {author} {\bibinfo {author} {\bibfnamefont {A.~M.}\ \bibnamefont
  {Turner}}, \bibinfo {author} {\bibfnamefont {F.}~\bibnamefont {Pollmann}}, \
  and\ \bibinfo {author} {\bibfnamefont {E.}~\bibnamefont {Berg}},\ }\href
  {\doibase 10.1103/PhysRevB.83.075102} {\bibfield  {journal} {\bibinfo
  {journal} {Phys. Rev. B}\ }\textbf {\bibinfo {volume} {83}},\ \bibinfo
  {pages} {075102} (\bibinfo {year} {2011})}\BibitemShut {NoStop}%
\bibitem [{\citenamefont {Fidkowski}\ and\ \citenamefont
  {Kitaev}(2011)}]{Fidkowski_1Dclassificatin_PRB11}%
  \BibitemOpen
  \bibfield  {author} {\bibinfo {author} {\bibfnamefont {L.}~\bibnamefont
  {Fidkowski}}\ and\ \bibinfo {author} {\bibfnamefont {A.}~\bibnamefont
  {Kitaev}},\ }\href {\doibase 10.1103/PhysRevB.83.075103} {\bibfield
  {journal} {\bibinfo  {journal} {Phys. Rev. B}\ }\textbf {\bibinfo {volume}
  {83}},\ \bibinfo {pages} {075103} (\bibinfo {year} {2011})}\BibitemShut
  {NoStop}%
\bibitem [{\citenamefont {Gu}\ and\ \citenamefont
  {Wen}(2014)}]{gu_supercohomology}%
  \BibitemOpen
  \bibfield  {author} {\bibinfo {author} {\bibfnamefont {Z.-C.}\ \bibnamefont
  {Gu}}\ and\ \bibinfo {author} {\bibfnamefont {X.-G.}\ \bibnamefont {Wen}},\
  }\href {\doibase 10.1103/PhysRevB.90.115141} {\bibfield  {journal} {\bibinfo
  {journal} {Phys. Rev. B}\ }\textbf {\bibinfo {volume} {90}},\ \bibinfo
  {pages} {115141} (\bibinfo {year} {2014})}\BibitemShut {NoStop}%
\bibitem [{\citenamefont {Yao}\ and\ \citenamefont
  {Ryu}(2013)}]{YaoRyu_Z_to_Z8_2013}%
  \BibitemOpen
  \bibfield  {author} {\bibinfo {author} {\bibfnamefont {H.}~\bibnamefont
  {Yao}}\ and\ \bibinfo {author} {\bibfnamefont {S.}~\bibnamefont {Ryu}},\
  }\href {\doibase 10.1103/PhysRevB.88.064507} {\bibfield  {journal} {\bibinfo
  {journal} {Phys. Rev. B}\ }\textbf {\bibinfo {volume} {88}},\ \bibinfo
  {pages} {064507} (\bibinfo {year} {2013})}\BibitemShut {NoStop}%
\bibitem [{\citenamefont {Ryu}\ and\ \citenamefont
  {Zhang}(2012)}]{Ryu_Z_to_Z8_2013}%
  \BibitemOpen
  \bibfield  {author} {\bibinfo {author} {\bibfnamefont {S.}~\bibnamefont
  {Ryu}}\ and\ \bibinfo {author} {\bibfnamefont {S.-C.}\ \bibnamefont
  {Zhang}},\ }\href {\doibase 10.1103/PhysRevB.85.245132} {\bibfield  {journal}
  {\bibinfo  {journal} {Phys. Rev. B}\ }\textbf {\bibinfo {volume} {85}},\
  \bibinfo {pages} {245132} (\bibinfo {year} {2012})}\BibitemShut {NoStop}%
\bibitem [{\citenamefont {Qi}(2013)}]{Qi_Z_to_Z8_2013}%
  \BibitemOpen
  \bibfield  {author} {\bibinfo {author} {\bibfnamefont {X.-L.}\ \bibnamefont
  {Qi}},\ }\href@noop {} {\bibfield  {journal} {\bibinfo  {journal} {New J.
  Phys.}\ }\textbf {\bibinfo {volume} {15}},\ \bibinfo {pages} {065002}
  (\bibinfo {year} {2013})}\BibitemShut {NoStop}%
\bibitem [{\citenamefont {Pesin}\ and\ \citenamefont
  {Balents}(2010)}]{Pesin_TMI_NatPhys2010}%
  \BibitemOpen
  \bibfield  {author} {\bibinfo {author} {\bibfnamefont {D.}~\bibnamefont
  {Pesin}}\ and\ \bibinfo {author} {\bibfnamefont {L.}~\bibnamefont
  {Balents}},\ }\href {\doibase 10.1038/nphys1606} {\bibfield  {journal}
  {\bibinfo  {journal} {Nature Physics}\ }\textbf {\bibinfo {volume} {6}},\
  \bibinfo {pages} {376} (\bibinfo {year} {2010})}\BibitemShut {NoStop}%
\bibitem [{\citenamefont {Manmana}\ \emph {et~al.}(2012)\citenamefont
  {Manmana}, \citenamefont {Essin}, \citenamefont {Noack},\ and\ \citenamefont
  {Gurarie}}]{Manmana_Chiral1D_PRB12}%
  \BibitemOpen
  \bibfield  {author} {\bibinfo {author} {\bibfnamefont {S.~R.}\ \bibnamefont
  {Manmana}}, \bibinfo {author} {\bibfnamefont {A.~M.}\ \bibnamefont {Essin}},
  \bibinfo {author} {\bibfnamefont {R.~M.}\ \bibnamefont {Noack}}, \ and\
  \bibinfo {author} {\bibfnamefont {V.}~\bibnamefont {Gurarie}},\ }\href
  {\doibase 10.1103/PhysRevB.86.205119} {\bibfield  {journal} {\bibinfo
  {journal} {Phys. Rev. B}\ }\textbf {\bibinfo {volume} {86}},\ \bibinfo
  {pages} {205119} (\bibinfo {year} {2012})}\BibitemShut {NoStop}%
\bibitem [{\citenamefont {Yoshida}\ \emph {et~al.}(2014)\citenamefont
  {Yoshida}, \citenamefont {Peters}, \citenamefont {Fujimoto},\ and\
  \citenamefont {Kawakami}}]{Yoshida_TMI1D_PRL14}%
  \BibitemOpen
  \bibfield  {author} {\bibinfo {author} {\bibfnamefont {T.}~\bibnamefont
  {Yoshida}}, \bibinfo {author} {\bibfnamefont {R.}~\bibnamefont {Peters}},
  \bibinfo {author} {\bibfnamefont {S.}~\bibnamefont {Fujimoto}}, \ and\
  \bibinfo {author} {\bibfnamefont {N.}~\bibnamefont {Kawakami}},\ }\href
  {\doibase 10.1103/PhysRevLett.112.196404} {\bibfield  {journal} {\bibinfo
  {journal} {Phys. Rev. Lett.}\ }\textbf {\bibinfo {volume} {112}},\ \bibinfo
  {pages} {196404} (\bibinfo {year} {2014})}\BibitemShut {NoStop}%
\bibitem [{\citenamefont {Bergholtz}\ \emph {et~al.}(2021)\citenamefont
  {Bergholtz}, \citenamefont {Budich},\ and\ \citenamefont
  {Kunst}}]{Bergholtz_Review19}%
  \BibitemOpen
  \bibfield  {author} {\bibinfo {author} {\bibfnamefont {E.~J.}\ \bibnamefont
  {Bergholtz}}, \bibinfo {author} {\bibfnamefont {J.~C.}\ \bibnamefont
  {Budich}}, \ and\ \bibinfo {author} {\bibfnamefont {F.~K.}\ \bibnamefont
  {Kunst}},\ }\href@noop {} {\bibfield  {journal} {\bibinfo  {journal} {Rev.
  Mod. Phys.}\ }\textbf {\bibinfo {volume} {93}},\ \bibinfo {pages} {015005}
  (\bibinfo {year} {2021})}\BibitemShut {NoStop}%
\bibitem [{\citenamefont {Ashida}\ \emph {et~al.}(2020)\citenamefont {Ashida},
  \citenamefont {Gong},\ and\ \citenamefont
  {Ueda}}]{Ashida_nHReview_AdvPhys20}%
  \BibitemOpen
  \bibfield  {author} {\bibinfo {author} {\bibfnamefont {Y.}~\bibnamefont
  {Ashida}}, \bibinfo {author} {\bibfnamefont {Z.}~\bibnamefont {Gong}}, \ and\
  \bibinfo {author} {\bibfnamefont {M.}~\bibnamefont {Ueda}},\ }\href {\doibase
  10.1080/00018732.2021.1876991} {\bibfield  {journal} {\bibinfo  {journal}
  {Advances in Physics}\ }\textbf {\bibinfo {volume} {69}},\ \bibinfo {pages}
  {249} (\bibinfo {year} {2020})}\BibitemShut {NoStop}%
\bibitem [{\citenamefont {Gong}\ \emph {et~al.}(2018)\citenamefont {Gong},
  \citenamefont {Ashida}, \citenamefont {Kawabata}, \citenamefont {Takasan},
  \citenamefont {Higashikawa},\ and\ \citenamefont {Ueda}}]{Gong_class_PRX18}%
  \BibitemOpen
  \bibfield  {author} {\bibinfo {author} {\bibfnamefont {Z.}~\bibnamefont
  {Gong}}, \bibinfo {author} {\bibfnamefont {Y.}~\bibnamefont {Ashida}},
  \bibinfo {author} {\bibfnamefont {K.}~\bibnamefont {Kawabata}}, \bibinfo
  {author} {\bibfnamefont {K.}~\bibnamefont {Takasan}}, \bibinfo {author}
  {\bibfnamefont {S.}~\bibnamefont {Higashikawa}}, \ and\ \bibinfo {author}
  {\bibfnamefont {M.}~\bibnamefont {Ueda}},\ }\href {\doibase
  10.1103/PhysRevX.8.031079} {\bibfield  {journal} {\bibinfo  {journal} {Phys.
  Rev. X}\ }\textbf {\bibinfo {volume} {8}},\ \bibinfo {pages} {031079}
  (\bibinfo {year} {2018})}\BibitemShut {NoStop}%
\bibitem [{\citenamefont {Kawabata}\ \emph
  {et~al.}(2019{\natexlab{a}})\citenamefont {Kawabata}, \citenamefont
  {Higashikawa}, \citenamefont {Gong}, \citenamefont {Ashida},\ and\
  \citenamefont {Ueda}}]{KKawabata_TopoUni_NatComm19}%
  \BibitemOpen
  \bibfield  {author} {\bibinfo {author} {\bibfnamefont {K.}~\bibnamefont
  {Kawabata}}, \bibinfo {author} {\bibfnamefont {S.}~\bibnamefont
  {Higashikawa}}, \bibinfo {author} {\bibfnamefont {Z.}~\bibnamefont {Gong}},
  \bibinfo {author} {\bibfnamefont {Y.}~\bibnamefont {Ashida}}, \ and\ \bibinfo
  {author} {\bibfnamefont {M.}~\bibnamefont {Ueda}},\ }\href {\doibase
  10.1038/s41467-018-08254-y} {\bibfield  {journal} {\bibinfo  {journal}
  {Nature Communications}\ }\textbf {\bibinfo {volume} {10}},\ \bibinfo {pages}
  {297} (\bibinfo {year} {2019}{\natexlab{a}})}\BibitemShut {NoStop}%
\bibitem [{\citenamefont {Kawabata}\ \emph
  {et~al.}(2019{\natexlab{b}})\citenamefont {Kawabata}, \citenamefont
  {Shiozaki}, \citenamefont {Ueda},\ and\ \citenamefont
  {Sato}}]{Kawabata_gapped_PRX19}%
  \BibitemOpen
  \bibfield  {author} {\bibinfo {author} {\bibfnamefont {K.}~\bibnamefont
  {Kawabata}}, \bibinfo {author} {\bibfnamefont {K.}~\bibnamefont {Shiozaki}},
  \bibinfo {author} {\bibfnamefont {M.}~\bibnamefont {Ueda}}, \ and\ \bibinfo
  {author} {\bibfnamefont {M.}~\bibnamefont {Sato}},\ }\href {\doibase
  10.1103/PhysRevX.9.041015} {\bibfield  {journal} {\bibinfo  {journal} {Phys.
  Rev. X}\ }\textbf {\bibinfo {volume} {9}},\ \bibinfo {pages} {041015}
  (\bibinfo {year} {2019}{\natexlab{b}})}\BibitemShut {NoStop}%
\bibitem [{\citenamefont {Zhou}\ and\ \citenamefont
  {Lee}(2019)}]{Zhou_gapped_class_PRB19}%
  \BibitemOpen
  \bibfield  {author} {\bibinfo {author} {\bibfnamefont {H.}~\bibnamefont
  {Zhou}}\ and\ \bibinfo {author} {\bibfnamefont {J.~Y.}\ \bibnamefont {Lee}},\
  }\href {\doibase 10.1103/PhysRevB.99.235112} {\bibfield  {journal} {\bibinfo
  {journal} {Phys. Rev. B}\ }\textbf {\bibinfo {volume} {99}},\ \bibinfo
  {pages} {235112} (\bibinfo {year} {2019})}\BibitemShut {NoStop}%
\bibitem [{\citenamefont {Lieu}\ \emph {et~al.}(2020)\citenamefont {Lieu},
  \citenamefont {McGinley},\ and\ \citenamefont
  {Cooper}}]{Lieu_Liouclass_PRL20}%
  \BibitemOpen
  \bibfield  {author} {\bibinfo {author} {\bibfnamefont {S.}~\bibnamefont
  {Lieu}}, \bibinfo {author} {\bibfnamefont {M.}~\bibnamefont {McGinley}}, \
  and\ \bibinfo {author} {\bibfnamefont {N.~R.}\ \bibnamefont {Cooper}},\
  }\href {\doibase 10.1103/PhysRevLett.124.040401} {\bibfield  {journal}
  {\bibinfo  {journal} {Phys. Rev. Lett.}\ }\textbf {\bibinfo {volume} {124}},\
  \bibinfo {pages} {040401} (\bibinfo {year} {2020})}\BibitemShut {NoStop}%
\bibitem [{\citenamefont {Hatano}\ and\ \citenamefont
  {Nelson}(1996)}]{Hatano_PRL96}%
  \BibitemOpen
  \bibfield  {author} {\bibinfo {author} {\bibfnamefont {N.}~\bibnamefont
  {Hatano}}\ and\ \bibinfo {author} {\bibfnamefont {D.~R.}\ \bibnamefont
  {Nelson}},\ }\href {\doibase 10.1103/PhysRevLett.77.570} {\bibfield
  {journal} {\bibinfo  {journal} {Phys. Rev. Lett.}\ }\textbf {\bibinfo
  {volume} {77}},\ \bibinfo {pages} {570} (\bibinfo {year} {1996})}\BibitemShut
  {NoStop}%
\bibitem [{\citenamefont {Hatano}\ and\ \citenamefont
  {Nelson}(1997)}]{Hatano_PRB97}%
  \BibitemOpen
  \bibfield  {author} {\bibinfo {author} {\bibfnamefont {N.}~\bibnamefont
  {Hatano}}\ and\ \bibinfo {author} {\bibfnamefont {D.~R.}\ \bibnamefont
  {Nelson}},\ }\href {\doibase 10.1103/PhysRevB.56.8651} {\bibfield  {journal}
  {\bibinfo  {journal} {Phys. Rev. B}\ }\textbf {\bibinfo {volume} {56}},\
  \bibinfo {pages} {8651} (\bibinfo {year} {1997})}\BibitemShut {NoStop}%
\bibitem [{\citenamefont {Bender}\ and\ \citenamefont
  {Boettcher}(1998)}]{CMBender_PRL98}%
  \BibitemOpen
  \bibfield  {author} {\bibinfo {author} {\bibfnamefont {C.~M.}\ \bibnamefont
  {Bender}}\ and\ \bibinfo {author} {\bibfnamefont {S.}~\bibnamefont
  {Boettcher}},\ }\href {\doibase 10.1103/PhysRevLett.80.5243} {\bibfield
  {journal} {\bibinfo  {journal} {Phys. Rev. Lett.}\ }\textbf {\bibinfo
  {volume} {80}},\ \bibinfo {pages} {5243} (\bibinfo {year}
  {1998})}\BibitemShut {NoStop}%
\bibitem [{\citenamefont {Fukui}\ and\ \citenamefont
  {Kawakami}(1998)}]{Fukui_nH_PRB98}%
  \BibitemOpen
  \bibfield  {author} {\bibinfo {author} {\bibfnamefont {T.}~\bibnamefont
  {Fukui}}\ and\ \bibinfo {author} {\bibfnamefont {N.}~\bibnamefont
  {Kawakami}},\ }\href@noop {} {\bibfield  {journal} {\bibinfo  {journal}
  {Physical Review B}\ }\textbf {\bibinfo {volume} {58}},\ \bibinfo {pages}
  {16051} (\bibinfo {year} {1998})}\BibitemShut {NoStop}%
\bibitem [{\citenamefont {Ashida}\ \emph {et~al.}(2016)\citenamefont {Ashida},
  \citenamefont {Furukawa},\ and\ \citenamefont {Ueda}}]{Ashida_nHbHubb_PRA16}%
  \BibitemOpen
  \bibfield  {author} {\bibinfo {author} {\bibfnamefont {Y.}~\bibnamefont
  {Ashida}}, \bibinfo {author} {\bibfnamefont {S.}~\bibnamefont {Furukawa}}, \
  and\ \bibinfo {author} {\bibfnamefont {M.}~\bibnamefont {Ueda}},\ }\href
  {\doibase 10.1103/PhysRevA.94.053615} {\bibfield  {journal} {\bibinfo
  {journal} {Phys. Rev. A}\ }\textbf {\bibinfo {volume} {94}},\ \bibinfo
  {pages} {053615} (\bibinfo {year} {2016})}\BibitemShut {NoStop}%
\bibitem [{\citenamefont {Hu}\ and\ \citenamefont
  {Hughes}(2011)}]{Hu_nH_PRB11}%
  \BibitemOpen
  \bibfield  {author} {\bibinfo {author} {\bibfnamefont {Y.~C.}\ \bibnamefont
  {Hu}}\ and\ \bibinfo {author} {\bibfnamefont {T.~L.}\ \bibnamefont
  {Hughes}},\ }\href {\doibase 10.1103/PhysRevB.84.153101} {\bibfield
  {journal} {\bibinfo  {journal} {Phys. Rev. B}\ }\textbf {\bibinfo {volume}
  {84}},\ \bibinfo {pages} {153101} (\bibinfo {year} {2011})}\BibitemShut
  {NoStop}%
\bibitem [{\citenamefont {Esaki}\ \emph {et~al.}(2011)\citenamefont {Esaki},
  \citenamefont {Sato}, \citenamefont {Hasebe},\ and\ \citenamefont
  {Kohmoto}}]{Esaki_nH_PRB11}%
  \BibitemOpen
  \bibfield  {author} {\bibinfo {author} {\bibfnamefont {K.}~\bibnamefont
  {Esaki}}, \bibinfo {author} {\bibfnamefont {M.}~\bibnamefont {Sato}},
  \bibinfo {author} {\bibfnamefont {K.}~\bibnamefont {Hasebe}}, \ and\ \bibinfo
  {author} {\bibfnamefont {M.}~\bibnamefont {Kohmoto}},\ }\href {\doibase
  10.1103/PhysRevB.84.205128} {\bibfield  {journal} {\bibinfo  {journal} {Phys.
  Rev. B}\ }\textbf {\bibinfo {volume} {84}},\ \bibinfo {pages} {205128}
  (\bibinfo {year} {2011})}\BibitemShut {NoStop}%
\bibitem [{\citenamefont {Sato}\ \emph {et~al.}(2012)\citenamefont {Sato},
  \citenamefont {Hasebe}, \citenamefont {Esaki},\ and\ \citenamefont
  {Kohmoto}}]{Sato_nHPTEP12}%
  \BibitemOpen
  \bibfield  {author} {\bibinfo {author} {\bibfnamefont {M.}~\bibnamefont
  {Sato}}, \bibinfo {author} {\bibfnamefont {K.}~\bibnamefont {Hasebe}},
  \bibinfo {author} {\bibfnamefont {K.}~\bibnamefont {Esaki}}, \ and\ \bibinfo
  {author} {\bibfnamefont {M.}~\bibnamefont {Kohmoto}},\ }\href {\doibase
  10.1143/PTP.127.937} {\bibfield  {journal} {\bibinfo  {journal} {Progress of
  Theoretical Physics}\ }\textbf {\bibinfo {volume} {127}},\ \bibinfo {pages}
  {937} (\bibinfo {year} {2012})}\BibitemShut {NoStop}%
\bibitem [{\citenamefont {Diehl}\ \emph {et~al.}(2011)\citenamefont {Diehl},
  \citenamefont {Rico}, \citenamefont {Baranov},\ and\ \citenamefont
  {Zoller}}]{Diehl_DissCher_NatPhys11}%
  \BibitemOpen
  \bibfield  {author} {\bibinfo {author} {\bibfnamefont {S.}~\bibnamefont
  {Diehl}}, \bibinfo {author} {\bibfnamefont {E.}~\bibnamefont {Rico}},
  \bibinfo {author} {\bibfnamefont {M.~A.}\ \bibnamefont {Baranov}}, \ and\
  \bibinfo {author} {\bibfnamefont {P.}~\bibnamefont {Zoller}},\ }\href
  {\doibase 10.1038/nphys2106} {\bibfield  {journal} {\bibinfo  {journal}
  {Nature Physics}\ }\textbf {\bibinfo {volume} {7}},\ \bibinfo {pages} {971}
  (\bibinfo {year} {2011})}\BibitemShut {NoStop}%
\bibitem [{\citenamefont {Bardyn}\ \emph {et~al.}(2013)\citenamefont {Bardyn},
  \citenamefont {Baranov}, \citenamefont {Kraus}, \citenamefont {Rico},
  \citenamefont {{\.{I}}mamo{\u{g}}lu}, \citenamefont {Zoller},\ and\
  \citenamefont {Diehl}}]{Bardyn_DissCher_NJP2013}%
  \BibitemOpen
  \bibfield  {author} {\bibinfo {author} {\bibfnamefont {C.-E.}\ \bibnamefont
  {Bardyn}}, \bibinfo {author} {\bibfnamefont {M.~A.}\ \bibnamefont {Baranov}},
  \bibinfo {author} {\bibfnamefont {C.~V.}\ \bibnamefont {Kraus}}, \bibinfo
  {author} {\bibfnamefont {E.}~\bibnamefont {Rico}}, \bibinfo {author}
  {\bibfnamefont {A.}~\bibnamefont {{\.{I}}mamo{\u{g}}lu}}, \bibinfo {author}
  {\bibfnamefont {P.}~\bibnamefont {Zoller}}, \ and\ \bibinfo {author}
  {\bibfnamefont {S.}~\bibnamefont {Diehl}},\ }\href@noop {} {\bibfield
  {journal} {\bibinfo  {journal} {New Journal of Physics}\ }\textbf {\bibinfo
  {volume} {15}},\ \bibinfo {pages} {085001} (\bibinfo {year}
  {2013})}\BibitemShut {NoStop}%
\bibitem [{\citenamefont {Rivas}\ \emph {et~al.}(2013)\citenamefont {Rivas},
  \citenamefont {Viyuela},\ and\ \citenamefont
  {Martin-Delgado}}]{Rivas_DissCher_PRB13}%
  \BibitemOpen
  \bibfield  {author} {\bibinfo {author} {\bibfnamefont {A.}~\bibnamefont
  {Rivas}}, \bibinfo {author} {\bibfnamefont {O.}~\bibnamefont {Viyuela}}, \
  and\ \bibinfo {author} {\bibfnamefont {M.~A.}\ \bibnamefont
  {Martin-Delgado}},\ }\href {\doibase 10.1103/PhysRevB.88.155141} {\bibfield
  {journal} {\bibinfo  {journal} {Phys. Rev. B}\ }\textbf {\bibinfo {volume}
  {88}},\ \bibinfo {pages} {155141} (\bibinfo {year} {2013})}\BibitemShut
  {NoStop}%
\bibitem [{\citenamefont {Zhu}\ \emph {et~al.}(2014)\citenamefont {Zhu},
  \citenamefont {L\"u},\ and\ \citenamefont {Chen}}]{Zhu_nHPT_PRA14}%
  \BibitemOpen
  \bibfield  {author} {\bibinfo {author} {\bibfnamefont {B.}~\bibnamefont
  {Zhu}}, \bibinfo {author} {\bibfnamefont {R.}~\bibnamefont {L\"u}}, \ and\
  \bibinfo {author} {\bibfnamefont {S.}~\bibnamefont {Chen}},\ }\href {\doibase
  10.1103/PhysRevA.89.062102} {\bibfield  {journal} {\bibinfo  {journal} {Phys.
  Rev. A}\ }\textbf {\bibinfo {volume} {89}},\ \bibinfo {pages} {062102}
  (\bibinfo {year} {2014})}\BibitemShut {NoStop}%
\bibitem [{\citenamefont {Budich}\ \emph {et~al.}(2015)\citenamefont {Budich},
  \citenamefont {Zoller},\ and\ \citenamefont {Diehl}}]{Budich_DissCher_PRA15}%
  \BibitemOpen
  \bibfield  {author} {\bibinfo {author} {\bibfnamefont {J.~C.}\ \bibnamefont
  {Budich}}, \bibinfo {author} {\bibfnamefont {P.}~\bibnamefont {Zoller}}, \
  and\ \bibinfo {author} {\bibfnamefont {S.}~\bibnamefont {Diehl}},\ }\href
  {\doibase 10.1103/PhysRevA.91.042117} {\bibfield  {journal} {\bibinfo
  {journal} {Phys. Rev. A}\ }\textbf {\bibinfo {volume} {91}},\ \bibinfo
  {pages} {042117} (\bibinfo {year} {2015})}\BibitemShut {NoStop}%
\bibitem [{\citenamefont {Lieu}(2018)}]{Lieu_nHSSH_PRB2018}%
  \BibitemOpen
  \bibfield  {author} {\bibinfo {author} {\bibfnamefont {S.}~\bibnamefont
  {Lieu}},\ }\href {\doibase 10.1103/PhysRevB.97.045106} {\bibfield  {journal}
  {\bibinfo  {journal} {Phys. Rev. B}\ }\textbf {\bibinfo {volume} {97}},\
  \bibinfo {pages} {045106} (\bibinfo {year} {2018})}\BibitemShut {NoStop}%
\bibitem [{\citenamefont {Rui}\ \emph {et~al.}(2019)\citenamefont {Rui},
  \citenamefont {Zhao},\ and\ \citenamefont {Schnyder}}]{Rui_nH_PRB19}%
  \BibitemOpen
  \bibfield  {author} {\bibinfo {author} {\bibfnamefont {W.~B.}\ \bibnamefont
  {Rui}}, \bibinfo {author} {\bibfnamefont {Y.~X.}\ \bibnamefont {Zhao}}, \
  and\ \bibinfo {author} {\bibfnamefont {A.~P.}\ \bibnamefont {Schnyder}},\
  }\href {\doibase 10.1103/PhysRevB.99.241110} {\bibfield  {journal} {\bibinfo
  {journal} {Phys. Rev. B}\ }\textbf {\bibinfo {volume} {99}},\ \bibinfo
  {pages} {241110} (\bibinfo {year} {2019})}\BibitemShut {NoStop}%
\bibitem [{\citenamefont {Denner}\ \emph {et~al.}(2021)\citenamefont {Denner},
  \citenamefont {Skurativska}, \citenamefont {Schindler}, \citenamefont
  {Fischer}, \citenamefont {Thomale}, \citenamefont {Bzdusek},\ and\
  \citenamefont {Neupert}}]{Denner_ETI_NatComm21}%
  \BibitemOpen
  \bibfield  {author} {\bibinfo {author} {\bibfnamefont {M.~M.}\ \bibnamefont
  {Denner}}, \bibinfo {author} {\bibfnamefont {A.}~\bibnamefont {Skurativska}},
  \bibinfo {author} {\bibfnamefont {F.}~\bibnamefont {Schindler}}, \bibinfo
  {author} {\bibfnamefont {M.~H.}\ \bibnamefont {Fischer}}, \bibinfo {author}
  {\bibfnamefont {R.}~\bibnamefont {Thomale}}, \bibinfo {author} {\bibfnamefont
  {T.}~\bibnamefont {Bzdusek}}, \ and\ \bibinfo {author} {\bibfnamefont
  {T.}~\bibnamefont {Neupert}},\ }\href@noop {} {\bibfield  {journal} {\bibinfo
   {journal} {Nature Communications}\ }\textbf {\bibinfo {volume} {12}}
  (\bibinfo {year} {2021})}\BibitemShut {NoStop}%
\bibitem [{\citenamefont {Nakamura}\ \emph {et~al.}(2022)\citenamefont
  {Nakamura}, \citenamefont {Bessho},\ and\ \citenamefont
  {Sato}}]{Nakamura_BBCptG_arXiv22}%
  \BibitemOpen
  \bibfield  {author} {\bibinfo {author} {\bibfnamefont {D.}~\bibnamefont
  {Nakamura}}, \bibinfo {author} {\bibfnamefont {T.}~\bibnamefont {Bessho}}, \
  and\ \bibinfo {author} {\bibfnamefont {M.}~\bibnamefont {Sato}},\ }\href@noop
  {} {\bibfield  {journal} {\bibinfo  {journal} {arXiv preprint
  arXiv:2205.15635}\ } (\bibinfo {year} {2022})}\BibitemShut {NoStop}%
\bibitem [{\citenamefont {Yokomizo}\ and\ \citenamefont
  {Murakami}(2019)}]{Yokomizo_BBC_PRL19}%
  \BibitemOpen
  \bibfield  {author} {\bibinfo {author} {\bibfnamefont {K.}~\bibnamefont
  {Yokomizo}}\ and\ \bibinfo {author} {\bibfnamefont {S.}~\bibnamefont
  {Murakami}},\ }\href {\doibase 10.1103/PhysRevLett.123.066404} {\bibfield
  {journal} {\bibinfo  {journal} {Phys. Rev. Lett.}\ }\textbf {\bibinfo
  {volume} {123}},\ \bibinfo {pages} {066404} (\bibinfo {year}
  {2019})}\BibitemShut {NoStop}%
\bibitem [{\citenamefont {Kawabata}\ \emph
  {et~al.}(2020{\natexlab{a}})\citenamefont {Kawabata}, \citenamefont {Okuma},\
  and\ \citenamefont {Sato}}]{kawabata_NBlochBBC_PRB20}%
  \BibitemOpen
  \bibfield  {author} {\bibinfo {author} {\bibfnamefont {K.}~\bibnamefont
  {Kawabata}}, \bibinfo {author} {\bibfnamefont {N.}~\bibnamefont {Okuma}}, \
  and\ \bibinfo {author} {\bibfnamefont {M.}~\bibnamefont {Sato}},\ }\href
  {\doibase 10.1103/PhysRevB.101.195147} {\bibfield  {journal} {\bibinfo
  {journal} {Phys. Rev. B}\ }\textbf {\bibinfo {volume} {101}},\ \bibinfo
  {pages} {195147} (\bibinfo {year} {2020}{\natexlab{a}})}\BibitemShut
  {NoStop}%
\bibitem [{\citenamefont {Tonielli}\ \emph {et~al.}(2020)\citenamefont
  {Tonielli}, \citenamefont {Budich}, \citenamefont {Altland},\ and\
  \citenamefont {Diehl}}]{Tonielli_nHTQFT_PRL20}%
  \BibitemOpen
  \bibfield  {author} {\bibinfo {author} {\bibfnamefont {F.}~\bibnamefont
  {Tonielli}}, \bibinfo {author} {\bibfnamefont {J.~C.}\ \bibnamefont
  {Budich}}, \bibinfo {author} {\bibfnamefont {A.}~\bibnamefont {Altland}}, \
  and\ \bibinfo {author} {\bibfnamefont {S.}~\bibnamefont {Diehl}},\ }\href
  {\doibase 10.1103/PhysRevLett.124.240404} {\bibfield  {journal} {\bibinfo
  {journal} {Phys. Rev. Lett.}\ }\textbf {\bibinfo {volume} {124}},\ \bibinfo
  {pages} {240404} (\bibinfo {year} {2020})}\BibitemShut {NoStop}%
\bibitem [{\citenamefont {Kawabata}\ \emph {et~al.}(2021)\citenamefont
  {Kawabata}, \citenamefont {Shiozaki},\ and\ \citenamefont
  {Ryu}}]{Kawabata_TQFTSkin_PRL20}%
  \BibitemOpen
  \bibfield  {author} {\bibinfo {author} {\bibfnamefont {K.}~\bibnamefont
  {Kawabata}}, \bibinfo {author} {\bibfnamefont {K.}~\bibnamefont {Shiozaki}},
  \ and\ \bibinfo {author} {\bibfnamefont {S.}~\bibnamefont {Ryu}},\ }\href
  {\doibase 10.1103/PhysRevLett.126.216405} {\bibfield  {journal} {\bibinfo
  {journal} {Phys. Rev. Lett.}\ }\textbf {\bibinfo {volume} {126}},\ \bibinfo
  {pages} {216405} (\bibinfo {year} {2021})}\BibitemShut {NoStop}%
\bibitem [{\citenamefont {Sayyad}\ \emph {et~al.}(2022)\citenamefont {Sayyad},
  \citenamefont {Hannukainen},\ and\ \citenamefont
  {Grushin}}]{Sayyad_nHTQFT_PRR22}%
  \BibitemOpen
  \bibfield  {author} {\bibinfo {author} {\bibfnamefont {S.}~\bibnamefont
  {Sayyad}}, \bibinfo {author} {\bibfnamefont {J.~D.}\ \bibnamefont
  {Hannukainen}}, \ and\ \bibinfo {author} {\bibfnamefont {A.~G.}\ \bibnamefont
  {Grushin}},\ }\href {\doibase 10.1103/PhysRevResearch.4.L042004} {\bibfield
  {journal} {\bibinfo  {journal} {Phys. Rev. Res.}\ }\textbf {\bibinfo {volume}
  {4}},\ \bibinfo {pages} {L042004} (\bibinfo {year} {2022})}\BibitemShut
  {NoStop}%
\bibitem [{\citenamefont {Chang}\ \emph {et~al.}(2020)\citenamefont {Chang},
  \citenamefont {You}, \citenamefont {Wen},\ and\ \citenamefont
  {Ryu}}]{Chang_nHEntSpec_PRR}%
  \BibitemOpen
  \bibfield  {author} {\bibinfo {author} {\bibfnamefont {P.-Y.}\ \bibnamefont
  {Chang}}, \bibinfo {author} {\bibfnamefont {J.-S.}\ \bibnamefont {You}},
  \bibinfo {author} {\bibfnamefont {X.}~\bibnamefont {Wen}}, \ and\ \bibinfo
  {author} {\bibfnamefont {S.}~\bibnamefont {Ryu}},\ }\href {\doibase
  10.1103/PhysRevResearch.2.033069} {\bibfield  {journal} {\bibinfo  {journal}
  {Phys. Rev. Res.}\ }\textbf {\bibinfo {volume} {2}},\ \bibinfo {pages}
  {033069} (\bibinfo {year} {2020})}\BibitemShut {NoStop}%
\bibitem [{\citenamefont {Hsieh}\ and\ \citenamefont
  {Chang}(2023)}]{Hsieh_nHCritical_SciPost23}%
  \BibitemOpen
  \bibfield  {author} {\bibinfo {author} {\bibfnamefont {C.-T.}\ \bibnamefont
  {Hsieh}}\ and\ \bibinfo {author} {\bibfnamefont {P.-Y.}\ \bibnamefont
  {Chang}},\ }\href {\doibase 10.21468/SciPostPhysCore.6.3.062} {\bibfield
  {journal} {\bibinfo  {journal} {SciPost Phys. Core}\ }\textbf {\bibinfo
  {volume} {6}},\ \bibinfo {pages} {062} (\bibinfo {year} {2023})}\BibitemShut
  {NoStop}%
\bibitem [{\citenamefont {Shen}\ \emph {et~al.}(2018)\citenamefont {Shen},
  \citenamefont {Zhen},\ and\ \citenamefont {Fu}}]{HShen2017_non-Hermi}%
  \BibitemOpen
  \bibfield  {author} {\bibinfo {author} {\bibfnamefont {H.}~\bibnamefont
  {Shen}}, \bibinfo {author} {\bibfnamefont {B.}~\bibnamefont {Zhen}}, \ and\
  \bibinfo {author} {\bibfnamefont {L.}~\bibnamefont {Fu}},\ }\href {\doibase
  10.1103/PhysRevLett.120.146402} {\bibfield  {journal} {\bibinfo  {journal}
  {Phys. Rev. Lett.}\ }\textbf {\bibinfo {volume} {120}},\ \bibinfo {pages}
  {146402} (\bibinfo {year} {2018})}\BibitemShut {NoStop}%
\bibitem [{\citenamefont {Kozii}\ and\ \citenamefont
  {Fu}(2017)}]{VKozii_nH_arXiv17}%
  \BibitemOpen
  \bibfield  {author} {\bibinfo {author} {\bibfnamefont {V.}~\bibnamefont
  {Kozii}}\ and\ \bibinfo {author} {\bibfnamefont {L.}~\bibnamefont {Fu}},\
  }\href@noop {} {\bibfield  {journal} {\bibinfo  {journal} {arXiv preprint
  arXiv:1708.05841}\ } (\bibinfo {year} {2017})}\BibitemShut {NoStop}%
\bibitem [{\citenamefont {Yoshida}\ \emph {et~al.}(2018)\citenamefont
  {Yoshida}, \citenamefont {Peters},\ and\ \citenamefont
  {Kawakami}}]{Yoshida_EP_DMFT_PRB18}%
  \BibitemOpen
  \bibfield  {author} {\bibinfo {author} {\bibfnamefont {T.}~\bibnamefont
  {Yoshida}}, \bibinfo {author} {\bibfnamefont {R.}~\bibnamefont {Peters}}, \
  and\ \bibinfo {author} {\bibfnamefont {N.}~\bibnamefont {Kawakami}},\ }\href
  {\doibase 10.1103/PhysRevB.98.035141} {\bibfield  {journal} {\bibinfo
  {journal} {Phys. Rev. B}\ }\textbf {\bibinfo {volume} {98}},\ \bibinfo
  {pages} {035141} (\bibinfo {year} {2018})}\BibitemShut {NoStop}%
\bibitem [{\citenamefont {Zhou}\ \emph {et~al.}(2018)\citenamefont {Zhou},
  \citenamefont {Peng}, \citenamefont {Yoon}, \citenamefont {Hsu},
  \citenamefont {Nelson}, \citenamefont {Fu}, \citenamefont {Joannopoulos},
  \citenamefont {Solja{\v c}i{\'c}},\ and\ \citenamefont
  {Zhen}}]{Zhou_ObEP_Science18}%
  \BibitemOpen
  \bibfield  {author} {\bibinfo {author} {\bibfnamefont {H.}~\bibnamefont
  {Zhou}}, \bibinfo {author} {\bibfnamefont {C.}~\bibnamefont {Peng}}, \bibinfo
  {author} {\bibfnamefont {Y.}~\bibnamefont {Yoon}}, \bibinfo {author}
  {\bibfnamefont {C.~W.}\ \bibnamefont {Hsu}}, \bibinfo {author} {\bibfnamefont
  {K.~A.}\ \bibnamefont {Nelson}}, \bibinfo {author} {\bibfnamefont
  {L.}~\bibnamefont {Fu}}, \bibinfo {author} {\bibfnamefont {J.~D.}\
  \bibnamefont {Joannopoulos}}, \bibinfo {author} {\bibfnamefont
  {M.}~\bibnamefont {Solja{\v c}i{\'c}}}, \ and\ \bibinfo {author}
  {\bibfnamefont {B.}~\bibnamefont {Zhen}},\ }\href {\doibase
  10.1126/science.aap9859} {\bibfield  {journal} {\bibinfo  {journal}
  {Science}\ }\textbf {\bibinfo {volume} {359}},\ \bibinfo {pages} {1009}
  (\bibinfo {year} {2018})}\BibitemShut {NoStop}%
\bibitem [{\citenamefont {Yang}\ \emph
  {et~al.}(2021{\natexlab{a}})\citenamefont {Yang}, \citenamefont {Schnyder},
  \citenamefont {Hu},\ and\ \citenamefont {Chiu}}]{Yang_Descri_PRL21}%
  \BibitemOpen
  \bibfield  {author} {\bibinfo {author} {\bibfnamefont {Z.}~\bibnamefont
  {Yang}}, \bibinfo {author} {\bibfnamefont {A.~P.}\ \bibnamefont {Schnyder}},
  \bibinfo {author} {\bibfnamefont {J.}~\bibnamefont {Hu}}, \ and\ \bibinfo
  {author} {\bibfnamefont {C.-K.}\ \bibnamefont {Chiu}},\ }\href {\doibase
  10.1103/PhysRevLett.126.086401} {\bibfield  {journal} {\bibinfo  {journal}
  {Phys. Rev. Lett.}\ }\textbf {\bibinfo {volume} {126}},\ \bibinfo {pages}
  {086401} (\bibinfo {year} {2021}{\natexlab{a}})}\BibitemShut {NoStop}%
\bibitem [{\citenamefont {Budich}\ \emph {et~al.}(2019)\citenamefont {Budich},
  \citenamefont {Carlstr\"om}, \citenamefont {Kunst},\ and\ \citenamefont
  {Bergholtz}}]{Budich_SPERs_PRB19}%
  \BibitemOpen
  \bibfield  {author} {\bibinfo {author} {\bibfnamefont {J.~C.}\ \bibnamefont
  {Budich}}, \bibinfo {author} {\bibfnamefont {J.}~\bibnamefont {Carlstr\"om}},
  \bibinfo {author} {\bibfnamefont {F.~K.}\ \bibnamefont {Kunst}}, \ and\
  \bibinfo {author} {\bibfnamefont {E.~J.}\ \bibnamefont {Bergholtz}},\ }\href
  {\doibase 10.1103/PhysRevB.99.041406} {\bibfield  {journal} {\bibinfo
  {journal} {Phys. Rev. B}\ }\textbf {\bibinfo {volume} {99}},\ \bibinfo
  {pages} {041406} (\bibinfo {year} {2019})}\BibitemShut {NoStop}%
\bibitem [{\citenamefont {Okugawa}\ and\ \citenamefont
  {Yokoyama}(2019)}]{Okugawa_SPERs_PRB19}%
  \BibitemOpen
  \bibfield  {author} {\bibinfo {author} {\bibfnamefont {R.}~\bibnamefont
  {Okugawa}}\ and\ \bibinfo {author} {\bibfnamefont {T.}~\bibnamefont
  {Yokoyama}},\ }\href {\doibase 10.1103/PhysRevB.99.041202} {\bibfield
  {journal} {\bibinfo  {journal} {Phys. Rev. B}\ }\textbf {\bibinfo {volume}
  {99}},\ \bibinfo {pages} {041202} (\bibinfo {year} {2019})}\BibitemShut
  {NoStop}%
\bibitem [{\citenamefont {Zhou}\ \emph {et~al.}(2019)\citenamefont {Zhou},
  \citenamefont {Lee}, \citenamefont {Liu},\ and\ \citenamefont
  {Zhen}}]{Zhou_SPERs_Optica19}%
  \BibitemOpen
  \bibfield  {author} {\bibinfo {author} {\bibfnamefont {H.}~\bibnamefont
  {Zhou}}, \bibinfo {author} {\bibfnamefont {J.~Y.}\ \bibnamefont {Lee}},
  \bibinfo {author} {\bibfnamefont {S.}~\bibnamefont {Liu}}, \ and\ \bibinfo
  {author} {\bibfnamefont {B.}~\bibnamefont {Zhen}},\ }\href@noop {} {\bibfield
   {journal} {\bibinfo  {journal} {Optica}\ }\textbf {\bibinfo {volume} {6}},\
  \bibinfo {pages} {190} (\bibinfo {year} {2019})}\BibitemShut {NoStop}%
\bibitem [{\citenamefont {Yoshida}\ \emph
  {et~al.}(2019{\natexlab{a}})\citenamefont {Yoshida}, \citenamefont {Peters},
  \citenamefont {Kawakami},\ and\ \citenamefont
  {Hatsugai}}]{Yoshida_SPERs_PRB19}%
  \BibitemOpen
  \bibfield  {author} {\bibinfo {author} {\bibfnamefont {T.}~\bibnamefont
  {Yoshida}}, \bibinfo {author} {\bibfnamefont {R.}~\bibnamefont {Peters}},
  \bibinfo {author} {\bibfnamefont {N.}~\bibnamefont {Kawakami}}, \ and\
  \bibinfo {author} {\bibfnamefont {Y.}~\bibnamefont {Hatsugai}},\ }\href
  {\doibase 10.1103/PhysRevB.99.121101} {\bibfield  {journal} {\bibinfo
  {journal} {Phys. Rev. B}\ }\textbf {\bibinfo {volume} {99}},\ \bibinfo
  {pages} {121101} (\bibinfo {year} {2019}{\natexlab{a}})}\BibitemShut
  {NoStop}%
\bibitem [{\citenamefont {Yoshida}\ \emph
  {et~al.}(2020{\natexlab{a}})\citenamefont {Yoshida}, \citenamefont {Peters},
  \citenamefont {Kawakami},\ and\ \citenamefont
  {Hatsugai}}]{Yoshida_nHReview_PTEP20}%
  \BibitemOpen
  \bibfield  {author} {\bibinfo {author} {\bibfnamefont {T.}~\bibnamefont
  {Yoshida}}, \bibinfo {author} {\bibfnamefont {R.}~\bibnamefont {Peters}},
  \bibinfo {author} {\bibfnamefont {N.}~\bibnamefont {Kawakami}}, \ and\
  \bibinfo {author} {\bibfnamefont {Y.}~\bibnamefont {Hatsugai}},\ }\href@noop
  {} {\bibfield  {journal} {\bibinfo  {journal} {Progress of Theoretical and
  Experimental Physics}\ }\textbf {\bibinfo {volume} {2020}},\ \bibinfo {pages}
  {12A109} (\bibinfo {year} {2020}{\natexlab{a}})}\BibitemShut {NoStop}%
\bibitem [{\citenamefont {Kawabata}\ \emph
  {et~al.}(2019{\natexlab{c}})\citenamefont {Kawabata}, \citenamefont
  {Bessho},\ and\ \citenamefont {Sato}}]{Kawabata_gapless_PRL19}%
  \BibitemOpen
  \bibfield  {author} {\bibinfo {author} {\bibfnamefont {K.}~\bibnamefont
  {Kawabata}}, \bibinfo {author} {\bibfnamefont {T.}~\bibnamefont {Bessho}}, \
  and\ \bibinfo {author} {\bibfnamefont {M.}~\bibnamefont {Sato}},\ }\href
  {\doibase 10.1103/PhysRevLett.123.066405} {\bibfield  {journal} {\bibinfo
  {journal} {Phys. Rev. Lett.}\ }\textbf {\bibinfo {volume} {123}},\ \bibinfo
  {pages} {066405} (\bibinfo {year} {2019}{\natexlab{c}})}\BibitemShut
  {NoStop}%
\bibitem [{\citenamefont {Carlstr\"om}\ \emph {et~al.}(2019)\citenamefont
  {Carlstr\"om}, \citenamefont {St\aa{}lhammar}, \citenamefont {Budich},\ and\
  \citenamefont {Bergholtz}}]{Carlstrom_nHknot_PRB19}%
  \BibitemOpen
  \bibfield  {author} {\bibinfo {author} {\bibfnamefont {J.}~\bibnamefont
  {Carlstr\"om}}, \bibinfo {author} {\bibfnamefont {M.}~\bibnamefont
  {St\aa{}lhammar}}, \bibinfo {author} {\bibfnamefont {J.~C.}\ \bibnamefont
  {Budich}}, \ and\ \bibinfo {author} {\bibfnamefont {E.~J.}\ \bibnamefont
  {Bergholtz}},\ }\href {\doibase 10.1103/PhysRevB.99.161115} {\bibfield
  {journal} {\bibinfo  {journal} {Phys. Rev. B}\ }\textbf {\bibinfo {volume}
  {99}},\ \bibinfo {pages} {161115} (\bibinfo {year} {2019})}\BibitemShut
  {NoStop}%
\bibitem [{\citenamefont {Xu}\ \emph {et~al.}(2017)\citenamefont {Xu},
  \citenamefont {Wang},\ and\ \citenamefont
  {Duan}}]{YXuPRL17_exceptional_ring}%
  \BibitemOpen
  \bibfield  {author} {\bibinfo {author} {\bibfnamefont {Y.}~\bibnamefont
  {Xu}}, \bibinfo {author} {\bibfnamefont {S.-T.}\ \bibnamefont {Wang}}, \ and\
  \bibinfo {author} {\bibfnamefont {L.-M.}\ \bibnamefont {Duan}},\ }\href
  {\doibase 10.1103/PhysRevLett.118.045701} {\bibfield  {journal} {\bibinfo
  {journal} {Phys. Rev. Lett.}\ }\textbf {\bibinfo {volume} {118}},\ \bibinfo
  {pages} {045701} (\bibinfo {year} {2017})}\BibitemShut {NoStop}%
\bibitem [{\citenamefont {Delplace}\ \emph {et~al.}(2021)\citenamefont
  {Delplace}, \citenamefont {Yoshida},\ and\ \citenamefont
  {Hatsugai}}]{Delplace_Resul_PRL21}%
  \BibitemOpen
  \bibfield  {author} {\bibinfo {author} {\bibfnamefont {P.}~\bibnamefont
  {Delplace}}, \bibinfo {author} {\bibfnamefont {T.}~\bibnamefont {Yoshida}}, \
  and\ \bibinfo {author} {\bibfnamefont {Y.}~\bibnamefont {Hatsugai}},\ }\href
  {\doibase 10.1103/PhysRevLett.127.186602} {\bibfield  {journal} {\bibinfo
  {journal} {Phys. Rev. Lett.}\ }\textbf {\bibinfo {volume} {127}},\ \bibinfo
  {pages} {186602} (\bibinfo {year} {2021})}\BibitemShut {NoStop}%
\bibitem [{\citenamefont {Mandal}\ and\ \citenamefont
  {Bergholtz}(2021)}]{Mandal_EP3_PRL21}%
  \BibitemOpen
  \bibfield  {author} {\bibinfo {author} {\bibfnamefont {I.}~\bibnamefont
  {Mandal}}\ and\ \bibinfo {author} {\bibfnamefont {E.~J.}\ \bibnamefont
  {Bergholtz}},\ }\href {\doibase 10.1103/PhysRevLett.127.186601} {\bibfield
  {journal} {\bibinfo  {journal} {Phys. Rev. Lett.}\ }\textbf {\bibinfo
  {volume} {127}},\ \bibinfo {pages} {186601} (\bibinfo {year}
  {2021})}\BibitemShut {NoStop}%
\bibitem [{\citenamefont {Lee}(2016)}]{TELeePRL16_Half_quantized}%
  \BibitemOpen
  \bibfield  {author} {\bibinfo {author} {\bibfnamefont {T.~E.}\ \bibnamefont
  {Lee}},\ }\href {\doibase 10.1103/PhysRevLett.116.133903} {\bibfield
  {journal} {\bibinfo  {journal} {Phys. Rev. Lett.}\ }\textbf {\bibinfo
  {volume} {116}},\ \bibinfo {pages} {133903} (\bibinfo {year}
  {2016})}\BibitemShut {NoStop}%
\bibitem [{\citenamefont {Martinez~Alvarez}\ \emph {et~al.}(2018)\citenamefont
  {Martinez~Alvarez}, \citenamefont {Barrios~Vargas},\ and\ \citenamefont
  {Foa~Torres}}]{Alvarez_nHSkin_PRB18}%
  \BibitemOpen
  \bibfield  {author} {\bibinfo {author} {\bibfnamefont {V.~M.}\ \bibnamefont
  {Martinez~Alvarez}}, \bibinfo {author} {\bibfnamefont {J.~E.}\ \bibnamefont
  {Barrios~Vargas}}, \ and\ \bibinfo {author} {\bibfnamefont {L.~E.~F.}\
  \bibnamefont {Foa~Torres}},\ }\href {\doibase 10.1103/PhysRevB.97.121401}
  {\bibfield  {journal} {\bibinfo  {journal} {Phys. Rev. B}\ }\textbf {\bibinfo
  {volume} {97}},\ \bibinfo {pages} {121401} (\bibinfo {year}
  {2018})}\BibitemShut {NoStop}%
\bibitem [{\citenamefont {Yao}\ and\ \citenamefont
  {Wang}(2018)}]{SYao_nHSkin-1D_PRL18}%
  \BibitemOpen
  \bibfield  {author} {\bibinfo {author} {\bibfnamefont {S.}~\bibnamefont
  {Yao}}\ and\ \bibinfo {author} {\bibfnamefont {Z.}~\bibnamefont {Wang}},\
  }\href {\doibase 10.1103/PhysRevLett.121.086803} {\bibfield  {journal}
  {\bibinfo  {journal} {Phys. Rev. Lett.}\ }\textbf {\bibinfo {volume} {121}},\
  \bibinfo {pages} {086803} (\bibinfo {year} {2018})}\BibitemShut {NoStop}%
\bibitem [{\citenamefont {Yao}\ \emph {et~al.}(2018)\citenamefont {Yao},
  \citenamefont {Song},\ and\ \citenamefont {Wang}}]{SYao_nHSkin-2D_PRL18}%
  \BibitemOpen
  \bibfield  {author} {\bibinfo {author} {\bibfnamefont {S.}~\bibnamefont
  {Yao}}, \bibinfo {author} {\bibfnamefont {F.}~\bibnamefont {Song}}, \ and\
  \bibinfo {author} {\bibfnamefont {Z.}~\bibnamefont {Wang}},\ }\href {\doibase
  10.1103/PhysRevLett.121.136802} {\bibfield  {journal} {\bibinfo  {journal}
  {Phys. Rev. Lett.}\ }\textbf {\bibinfo {volume} {121}},\ \bibinfo {pages}
  {136802} (\bibinfo {year} {2018})}\BibitemShut {NoStop}%
\bibitem [{\citenamefont {Kunst}\ \emph {et~al.}(2018)\citenamefont {Kunst},
  \citenamefont {Edvardsson}, \citenamefont {Budich},\ and\ \citenamefont
  {Bergholtz}}]{KFlore_nHSkin_PRL18}%
  \BibitemOpen
  \bibfield  {author} {\bibinfo {author} {\bibfnamefont {F.~K.}\ \bibnamefont
  {Kunst}}, \bibinfo {author} {\bibfnamefont {E.}~\bibnamefont {Edvardsson}},
  \bibinfo {author} {\bibfnamefont {J.~C.}\ \bibnamefont {Budich}}, \ and\
  \bibinfo {author} {\bibfnamefont {E.~J.}\ \bibnamefont {Bergholtz}},\ }\href
  {\doibase 10.1103/PhysRevLett.121.026808} {\bibfield  {journal} {\bibinfo
  {journal} {Phys. Rev. Lett.}\ }\textbf {\bibinfo {volume} {121}},\ \bibinfo
  {pages} {026808} (\bibinfo {year} {2018})}\BibitemShut {NoStop}%
\bibitem [{\citenamefont {Edvardsson}\ \emph {et~al.}(2019)\citenamefont
  {Edvardsson}, \citenamefont {Kunst},\ and\ \citenamefont
  {Bergholtz}}]{EEdvardsson_PRBnHSkinHOTI_PRB19}%
  \BibitemOpen
  \bibfield  {author} {\bibinfo {author} {\bibfnamefont {E.}~\bibnamefont
  {Edvardsson}}, \bibinfo {author} {\bibfnamefont {F.~K.}\ \bibnamefont
  {Kunst}}, \ and\ \bibinfo {author} {\bibfnamefont {E.~J.}\ \bibnamefont
  {Bergholtz}},\ }\href {\doibase 10.1103/PhysRevB.99.081302} {\bibfield
  {journal} {\bibinfo  {journal} {Phys. Rev. B}\ }\textbf {\bibinfo {volume}
  {99}},\ \bibinfo {pages} {081302} (\bibinfo {year} {2019})}\BibitemShut
  {NoStop}%
\bibitem [{\citenamefont {Borgnia}\ \emph {et~al.}(2020)\citenamefont
  {Borgnia}, \citenamefont {Kruchkov},\ and\ \citenamefont
  {Slager}}]{Borgnia_ptGapPRL2020}%
  \BibitemOpen
  \bibfield  {author} {\bibinfo {author} {\bibfnamefont {D.~S.}\ \bibnamefont
  {Borgnia}}, \bibinfo {author} {\bibfnamefont {A.~J.}\ \bibnamefont
  {Kruchkov}}, \ and\ \bibinfo {author} {\bibfnamefont {R.-J.}\ \bibnamefont
  {Slager}},\ }\href {\doibase 10.1103/PhysRevLett.124.056802} {\bibfield
  {journal} {\bibinfo  {journal} {Phys. Rev. Lett.}\ }\textbf {\bibinfo
  {volume} {124}},\ \bibinfo {pages} {056802} (\bibinfo {year}
  {2020})}\BibitemShut {NoStop}%
\bibitem [{\citenamefont {Lee}\ and\ \citenamefont
  {Thomale}(2019)}]{Lee_Skin19}%
  \BibitemOpen
  \bibfield  {author} {\bibinfo {author} {\bibfnamefont {C.~H.}\ \bibnamefont
  {Lee}}\ and\ \bibinfo {author} {\bibfnamefont {R.}~\bibnamefont {Thomale}},\
  }\href {\doibase 10.1103/PhysRevB.99.201103} {\bibfield  {journal} {\bibinfo
  {journal} {Phys. Rev. B}\ }\textbf {\bibinfo {volume} {99}},\ \bibinfo
  {pages} {201103} (\bibinfo {year} {2019})}\BibitemShut {NoStop}%
\bibitem [{\citenamefont {Zhang}\ \emph
  {et~al.}(2020{\natexlab{a}})\citenamefont {Zhang}, \citenamefont {Yang},\
  and\ \citenamefont {Fang}}]{Zhang_BECskin19}%
  \BibitemOpen
  \bibfield  {author} {\bibinfo {author} {\bibfnamefont {K.}~\bibnamefont
  {Zhang}}, \bibinfo {author} {\bibfnamefont {Z.}~\bibnamefont {Yang}}, \ and\
  \bibinfo {author} {\bibfnamefont {C.}~\bibnamefont {Fang}},\ }\href@noop {}
  {\bibfield  {journal} {\bibinfo  {journal} {Phys. Rev. Lett.}\ }\textbf
  {\bibinfo {volume} {125}},\ \bibinfo {pages} {126402} (\bibinfo {year}
  {2020}{\natexlab{a}})}\BibitemShut {NoStop}%
\bibitem [{\citenamefont {Okuma}\ \emph {et~al.}(2020)\citenamefont {Okuma},
  \citenamefont {Kawabata}, \citenamefont {Shiozaki},\ and\ \citenamefont
  {Sato}}]{Okuma_BECskin19}%
  \BibitemOpen
  \bibfield  {author} {\bibinfo {author} {\bibfnamefont {N.}~\bibnamefont
  {Okuma}}, \bibinfo {author} {\bibfnamefont {K.}~\bibnamefont {Kawabata}},
  \bibinfo {author} {\bibfnamefont {K.}~\bibnamefont {Shiozaki}}, \ and\
  \bibinfo {author} {\bibfnamefont {M.}~\bibnamefont {Sato}},\ }\href {\doibase
  10.1103/PhysRevLett.124.086801} {\bibfield  {journal} {\bibinfo  {journal}
  {Phys. Rev. Lett.}\ }\textbf {\bibinfo {volume} {124}},\ \bibinfo {pages}
  {086801} (\bibinfo {year} {2020})}\BibitemShut {NoStop}%
\bibitem [{\citenamefont {Okuma}\ and\ \citenamefont
  {Sato}(2023)}]{Okuma_nHSkinReview_AnnRev23}%
  \BibitemOpen
  \bibfield  {author} {\bibinfo {author} {\bibfnamefont {N.}~\bibnamefont
  {Okuma}}\ and\ \bibinfo {author} {\bibfnamefont {M.}~\bibnamefont {Sato}},\
  }\href {\doibase 10.1146/annurev-conmatphys-040521-033133} {\bibfield
  {journal} {\bibinfo  {journal} {Annual Review of Condensed Matter Physics}\
  }\textbf {\bibinfo {volume} {14}},\ \bibinfo {pages} {83} (\bibinfo {year}
  {2023})}\BibitemShut {NoStop}%
\bibitem [{\citenamefont {Xiao}\ \emph {et~al.}(2020)\citenamefont {Xiao},
  \citenamefont {Deng}, \citenamefont {Wang}, \citenamefont {Zhu},
  \citenamefont {Wang}, \citenamefont {Yi},\ and\ \citenamefont
  {Xue}}]{Xiao_nHSkin_Exp_NatPhys19}%
  \BibitemOpen
  \bibfield  {author} {\bibinfo {author} {\bibfnamefont {L.}~\bibnamefont
  {Xiao}}, \bibinfo {author} {\bibfnamefont {T.}~\bibnamefont {Deng}}, \bibinfo
  {author} {\bibfnamefont {K.}~\bibnamefont {Wang}}, \bibinfo {author}
  {\bibfnamefont {G.}~\bibnamefont {Zhu}}, \bibinfo {author} {\bibfnamefont
  {Z.}~\bibnamefont {Wang}}, \bibinfo {author} {\bibfnamefont {W.}~\bibnamefont
  {Yi}}, \ and\ \bibinfo {author} {\bibfnamefont {P.}~\bibnamefont {Xue}},\
  }\href {\doibase 10.1038/s41567-020-0836-6} {\bibfield  {journal} {\bibinfo
  {journal} {Nature Physics}\ }\textbf {\bibinfo {volume} {16}},\ \bibinfo
  {pages} {761} (\bibinfo {year} {2020})}\BibitemShut {NoStop}%
\bibitem [{\citenamefont {Helbig}\ \emph {et~al.}(2020)\citenamefont {Helbig},
  \citenamefont {Hofmann}, \citenamefont {Imhof}, \citenamefont {Abdelghany},
  \citenamefont {Kiessling}, \citenamefont {Molenkamp}, \citenamefont {Lee},
  \citenamefont {Szameit}, \citenamefont {Greiter},\ and\ \citenamefont
  {Thomale}}]{Helbig_ExpSkin_19}%
  \BibitemOpen
  \bibfield  {author} {\bibinfo {author} {\bibfnamefont {T.}~\bibnamefont
  {Helbig}}, \bibinfo {author} {\bibfnamefont {T.}~\bibnamefont {Hofmann}},
  \bibinfo {author} {\bibfnamefont {S.}~\bibnamefont {Imhof}}, \bibinfo
  {author} {\bibfnamefont {M.}~\bibnamefont {Abdelghany}}, \bibinfo {author}
  {\bibfnamefont {T.}~\bibnamefont {Kiessling}}, \bibinfo {author}
  {\bibfnamefont {L.~W.}\ \bibnamefont {Molenkamp}}, \bibinfo {author}
  {\bibfnamefont {C.~H.}\ \bibnamefont {Lee}}, \bibinfo {author} {\bibfnamefont
  {A.}~\bibnamefont {Szameit}}, \bibinfo {author} {\bibfnamefont
  {M.}~\bibnamefont {Greiter}}, \ and\ \bibinfo {author} {\bibfnamefont
  {R.}~\bibnamefont {Thomale}},\ }\href {\doibase 10.1038/s41567-020-0922-9}
  {\bibfield  {journal} {\bibinfo  {journal} {Nature Physics}\ }\textbf
  {\bibinfo {volume} {16}},\ \bibinfo {pages} {747} (\bibinfo {year}
  {2020})}\BibitemShut {NoStop}%
\bibitem [{\citenamefont {Liang}\ \emph {et~al.}(2022)\citenamefont {Liang},
  \citenamefont {Xie}, \citenamefont {Dong}, \citenamefont {Li}, \citenamefont
  {Li}, \citenamefont {Gadway}, \citenamefont {Yi},\ and\ \citenamefont
  {Yan}}]{Liang_nHSkinColdAtom_PRL22}%
  \BibitemOpen
  \bibfield  {author} {\bibinfo {author} {\bibfnamefont {Q.}~\bibnamefont
  {Liang}}, \bibinfo {author} {\bibfnamefont {D.}~\bibnamefont {Xie}}, \bibinfo
  {author} {\bibfnamefont {Z.}~\bibnamefont {Dong}}, \bibinfo {author}
  {\bibfnamefont {H.}~\bibnamefont {Li}}, \bibinfo {author} {\bibfnamefont
  {H.}~\bibnamefont {Li}}, \bibinfo {author} {\bibfnamefont {B.}~\bibnamefont
  {Gadway}}, \bibinfo {author} {\bibfnamefont {W.}~\bibnamefont {Yi}}, \ and\
  \bibinfo {author} {\bibfnamefont {B.}~\bibnamefont {Yan}},\ }\href {\doibase
  10.1103/PhysRevLett.129.070401} {\bibfield  {journal} {\bibinfo  {journal}
  {Phys. Rev. Lett.}\ }\textbf {\bibinfo {volume} {129}},\ \bibinfo {pages}
  {070401} (\bibinfo {year} {2022})}\BibitemShut {NoStop}%
\bibitem [{\citenamefont {Yoshida}\ \emph
  {et~al.}(2020{\natexlab{b}})\citenamefont {Yoshida}, \citenamefont
  {Mizoguchi},\ and\ \citenamefont {Hatsugai}}]{Yoshida_MSkinPRR20}%
  \BibitemOpen
  \bibfield  {author} {\bibinfo {author} {\bibfnamefont {T.}~\bibnamefont
  {Yoshida}}, \bibinfo {author} {\bibfnamefont {T.}~\bibnamefont {Mizoguchi}},
  \ and\ \bibinfo {author} {\bibfnamefont {Y.}~\bibnamefont {Hatsugai}},\
  }\href {\doibase 10.1103/PhysRevResearch.2.022062} {\bibfield  {journal}
  {\bibinfo  {journal} {Phys. Rev. Research}\ }\textbf {\bibinfo {volume}
  {2}},\ \bibinfo {pages} {022062} (\bibinfo {year}
  {2020}{\natexlab{b}})}\BibitemShut {NoStop}%
\bibitem [{\citenamefont {Okugawa}\ \emph {et~al.}(2020)\citenamefont
  {Okugawa}, \citenamefont {Takahashi},\ and\ \citenamefont
  {Yokomizo}}]{Okugawa_HOSkin_PRB20}%
  \BibitemOpen
  \bibfield  {author} {\bibinfo {author} {\bibfnamefont {R.}~\bibnamefont
  {Okugawa}}, \bibinfo {author} {\bibfnamefont {R.}~\bibnamefont {Takahashi}},
  \ and\ \bibinfo {author} {\bibfnamefont {K.}~\bibnamefont {Yokomizo}},\
  }\href {\doibase 10.1103/PhysRevB.102.241202} {\bibfield  {journal} {\bibinfo
   {journal} {Phys. Rev. B}\ }\textbf {\bibinfo {volume} {102}},\ \bibinfo
  {pages} {241202} (\bibinfo {year} {2020})}\BibitemShut {NoStop}%
\bibitem [{\citenamefont {Kawabata}\ \emph
  {et~al.}(2020{\natexlab{b}})\citenamefont {Kawabata}, \citenamefont {Sato},\
  and\ \citenamefont {Shiozaki}}]{Kawabata_HOSkin_PRB20}%
  \BibitemOpen
  \bibfield  {author} {\bibinfo {author} {\bibfnamefont {K.}~\bibnamefont
  {Kawabata}}, \bibinfo {author} {\bibfnamefont {M.}~\bibnamefont {Sato}}, \
  and\ \bibinfo {author} {\bibfnamefont {K.}~\bibnamefont {Shiozaki}},\
  }\href@noop {} {\bibfield  {journal} {\bibinfo  {journal} {Phys. Rev. B}\
  }\textbf {\bibinfo {volume} {102}},\ \bibinfo {pages} {205118} (\bibinfo
  {year} {2020}{\natexlab{b}})}\BibitemShut {NoStop}%
\bibitem [{\citenamefont {Fu}\ and\ \citenamefont
  {Wan}(2020)}]{Fu_HOSkin_arXiv2020}%
  \BibitemOpen
  \bibfield  {author} {\bibinfo {author} {\bibfnamefont {Y.}~\bibnamefont
  {Fu}}\ and\ \bibinfo {author} {\bibfnamefont {S.}~\bibnamefont {Wan}},\
  }\href@noop {} {\bibfield  {journal} {\bibinfo  {journal} {arXiv preprint
  arXiv:2008.09033}\ } (\bibinfo {year} {2020})}\BibitemShut {NoStop}%
\bibitem [{\citenamefont {Okuma}\ and\ \citenamefont
  {Sato}(2019)}]{Okuma_BECpg_PRL19}%
  \BibitemOpen
  \bibfield  {author} {\bibinfo {author} {\bibfnamefont {N.}~\bibnamefont
  {Okuma}}\ and\ \bibinfo {author} {\bibfnamefont {M.}~\bibnamefont {Sato}},\
  }\href {\doibase 10.1103/PhysRevLett.123.097701} {\bibfield  {journal}
  {\bibinfo  {journal} {Phys. Rev. Lett.}\ }\textbf {\bibinfo {volume} {123}},\
  \bibinfo {pages} {097701} (\bibinfo {year} {2019})}\BibitemShut {NoStop}%
\bibitem [{\citenamefont {Song}\ \emph {et~al.}(2019)\citenamefont {Song},
  \citenamefont {Yao},\ and\ \citenamefont {Wang}}]{Song_LSkin_PRL19}%
  \BibitemOpen
  \bibfield  {author} {\bibinfo {author} {\bibfnamefont {F.}~\bibnamefont
  {Song}}, \bibinfo {author} {\bibfnamefont {S.}~\bibnamefont {Yao}}, \ and\
  \bibinfo {author} {\bibfnamefont {Z.}~\bibnamefont {Wang}},\ }\href {\doibase
  10.1103/PhysRevLett.123.170401} {\bibfield  {journal} {\bibinfo  {journal}
  {Phys. Rev. Lett.}\ }\textbf {\bibinfo {volume} {123}},\ \bibinfo {pages}
  {170401} (\bibinfo {year} {2019})}\BibitemShut {NoStop}%
\bibitem [{\citenamefont {Haga}\ \emph {et~al.}(2021)\citenamefont {Haga},
  \citenamefont {Nakagawa}, \citenamefont {Hamazaki},\ and\ \citenamefont
  {Ueda}}]{Haga_LSkin_PRL21}%
  \BibitemOpen
  \bibfield  {author} {\bibinfo {author} {\bibfnamefont {T.}~\bibnamefont
  {Haga}}, \bibinfo {author} {\bibfnamefont {M.}~\bibnamefont {Nakagawa}},
  \bibinfo {author} {\bibfnamefont {R.}~\bibnamefont {Hamazaki}}, \ and\
  \bibinfo {author} {\bibfnamefont {M.}~\bibnamefont {Ueda}},\ }\href {\doibase
  10.1103/PhysRevLett.127.070402} {\bibfield  {journal} {\bibinfo  {journal}
  {Phys. Rev. Lett.}\ }\textbf {\bibinfo {volume} {127}},\ \bibinfo {pages}
  {070402} (\bibinfo {year} {2021})}\BibitemShut {NoStop}%
\bibitem [{\citenamefont {Yang}\ \emph {et~al.}(2022)\citenamefont {Yang},
  \citenamefont {Jiang},\ and\ \citenamefont {Bergholtz}}]{Yang_LSkin_PRR22}%
  \BibitemOpen
  \bibfield  {author} {\bibinfo {author} {\bibfnamefont {F.}~\bibnamefont
  {Yang}}, \bibinfo {author} {\bibfnamefont {Q.-D.}\ \bibnamefont {Jiang}}, \
  and\ \bibinfo {author} {\bibfnamefont {E.~J.}\ \bibnamefont {Bergholtz}},\
  }\href {\doibase 10.1103/PhysRevResearch.4.023160} {\bibfield  {journal}
  {\bibinfo  {journal} {Phys. Rev. Res.}\ }\textbf {\bibinfo {volume} {4}},\
  \bibinfo {pages} {023160} (\bibinfo {year} {2022})}\BibitemShut {NoStop}%
\bibitem [{\citenamefont {Hwang}\ and\ \citenamefont
  {Obuse}(2023)}]{Hwang_SkinJunction_arXiv23}%
  \BibitemOpen
  \bibfield  {author} {\bibinfo {author} {\bibfnamefont {G.}~\bibnamefont
  {Hwang}}\ and\ \bibinfo {author} {\bibfnamefont {H.}~\bibnamefont {Obuse}},\
  }\href@noop {} {\bibfield  {journal} {\bibinfo  {journal} {arXiv preprint
  arXiv:2305.08548}\ } (\bibinfo {year} {2023})}\BibitemShut {NoStop}%
\bibitem [{\citenamefont {Tomita}\ \emph {et~al.}(2017)\citenamefont {Tomita},
  \citenamefont {Nakajima}, \citenamefont {Danshita}, \citenamefont {Takasu},\
  and\ \citenamefont {Takahashi}}]{Tomita_Zeno_SciAdv17}%
  \BibitemOpen
  \bibfield  {author} {\bibinfo {author} {\bibfnamefont {T.}~\bibnamefont
  {Tomita}}, \bibinfo {author} {\bibfnamefont {S.}~\bibnamefont {Nakajima}},
  \bibinfo {author} {\bibfnamefont {I.}~\bibnamefont {Danshita}}, \bibinfo
  {author} {\bibfnamefont {Y.}~\bibnamefont {Takasu}}, \ and\ \bibinfo {author}
  {\bibfnamefont {Y.}~\bibnamefont {Takahashi}},\ }\href@noop {} {\bibfield
  {journal} {\bibinfo  {journal} {Science Advances}\ }\textbf {\bibinfo
  {volume} {3}},\ \bibinfo {pages} {e1701513} (\bibinfo {year}
  {2017})}\BibitemShut {NoStop}%
\bibitem [{\citenamefont {Tomita}\ \emph {et~al.}(2019)\citenamefont {Tomita},
  \citenamefont {Nakajima}, \citenamefont {Takasu},\ and\ \citenamefont
  {Takahashi}}]{Tomita_2BdyLoss_PRA19}%
  \BibitemOpen
  \bibfield  {author} {\bibinfo {author} {\bibfnamefont {T.}~\bibnamefont
  {Tomita}}, \bibinfo {author} {\bibfnamefont {S.}~\bibnamefont {Nakajima}},
  \bibinfo {author} {\bibfnamefont {Y.}~\bibnamefont {Takasu}}, \ and\ \bibinfo
  {author} {\bibfnamefont {Y.}~\bibnamefont {Takahashi}},\ }\href {\doibase
  10.1103/PhysRevA.99.031601} {\bibfield  {journal} {\bibinfo  {journal} {Phys.
  Rev. A}\ }\textbf {\bibinfo {volume} {99}},\ \bibinfo {pages} {031601}
  (\bibinfo {year} {2019})}\BibitemShut {NoStop}%
\bibitem [{\citenamefont {Takasu}\ \emph {et~al.}(2020)\citenamefont {Takasu},
  \citenamefont {Yagami}, \citenamefont {Ashida}, \citenamefont {Hamazaki},
  \citenamefont {Kuno},\ and\ \citenamefont
  {Takahashi}}]{Takasu_nHPTcoldAtom_PTEP2020}%
  \BibitemOpen
  \bibfield  {author} {\bibinfo {author} {\bibfnamefont {Y.}~\bibnamefont
  {Takasu}}, \bibinfo {author} {\bibfnamefont {T.}~\bibnamefont {Yagami}},
  \bibinfo {author} {\bibfnamefont {Y.}~\bibnamefont {Ashida}}, \bibinfo
  {author} {\bibfnamefont {R.}~\bibnamefont {Hamazaki}}, \bibinfo {author}
  {\bibfnamefont {Y.}~\bibnamefont {Kuno}}, \ and\ \bibinfo {author}
  {\bibfnamefont {Y.}~\bibnamefont {Takahashi}},\ }\href@noop {} {\bibfield
  {journal} {\bibinfo  {journal} {Progress of Theoretical and Experimental
  Physics}\ }\textbf {\bibinfo {volume} {2020}},\ \bibinfo {pages} {12A110}
  (\bibinfo {year} {2020})}\BibitemShut {NoStop}%
\bibitem [{\citenamefont {Ma}\ \emph {et~al.}(2019)\citenamefont {Ma},
  \citenamefont {Saxberg}, \citenamefont {Owens}, \citenamefont {Leung},
  \citenamefont {Lu}, \citenamefont {Simon},\ and\ \citenamefont
  {Schuster}}]{Ma_LossQuantumCircuits_Nature2019}%
  \BibitemOpen
  \bibfield  {author} {\bibinfo {author} {\bibfnamefont {R.}~\bibnamefont
  {Ma}}, \bibinfo {author} {\bibfnamefont {B.}~\bibnamefont {Saxberg}},
  \bibinfo {author} {\bibfnamefont {C.}~\bibnamefont {Owens}}, \bibinfo
  {author} {\bibfnamefont {N.}~\bibnamefont {Leung}}, \bibinfo {author}
  {\bibfnamefont {Y.}~\bibnamefont {Lu}}, \bibinfo {author} {\bibfnamefont
  {J.}~\bibnamefont {Simon}}, \ and\ \bibinfo {author} {\bibfnamefont {D.~I.}\
  \bibnamefont {Schuster}},\ }\href {\doibase 10.1038/s41586-019-0897-9}
  {\bibfield  {journal} {\bibinfo  {journal} {Nature}\ }\textbf {\bibinfo
  {volume} {566}},\ \bibinfo {pages} {51} (\bibinfo {year} {2019})}\BibitemShut
  {NoStop}%
\bibitem [{\citenamefont {Yoshida}\ \emph
  {et~al.}(2019{\natexlab{b}})\citenamefont {Yoshida}, \citenamefont {Kudo},\
  and\ \citenamefont {Hatsugai}}]{Yoshida_nHFQH19}%
  \BibitemOpen
  \bibfield  {author} {\bibinfo {author} {\bibfnamefont {T.}~\bibnamefont
  {Yoshida}}, \bibinfo {author} {\bibfnamefont {K.}~\bibnamefont {Kudo}}, \
  and\ \bibinfo {author} {\bibfnamefont {Y.}~\bibnamefont {Hatsugai}},\ }\href
  {\doibase 10.1038/s41598-019-53253-8} {\bibfield  {journal} {\bibinfo
  {journal} {Scientific Reports}\ }\textbf {\bibinfo {volume} {9}},\ \bibinfo
  {pages} {16895} (\bibinfo {year} {2019}{\natexlab{b}})}\BibitemShut {NoStop}%
\bibitem [{\citenamefont {Yoshida}\ \emph
  {et~al.}(2020{\natexlab{c}})\citenamefont {Yoshida}, \citenamefont {Kudo},
  \citenamefont {Katsura},\ and\ \citenamefont
  {Hatsugai}}]{Yoshida_nHFQHJ_PRR20}%
  \BibitemOpen
  \bibfield  {author} {\bibinfo {author} {\bibfnamefont {T.}~\bibnamefont
  {Yoshida}}, \bibinfo {author} {\bibfnamefont {K.}~\bibnamefont {Kudo}},
  \bibinfo {author} {\bibfnamefont {H.}~\bibnamefont {Katsura}}, \ and\
  \bibinfo {author} {\bibfnamefont {Y.}~\bibnamefont {Hatsugai}},\ }\href
  {\doibase 10.1103/PhysRevResearch.2.033428} {\bibfield  {journal} {\bibinfo
  {journal} {Phys. Rev. Research}\ }\textbf {\bibinfo {volume} {2}},\ \bibinfo
  {pages} {033428} (\bibinfo {year} {2020}{\natexlab{c}})}\BibitemShut
  {NoStop}%
\bibitem [{\citenamefont {Guo}\ \emph {et~al.}(2020{\natexlab{a}})\citenamefont
  {Guo}, \citenamefont {Wang}, \citenamefont {Wang},\ and\ \citenamefont
  {Kou}}]{Guo_nHToric_PRB20}%
  \BibitemOpen
  \bibfield  {author} {\bibinfo {author} {\bibfnamefont {C.-X.}\ \bibnamefont
  {Guo}}, \bibinfo {author} {\bibfnamefont {X.-R.}\ \bibnamefont {Wang}},
  \bibinfo {author} {\bibfnamefont {C.}~\bibnamefont {Wang}}, \ and\ \bibinfo
  {author} {\bibfnamefont {S.-P.}\ \bibnamefont {Kou}},\ }\href {\doibase
  10.1103/PhysRevB.101.144439} {\bibfield  {journal} {\bibinfo  {journal}
  {Phys. Rev. B}\ }\textbf {\bibinfo {volume} {101}},\ \bibinfo {pages}
  {144439} (\bibinfo {year} {2020}{\natexlab{a}})}\BibitemShut {NoStop}%
\bibitem [{\citenamefont {Matsumoto}\ \emph {et~al.}(2020)\citenamefont
  {Matsumoto}, \citenamefont {Kawabata}, \citenamefont {Ashida}, \citenamefont
  {Furukawa},\ and\ \citenamefont {Ueda}}]{Matsumoto_nHToric_PRL20}%
  \BibitemOpen
  \bibfield  {author} {\bibinfo {author} {\bibfnamefont {N.}~\bibnamefont
  {Matsumoto}}, \bibinfo {author} {\bibfnamefont {K.}~\bibnamefont {Kawabata}},
  \bibinfo {author} {\bibfnamefont {Y.}~\bibnamefont {Ashida}}, \bibinfo
  {author} {\bibfnamefont {S.}~\bibnamefont {Furukawa}}, \ and\ \bibinfo
  {author} {\bibfnamefont {M.}~\bibnamefont {Ueda}},\ }\href {\doibase
  10.1103/PhysRevLett.125.260601} {\bibfield  {journal} {\bibinfo  {journal}
  {Phys. Rev. Lett.}\ }\textbf {\bibinfo {volume} {125}},\ \bibinfo {pages}
  {260601} (\bibinfo {year} {2020})}\BibitemShut {NoStop}%
\bibitem [{\citenamefont {Zhang}\ \emph
  {et~al.}(2020{\natexlab{b}})\citenamefont {Zhang}, \citenamefont {Xu},
  \citenamefont {Wang},\ and\ \citenamefont {Zhang}}]{Zhang_nHToric_Natcomm20}%
  \BibitemOpen
  \bibfield  {author} {\bibinfo {author} {\bibfnamefont {Q.}~\bibnamefont
  {Zhang}}, \bibinfo {author} {\bibfnamefont {W.-T.}\ \bibnamefont {Xu}},
  \bibinfo {author} {\bibfnamefont {Z.-Q.}\ \bibnamefont {Wang}}, \ and\
  \bibinfo {author} {\bibfnamefont {G.-M.}\ \bibnamefont {Zhang}},\ }\href
  {\doibase 10.1038/s42005-020-00479-y} {\bibfield  {journal} {\bibinfo
  {journal} {Communications Physics}\ }\textbf {\bibinfo {volume} {3}},\
  \bibinfo {pages} {209} (\bibinfo {year} {2020}{\natexlab{b}})}\BibitemShut
  {NoStop}%
\bibitem [{\citenamefont {Guo}\ \emph {et~al.}(2020{\natexlab{b}})\citenamefont
  {Guo}, \citenamefont {Wang},\ and\ \citenamefont
  {Kou}}]{Guo_nHToric_EPL2020}%
  \BibitemOpen
  \bibfield  {author} {\bibinfo {author} {\bibfnamefont {C.-X.}\ \bibnamefont
  {Guo}}, \bibinfo {author} {\bibfnamefont {X.-R.}\ \bibnamefont {Wang}}, \
  and\ \bibinfo {author} {\bibfnamefont {S.-P.}\ \bibnamefont {Kou}},\
  }\href@noop {} {\bibfield  {journal} {\bibinfo  {journal} {{EPL} (Europhysics
  Letters)}\ }\textbf {\bibinfo {volume} {131}},\ \bibinfo {pages} {27002}
  (\bibinfo {year} {2020}{\natexlab{b}})}\BibitemShut {NoStop}%
\bibitem [{\citenamefont {Shackleton}\ and\ \citenamefont
  {Scheurer}(2020)}]{Shackleton_nHFracton_PRR20}%
  \BibitemOpen
  \bibfield  {author} {\bibinfo {author} {\bibfnamefont {H.}~\bibnamefont
  {Shackleton}}\ and\ \bibinfo {author} {\bibfnamefont {M.~S.}\ \bibnamefont
  {Scheurer}},\ }\href {\doibase 10.1103/PhysRevResearch.2.033022} {\bibfield
  {journal} {\bibinfo  {journal} {Phys. Rev. Research}\ }\textbf {\bibinfo
  {volume} {2}},\ \bibinfo {pages} {033022} (\bibinfo {year}
  {2020})}\BibitemShut {NoStop}%
\bibitem [{\citenamefont {Yang}\ \emph
  {et~al.}(2021{\natexlab{b}})\citenamefont {Yang}, \citenamefont {Morampudi},\
  and\ \citenamefont {Bergholtz}}]{Yang_EPKitaev_PRL21}%
  \BibitemOpen
  \bibfield  {author} {\bibinfo {author} {\bibfnamefont {K.}~\bibnamefont
  {Yang}}, \bibinfo {author} {\bibfnamefont {S.~C.}\ \bibnamefont {Morampudi}},
  \ and\ \bibinfo {author} {\bibfnamefont {E.~J.}\ \bibnamefont {Bergholtz}},\
  }\href {\doibase 10.1103/PhysRevLett.126.077201} {\bibfield  {journal}
  {\bibinfo  {journal} {Phys. Rev. Lett.}\ }\textbf {\bibinfo {volume} {126}},\
  \bibinfo {pages} {077201} (\bibinfo {year} {2021}{\natexlab{b}})}\BibitemShut
  {NoStop}%
\bibitem [{\citenamefont {Wang}\ \emph {et~al.}(2023)\citenamefont {Wang},
  \citenamefont {Dai}, \citenamefont {Wang},\ and\ \citenamefont
  {Wang}}]{Wang_SteadyToricCode_arXiv23}%
  \BibitemOpen
  \bibfield  {author} {\bibinfo {author} {\bibfnamefont {Z.}~\bibnamefont
  {Wang}}, \bibinfo {author} {\bibfnamefont {X.-D.}\ \bibnamefont {Dai}},
  \bibinfo {author} {\bibfnamefont {H.-R.}\ \bibnamefont {Wang}}, \ and\
  \bibinfo {author} {\bibfnamefont {Z.}~\bibnamefont {Wang}},\ }\href@noop {}
  {\bibfield  {journal} {\bibinfo  {journal} {arXiv preprint arXiv:2306.12482}\
  } (\bibinfo {year} {2023})}\BibitemShut {NoStop}%
\bibitem [{\citenamefont {Zhang}\ \emph
  {et~al.}(2020{\natexlab{c}})\citenamefont {Zhang}, \citenamefont {Chen},
  \citenamefont {Zhang}, \citenamefont {Lang}, \citenamefont {Li},\ and\
  \citenamefont {Zhu}}]{Zhang_nHTMI_PRB20}%
  \BibitemOpen
  \bibfield  {author} {\bibinfo {author} {\bibfnamefont {D.-W.}\ \bibnamefont
  {Zhang}}, \bibinfo {author} {\bibfnamefont {Y.-L.}\ \bibnamefont {Chen}},
  \bibinfo {author} {\bibfnamefont {G.-Q.}\ \bibnamefont {Zhang}}, \bibinfo
  {author} {\bibfnamefont {L.-J.}\ \bibnamefont {Lang}}, \bibinfo {author}
  {\bibfnamefont {Z.}~\bibnamefont {Li}}, \ and\ \bibinfo {author}
  {\bibfnamefont {S.-L.}\ \bibnamefont {Zhu}},\ }\href@noop {} {\bibfield
  {journal} {\bibinfo  {journal} {Phys. Rev. B}\ }\textbf {\bibinfo {volume}
  {101}},\ \bibinfo {pages} {235150} (\bibinfo {year}
  {2020}{\natexlab{c}})}\BibitemShut {NoStop}%
\bibitem [{\citenamefont {Liu}\ \emph {et~al.}(2020)\citenamefont {Liu},
  \citenamefont {He}, \citenamefont {Yoshida}, \citenamefont {Xiang},\ and\
  \citenamefont {Nori}}]{Liu_nHTMI_RPB20}%
  \BibitemOpen
  \bibfield  {author} {\bibinfo {author} {\bibfnamefont {T.}~\bibnamefont
  {Liu}}, \bibinfo {author} {\bibfnamefont {J.~J.}\ \bibnamefont {He}},
  \bibinfo {author} {\bibfnamefont {T.}~\bibnamefont {Yoshida}}, \bibinfo
  {author} {\bibfnamefont {Z.-L.}\ \bibnamefont {Xiang}}, \ and\ \bibinfo
  {author} {\bibfnamefont {F.}~\bibnamefont {Nori}},\ }\href {\doibase
  10.1103/PhysRevB.102.235151} {\bibfield  {journal} {\bibinfo  {journal}
  {Phys. Rev. B}\ }\textbf {\bibinfo {volume} {102}},\ \bibinfo {pages}
  {235151} (\bibinfo {year} {2020})}\BibitemShut {NoStop}%
\bibitem [{\citenamefont {Xu}\ and\ \citenamefont
  {Chen}(2020)}]{Xu_nHBM_PRB20}%
  \BibitemOpen
  \bibfield  {author} {\bibinfo {author} {\bibfnamefont {Z.}~\bibnamefont
  {Xu}}\ and\ \bibinfo {author} {\bibfnamefont {S.}~\bibnamefont {Chen}},\
  }\href@noop {} {\bibfield  {journal} {\bibinfo  {journal} {Phys. Rev. B}\
  }\textbf {\bibinfo {volume} {102}},\ \bibinfo {pages} {035153} (\bibinfo
  {year} {2020})}\BibitemShut {NoStop}%
\bibitem [{\citenamefont {Pan}\ \emph {et~al.}(2020)\citenamefont {Pan},
  \citenamefont {Wang}, \citenamefont {Cui},\ and\ \citenamefont
  {Chen}}]{Pan_PTHubb_oQS_PRA20}%
  \BibitemOpen
  \bibfield  {author} {\bibinfo {author} {\bibfnamefont {L.}~\bibnamefont
  {Pan}}, \bibinfo {author} {\bibfnamefont {X.}~\bibnamefont {Wang}}, \bibinfo
  {author} {\bibfnamefont {X.}~\bibnamefont {Cui}}, \ and\ \bibinfo {author}
  {\bibfnamefont {S.}~\bibnamefont {Chen}},\ }\href {\doibase
  10.1103/PhysRevA.102.023306} {\bibfield  {journal} {\bibinfo  {journal}
  {Phys. Rev. A}\ }\textbf {\bibinfo {volume} {102}},\ \bibinfo {pages}
  {023306} (\bibinfo {year} {2020})}\BibitemShut {NoStop}%
\bibitem [{\citenamefont {Xi}\ \emph {et~al.}(2021)\citenamefont {Xi},
  \citenamefont {Zhang}, \citenamefont {Gu},\ and\ \citenamefont
  {Chen}}]{Xi_nHcohomology_SciBull21}%
  \BibitemOpen
  \bibfield  {author} {\bibinfo {author} {\bibfnamefont {W.}~\bibnamefont
  {Xi}}, \bibinfo {author} {\bibfnamefont {Z.-H.}\ \bibnamefont {Zhang}},
  \bibinfo {author} {\bibfnamefont {Z.-C.}\ \bibnamefont {Gu}}, \ and\ \bibinfo
  {author} {\bibfnamefont {W.-Q.}\ \bibnamefont {Chen}},\ }\href {\doibase
  https://doi.org/10.1016/j.scib.2021.04.027} {\bibfield  {journal} {\bibinfo
  {journal} {Science Bulletin}\ }\textbf {\bibinfo {volume} {66}},\ \bibinfo
  {pages} {1731} (\bibinfo {year} {2021})}\BibitemShut {NoStop}%
\bibitem [{\citenamefont {Yoshida}\ and\ \citenamefont
  {Hatsugai}(2021)}]{Yoshida_PtGpZtoZ2_PRB21}%
  \BibitemOpen
  \bibfield  {author} {\bibinfo {author} {\bibfnamefont {T.}~\bibnamefont
  {Yoshida}}\ and\ \bibinfo {author} {\bibfnamefont {Y.}~\bibnamefont
  {Hatsugai}},\ }\href {\doibase 10.1103/PhysRevB.104.075106} {\bibfield
  {journal} {\bibinfo  {journal} {Phys. Rev. B}\ }\textbf {\bibinfo {volume}
  {104}},\ \bibinfo {pages} {075106} (\bibinfo {year} {2021})}\BibitemShut
  {NoStop}%
\bibitem [{\citenamefont {Yoshida}\ and\ \citenamefont
  {Hatsugai}(2022)}]{Yoshida_reduction1Dptgp_PRB22}%
  \BibitemOpen
  \bibfield  {author} {\bibinfo {author} {\bibfnamefont {T.}~\bibnamefont
  {Yoshida}}\ and\ \bibinfo {author} {\bibfnamefont {Y.}~\bibnamefont
  {Hatsugai}},\ }\href {\doibase 10.1103/PhysRevB.106.205147} {\bibfield
  {journal} {\bibinfo  {journal} {Phys. Rev. B}\ }\textbf {\bibinfo {volume}
  {106}},\ \bibinfo {pages} {205147} (\bibinfo {year} {2022})}\BibitemShut
  {NoStop}%
\bibitem [{\citenamefont {Yoshida}\ and\ \citenamefont
  {Hatsugai}(2023)}]{Yoshida_reductionEP_PRB23}%
  \BibitemOpen
  \bibfield  {author} {\bibinfo {author} {\bibfnamefont {T.}~\bibnamefont
  {Yoshida}}\ and\ \bibinfo {author} {\bibfnamefont {Y.}~\bibnamefont
  {Hatsugai}},\ }\href {\doibase 10.1103/PhysRevB.107.075118} {\bibfield
  {journal} {\bibinfo  {journal} {Phys. Rev. B}\ }\textbf {\bibinfo {volume}
  {107}},\ \bibinfo {pages} {075118} (\bibinfo {year} {2023})}\BibitemShut
  {NoStop}%
\bibitem [{\citenamefont {Orito}\ and\ \citenamefont
  {Imura}(2022)}]{Orito_CorrSkin_PRB22}%
  \BibitemOpen
  \bibfield  {author} {\bibinfo {author} {\bibfnamefont {T.}~\bibnamefont
  {Orito}}\ and\ \bibinfo {author} {\bibfnamefont {K.-I.}\ \bibnamefont
  {Imura}},\ }\href {\doibase 10.1103/PhysRevB.105.024303} {\bibfield
  {journal} {\bibinfo  {journal} {Phys. Rev. B}\ }\textbf {\bibinfo {volume}
  {105}},\ \bibinfo {pages} {024303} (\bibinfo {year} {2022})}\BibitemShut
  {NoStop}%
\bibitem [{\citenamefont {Shen}\ and\ \citenamefont
  {Lee}(2022)}]{Shen_CorrSkin_CommPhys22}%
  \BibitemOpen
  \bibfield  {author} {\bibinfo {author} {\bibfnamefont {R.}~\bibnamefont
  {Shen}}\ and\ \bibinfo {author} {\bibfnamefont {C.~H.}\ \bibnamefont {Lee}},\
  }\href {\doibase 10.1038/s42005-022-01015-w} {\bibfield  {journal} {\bibinfo
  {journal} {Communications Physics}\ }\textbf {\bibinfo {volume} {5}},\
  \bibinfo {pages} {238} (\bibinfo {year} {2022})}\BibitemShut {NoStop}%
\bibitem [{\citenamefont {Mu}\ \emph {et~al.}(2020)\citenamefont {Mu},
  \citenamefont {Lee}, \citenamefont {Li},\ and\ \citenamefont
  {Gong}}]{Mu_MbdySkin_PRB20}%
  \BibitemOpen
  \bibfield  {author} {\bibinfo {author} {\bibfnamefont {S.}~\bibnamefont
  {Mu}}, \bibinfo {author} {\bibfnamefont {C.~H.}\ \bibnamefont {Lee}},
  \bibinfo {author} {\bibfnamefont {L.}~\bibnamefont {Li}}, \ and\ \bibinfo
  {author} {\bibfnamefont {J.}~\bibnamefont {Gong}},\ }\href {\doibase
  10.1103/PhysRevB.102.081115} {\bibfield  {journal} {\bibinfo  {journal}
  {Phys. Rev. B}\ }\textbf {\bibinfo {volume} {102}},\ \bibinfo {pages}
  {081115} (\bibinfo {year} {2020})}\BibitemShut {NoStop}%
\bibitem [{\citenamefont {Lee}(2021)}]{Lee_MbdySkin_PRB21}%
  \BibitemOpen
  \bibfield  {author} {\bibinfo {author} {\bibfnamefont {C.~H.}\ \bibnamefont
  {Lee}},\ }\href {\doibase 10.1103/PhysRevB.104.195102} {\bibfield  {journal}
  {\bibinfo  {journal} {Phys. Rev. B}\ }\textbf {\bibinfo {volume} {104}},\
  \bibinfo {pages} {195102} (\bibinfo {year} {2021})}\BibitemShut {NoStop}%
\bibitem [{\citenamefont {Zhang}\ \emph {et~al.}(2022)\citenamefont {Zhang},
  \citenamefont {Denner}, \citenamefont {Bzdu\ifmmode~\check{s}\else
  \v{s}\fi{}ek}, \citenamefont {Sentef},\ and\ \citenamefont
  {Neupert}}]{Zhang_CorrSkin_PRB22}%
  \BibitemOpen
  \bibfield  {author} {\bibinfo {author} {\bibfnamefont {S.-B.}\ \bibnamefont
  {Zhang}}, \bibinfo {author} {\bibfnamefont {M.~M.}\ \bibnamefont {Denner}},
  \bibinfo {author} {\bibfnamefont {T.~c.~v.}\ \bibnamefont
  {Bzdu\ifmmode~\check{s}\else \v{s}\fi{}ek}}, \bibinfo {author} {\bibfnamefont
  {M.~A.}\ \bibnamefont {Sentef}}, \ and\ \bibinfo {author} {\bibfnamefont
  {T.}~\bibnamefont {Neupert}},\ }\href {\doibase 10.1103/PhysRevB.106.L121102}
  {\bibfield  {journal} {\bibinfo  {journal} {Phys. Rev. B}\ }\textbf {\bibinfo
  {volume} {106}},\ \bibinfo {pages} {L121102} (\bibinfo {year}
  {2022})}\BibitemShut {NoStop}%
\bibitem [{\citenamefont {Kawabata}\ \emph {et~al.}(2022)\citenamefont
  {Kawabata}, \citenamefont {Shiozaki},\ and\ \citenamefont
  {Ryu}}]{Kawabata_CorrSkin_PRB22}%
  \BibitemOpen
  \bibfield  {author} {\bibinfo {author} {\bibfnamefont {K.}~\bibnamefont
  {Kawabata}}, \bibinfo {author} {\bibfnamefont {K.}~\bibnamefont {Shiozaki}},
  \ and\ \bibinfo {author} {\bibfnamefont {S.}~\bibnamefont {Ryu}},\ }\href
  {\doibase 10.1103/PhysRevB.105.165137} {\bibfield  {journal} {\bibinfo
  {journal} {Phys. Rev. B}\ }\textbf {\bibinfo {volume} {105}},\ \bibinfo
  {pages} {165137} (\bibinfo {year} {2022})}\BibitemShut {NoStop}%
\bibitem [{\citenamefont {Alsallom}\ \emph {et~al.}(2022)\citenamefont
  {Alsallom}, \citenamefont {Herviou}, \citenamefont {Yazyev},\ and\
  \citenamefont {Brzezi\ifmmode~\acute{n}\else
  \'{n}\fi{}ska}}]{Alsallom_CorrSkin_PRR22}%
  \BibitemOpen
  \bibfield  {author} {\bibinfo {author} {\bibfnamefont {F.}~\bibnamefont
  {Alsallom}}, \bibinfo {author} {\bibfnamefont {L.}~\bibnamefont {Herviou}},
  \bibinfo {author} {\bibfnamefont {O.~V.}\ \bibnamefont {Yazyev}}, \ and\
  \bibinfo {author} {\bibfnamefont {M.}~\bibnamefont
  {Brzezi\ifmmode~\acute{n}\else \'{n}\fi{}ska}},\ }\href {\doibase
  10.1103/PhysRevResearch.4.033122} {\bibfield  {journal} {\bibinfo  {journal}
  {Phys. Rev. Res.}\ }\textbf {\bibinfo {volume} {4}},\ \bibinfo {pages}
  {033122} (\bibinfo {year} {2022})}\BibitemShut {NoStop}%
\bibitem [{\citenamefont {Faugno}\ and\ \citenamefont
  {Ozawa}(2022)}]{Faugno_corrnHSkin_PRL22}%
  \BibitemOpen
  \bibfield  {author} {\bibinfo {author} {\bibfnamefont {W.~N.}\ \bibnamefont
  {Faugno}}\ and\ \bibinfo {author} {\bibfnamefont {T.}~\bibnamefont {Ozawa}},\
  }\href {\doibase 10.1103/PhysRevLett.129.180401} {\bibfield  {journal}
  {\bibinfo  {journal} {Phys. Rev. Lett.}\ }\textbf {\bibinfo {volume} {129}},\
  \bibinfo {pages} {180401} (\bibinfo {year} {2022})}\BibitemShut {NoStop}%
\bibitem [{\citenamefont {Hamanaka}\ \emph {et~al.}(2023)\citenamefont
  {Hamanaka}, \citenamefont {Yamamoto},\ and\ \citenamefont
  {Yoshida}}]{Hamanaka_InteractionLSE_arXiv23}%
  \BibitemOpen
  \bibfield  {author} {\bibinfo {author} {\bibfnamefont {S.}~\bibnamefont
  {Hamanaka}}, \bibinfo {author} {\bibfnamefont {K.}~\bibnamefont {Yamamoto}},
  \ and\ \bibinfo {author} {\bibfnamefont {T.}~\bibnamefont {Yoshida}},\
  }\href@noop {} {\bibfield  {journal} {\bibinfo  {journal} {arXiv preprint
  arXiv:2305.19697}\ } (\bibinfo {year} {2023})}\BibitemShut {NoStop}%
\bibitem [{\citenamefont {Tsubota}\ \emph {et~al.}(2022)\citenamefont
  {Tsubota}, \citenamefont {Yang}, \citenamefont {Akagi},\ and\ \citenamefont
  {Katsura}}]{Tsubota_CorrInv_PRB22}%
  \BibitemOpen
  \bibfield  {author} {\bibinfo {author} {\bibfnamefont {S.}~\bibnamefont
  {Tsubota}}, \bibinfo {author} {\bibfnamefont {H.}~\bibnamefont {Yang}},
  \bibinfo {author} {\bibfnamefont {Y.}~\bibnamefont {Akagi}}, \ and\ \bibinfo
  {author} {\bibfnamefont {H.}~\bibnamefont {Katsura}},\ }\href {\doibase
  10.1103/PhysRevB.105.L201113} {\bibfield  {journal} {\bibinfo  {journal}
  {Phys. Rev. B}\ }\textbf {\bibinfo {volume} {105}},\ \bibinfo {pages}
  {L201113} (\bibinfo {year} {2022})}\BibitemShut {NoStop}%
\bibitem [{\citenamefont {Sayyad}\ and\ \citenamefont
  {Lado}(2023{\natexlab{a}})}]{Sayyad_corr1DKitaev_PRR2023}%
  \BibitemOpen
  \bibfield  {author} {\bibinfo {author} {\bibfnamefont {S.}~\bibnamefont
  {Sayyad}}\ and\ \bibinfo {author} {\bibfnamefont {J.~L.}\ \bibnamefont
  {Lado}},\ }\href {\doibase 10.1103/PhysRevResearch.5.L022046} {\bibfield
  {journal} {\bibinfo  {journal} {Phys. Rev. Res.}\ }\textbf {\bibinfo {volume}
  {5}},\ \bibinfo {pages} {L022046} (\bibinfo {year}
  {2023}{\natexlab{a}})}\BibitemShut {NoStop}%
\bibitem [{\citenamefont {Chen}\ \emph {et~al.}(2023)\citenamefont {Chen},
  \citenamefont {Song},\ and\ \citenamefont {Lado}}]{Chen_nHSpinChainPRL2023}%
  \BibitemOpen
  \bibfield  {author} {\bibinfo {author} {\bibfnamefont {G.}~\bibnamefont
  {Chen}}, \bibinfo {author} {\bibfnamefont {F.}~\bibnamefont {Song}}, \ and\
  \bibinfo {author} {\bibfnamefont {J.~L.}\ \bibnamefont {Lado}},\ }\href
  {\doibase 10.1103/PhysRevLett.130.100401} {\bibfield  {journal} {\bibinfo
  {journal} {Phys. Rev. Lett.}\ }\textbf {\bibinfo {volume} {130}},\ \bibinfo
  {pages} {100401} (\bibinfo {year} {2023})}\BibitemShut {NoStop}%
\bibitem [{\citenamefont {Sayyad}(2023)}]{Sayyad_nHChiral_arXiv2023}%
  \BibitemOpen
  \bibfield  {author} {\bibinfo {author} {\bibfnamefont {S.}~\bibnamefont
  {Sayyad}},\ }\href@noop {} {\bibfield  {journal} {\bibinfo  {journal} {arXiv
  preprint arXiv:2306.14766}\ } (\bibinfo {year} {2023})}\BibitemShut {NoStop}%
\bibitem [{\citenamefont {Sayyad}\ and\ \citenamefont
  {Lado}(2023{\natexlab{b}})}]{Sayyad_Transfe_arXiv2023}%
  \BibitemOpen
  \bibfield  {author} {\bibinfo {author} {\bibfnamefont {S.}~\bibnamefont
  {Sayyad}}\ and\ \bibinfo {author} {\bibfnamefont {J.~L.}\ \bibnamefont
  {Lado}},\ }\href@noop {} {\bibfield  {journal} {\bibinfo  {journal} {arXiv
  preprint arXiv:2309.06303}\ } (\bibinfo {year}
  {2023}{\natexlab{b}})}\BibitemShut {NoStop}%
\bibitem [{pse()}]{pseudo-spin_ftnt}%
  \BibitemOpen
  \href@noop {} {}\bibinfo {note} {To be precise, ``spin" denotes pseudo-spin
  because the system is bosonic. However, for simplicity, we denote it by
  ``spin" \hspace{-2mm}}\BibitemShut {NoStop}%
\bibitem [{Def()}]{DefsOfptGap_ftnt}%
  \BibitemOpen
  \href@noop {} {}\bibinfo {note} {The point-gap of the many-body Hamiltonian
  opens for $\mathrm{det} \left(
  \hat{H}_{[N_\uparrow,N_\downarrow]}(\theta_{\mathrm{s}})-E_{\mathrm{ref}}
  \right)$ \hspace{-2mm}}\BibitemShut {NoStop}%
\bibitem [{stw()}]{stwist_ftnt}%
  \BibitemOpen
  \href@noop {} {}\bibinfo {note} {This is because the Hamiltonian~(\ref{eq:
  H}) is decomposed into two bosonic Hatano-Nelson model where the hopping to
  the right (left) is larger than the other for $\sigma=\uparrow$
  ($\sigma=\downarrow$)\hspace{-2mm}}\BibitemShut {NoStop}%
\bibitem [{sup()}]{supple}%
  \BibitemOpen
  \href@noop {} {}\bibinfo {note} {Supplemental Material for details of the
  perturbation theory, detailed results of the bosonic Hubbard
  model\hspace{-2mm}}\BibitemShut {NoStop}%
\bibitem [{com()}]{complexHN_ftnt}%
  \BibitemOpen
  \href@noop {} {}\bibinfo {note} {It is known that when hopping integrals of
  the Hatano-Nelson model are real, all of the eigenvalues take real values,
  which can be seen by applying an imaginary gauge transformation (see
  Sec.~\ref{sec: 1bdy app} of Supplemental Material~\cite{supple}). Combining
  this fact and realness of $iJ_+$ and $iJ_-$, we can see that all of the
  eigenvalues of the effective spin model~(\ref{eq: Hspin_f simple}) are purely
  imaginary\hspace{-2mm}}\BibitemShut {NoStop}%
\bibitem [{lab()}]{labelE_ftnt}%
  \BibitemOpen
  \href@noop {} {}\bibinfo {note} {The eigenvalues are labeled by $m$
  ($m=0,1,2,\ldots$) so that the following conditions are satisfied:
  $\mathrm{Im} E_{m+1}< \mathrm{Im} E_{m}$; if $\mathrm{Im} E_{m} = \mathrm{Im}
  E_{m+1}$, $m$ satisfy $\mathrm{Re}{E}_{m}\leq \mathrm{Re}{E}_{m+1}$
  \hspace{-2mm}}\BibitemShut {NoStop}%
\bibitem [{\citenamefont {Garbe}\ \emph {et~al.}(2023)\citenamefont {Garbe},
  \citenamefont {Minoguchi}, \citenamefont {Huber},\ and\ \citenamefont
  {Rabl}}]{Garbe_BosonicSkin_arXiv23}%
  \BibitemOpen
  \bibfield  {author} {\bibinfo {author} {\bibfnamefont {L.}~\bibnamefont
  {Garbe}}, \bibinfo {author} {\bibfnamefont {Y.}~\bibnamefont {Minoguchi}},
  \bibinfo {author} {\bibfnamefont {J.}~\bibnamefont {Huber}}, \ and\ \bibinfo
  {author} {\bibfnamefont {P.}~\bibnamefont {Rabl}},\ }\href@noop {} {\bibfield
   {journal} {\bibinfo  {journal} {arXiv preprint arXiv:2301.11339}\ }
  (\bibinfo {year} {2023})}\BibitemShut {NoStop}%
\bibitem [{Nup()}]{Nup2Ndow4_ftnt}%
  \BibitemOpen
  \href@noop {} {}\bibinfo {note} {The Mott skin effect is also observed for
  other subspaces of half-filling $N_{\uparrow}+ N_{\downarrow}=L$
  \hspace{-2mm}}\BibitemShut {NoStop}%
\bibitem [{Ptb()}]{Ptb_param_ftnt}%
  \BibitemOpen
  \href@noop {} {}\bibinfo {note} {The results of eigenvalues are consistent
  with the results of perturbation theory, which can be seen by noting that the
  parameters of $\hat{H}_{\mathrm{spin}}$ take the following values: $
  (J_+,J_-)=(-0.1i,-0.001i) $ and $ J_z(\frac{L}{4}-1)+E_0=-0.028i $ for
  $(t_+,t_-,V,U)=(-1,0.1,100,20)$ and $L=6$ \hspace{-2mm}}\BibitemShut
  {NoStop}%
\bibitem [{nHS()}]{nHSeq_ftnt}%
  \BibitemOpen
  \href@noop {} {}\bibinfo {note} {We note that post-selected real-time
  dynamics of open quantum systems are described by $i\partial_t
  |\Phi(t)\rangle=\hat{H}|\Phi(t)\rangle
  $~\cite{Ashida_nHbHubb_PRA16}\hspace{-2mm}}\BibitemShut {NoStop}%
\bibitem [{osc()}]{osci_ftnt}%
  \BibitemOpen
  \href@noop {} {}\bibinfo {note} {We note that Bloch oscillation is observed
  around the left edge ($j=0$) as illustrated by the arrow in Fig.~\ref{fig:
  TimeEvol OBC}(a). This oscillation is due to the interference between skin
  modes whose time-evolution is given by
  $e^{-i\hat{H}t}|\Phi_m\rangle_{\mathrm{R}} = e^{-iE_m
  t}|\Phi_m\rangle_{\mathrm{R}}$\hspace{-2mm}}\BibitemShut {NoStop}%
\bibitem [{\citenamefont {Lindblad}(1976)}]{Lindblad_CommMathPhys1976}%
  \BibitemOpen
  \bibfield  {author} {\bibinfo {author} {\bibfnamefont {G.}~\bibnamefont
  {Lindblad}},\ }\href {\doibase 10.1007/BF01608499} {\bibfield  {journal}
  {\bibinfo  {journal} {Communications in Mathematical Physics}\ }\textbf
  {\bibinfo {volume} {48}},\ \bibinfo {pages} {119} (\bibinfo {year}
  {1976})}\BibitemShut {NoStop}%
\bibitem [{\citenamefont {Gorini}\ \emph {et~al.}(2008)\citenamefont {Gorini},
  \citenamefont {Kossakowski},\ and\ \citenamefont
  {Sudarshan}}]{Gorini_JMathPhys1976}%
  \BibitemOpen
  \bibfield  {author} {\bibinfo {author} {\bibfnamefont {V.}~\bibnamefont
  {Gorini}}, \bibinfo {author} {\bibfnamefont {A.}~\bibnamefont {Kossakowski}},
  \ and\ \bibinfo {author} {\bibfnamefont {E.~C.~G.}\ \bibnamefont
  {Sudarshan}},\ }\href {\doibase 10.1063/1.522979} {\bibfield  {journal}
  {\bibinfo  {journal} {Journal of Mathematical Physics}\ }\textbf {\bibinfo
  {volume} {17}},\ \bibinfo {pages} {821} (\bibinfo {year} {2008})},\ \Eprint
  {http://arxiv.org/abs/https://pubs.aip.org/aip/jmp/article-pdf/17/5/821/8148306/821\_1\_online.pdf}
  {https://pubs.aip.org/aip/jmp/article-pdf/17/5/821/8148306/821\_1\_online.pdf}
  \BibitemShut {NoStop}%
\bibitem [{\citenamefont {Breuer}\ and\ \citenamefont
  {Petruccione}(2002)}]{Breuer_textbook2007}%
  \BibitemOpen
  \bibfield  {author} {\bibinfo {author} {\bibfnamefont {H.-P.}\ \bibnamefont
  {Breuer}}\ and\ \bibinfo {author} {\bibfnamefont {F.}~\bibnamefont
  {Petruccione}},\ }\href@noop {} {\emph {\bibinfo {title} {The theory of open
  quantum systems}}}\ (\bibinfo  {publisher} {Oxford University Press, USA},\
  \bibinfo {year} {2002})\BibitemShut {NoStop}%
\bibitem [{\citenamefont {Kim}\ \emph {et~al.}(2023)\citenamefont {Kim},
  \citenamefont {Han},\ and\ \citenamefont
  {Park}}]{Kim_CollectiveSkin_arXiv23}%
  \BibitemOpen
  \bibfield  {author} {\bibinfo {author} {\bibfnamefont {B.~H.}\ \bibnamefont
  {Kim}}, \bibinfo {author} {\bibfnamefont {J.-H.}\ \bibnamefont {Han}}, \ and\
  \bibinfo {author} {\bibfnamefont {M.~J.}\ \bibnamefont {Park}},\ }\href@noop
  {} {\bibfield  {journal} {\bibinfo  {journal} {arXiv preprint
  arXiv:2309.07894}\ } (\bibinfo {year} {2023})}\BibitemShut {NoStop}%
\end{thebibliography}
\end{document}